\pdfoutput=1
\documentclass[aps,prd,amsmath,floats,floatfix, twocolumn,
superscriptaddress,nofootinbib,showpacs,longbibliography]{revtex4-1}

\usepackage[T1]{fontenc}
\usepackage[utf8]{inputenc}
\usepackage{lmodern}

\usepackage{verbatim}

\usepackage[dvipsnames, usenames]{xcolor}
\definecolor{linkcolor}{rgb}{0.0,0.3,0.5}
\usepackage[hypertexnames=false, unicode, colorlinks=true, linkcolor=linkcolor,
citecolor=linkcolor, filecolor=linkcolor,urlcolor=linkcolor,
pdfusetitle]{hyperref}

\usepackage[all]{hypcap}
\usepackage{graphicx}
\usepackage{xspace}
\usepackage{amssymb}
\usepackage[normalem]{ulem}
\usepackage{bm}
\usepackage{etoolbox}
\usepackage{booktabs}

\usepackage{microtype}

\usepackage{pbox}

\usepackage[english]{babel}
\usepackage{blindtext}

\graphicspath{%
  {figs/}%
}

\DeclareMathAlphabet{\mathpzc}{OT1}{pzc}{m}{it}

\newtoggle{commentsoff}
\togglefalse{commentsoff}

\ifdefined\nocomments
    \toggletrue{commentsoff}
\fi

\iftoggle{commentsoff}{

    \newcommand{\Note}[1]{}
    \newcommand{\TODO}[1]{}
    \newcommand{\AddCite}[1]{}
}{

    \newcommand{\Note}[1]{\textcolor{blue}{\textbf{[#1]}}}
    \newcommand{\TODO}[1]{\red{TODO: #1}}
    \newcommand{\AddCite}{\red{[Needs citation]}}
}

\newcommand{\red}{\textcolor{red}}

\newcommand{\tref}{t_{\mathrm{ref}}}
\newcommand{\trefmHundredM}{t_{\mathrm{ref}}/M\!=\!-100}
\newcommand{\tmHundredM}{t/M\!=\!-100}
\newcommand{\trefmFourthousandM}{t_{\mathrm{ref}}/M\!=\!-4000}
\newcommand{\fref}{f_{\mathrm{ref}}}
\newcommand{\frefTwentyHz}{f_{\mathrm{ref}}\!=\!20 \, \mathrm{Hz}}
\newcommand{\frefISCO}{M f_{\mathrm{ref}}\!=\!6^{-3/2}/\pi}
\newcommand{\frefISCOLong}{M\fref\!=\!M f_{\text{ISCO}}\!=\!6^{-3/2}/\pi}

\newcommand{\delphi}{\Delta \phi}

\newcommand{\bchi}{\bm{\chi}}
\newcommand{\bL}{\bm{L}}

\newcommand{\blambda}{\bm{\lambda}}

\newcommand{\Li}{\mathcal{L}}

\newcommand{\und}{\!{\tiny {\bf \_}}}
\newcommand{\myspace}{\hspace{1.12cm}}
\newcommand{\colspace}{\hspace{0.6cm}}

\newcommand{\fullspincaption}[1]{
Full spin posteriors at $\trefmHundredM$ and $\frefTwentyHz$ using \NRSur for
the events listed in Tab~\ref{tab:sur_events}.  Set #1 out of 6.  The
lower-triangle subplots show central 90\% and 50\% credible regions of joint 2D
posteriors, while the diagonal subplots show marginalized 1D posteriors.\xspace
}

\newcommand{\NRSur}{\texttt{NRSur7dq4}\xspace}

\newcommand{\PhenomT}{\texttt{IMRPhenomTPHM}\xspace}
\newcommand{\PhenomX}{\texttt{IMRPhenomXPHM}\xspace}
\newcommand{\EOB}{\texttt{SEOBNRv4PHM}\xspace}

\newcommand{\numEv}{31\xspace}

\newcommand{\Cornell}{\affiliation{Cornell Center for Astrophysics
    and Planetary Science, Cornell University, Ithaca, New York 14853, USA}}
\newcommand\CornellPhys{\affiliation{Department of Physics, Cornell
    University, Ithaca, New York 14853, USA}}

\newcommand{\AEI}{\affiliation{Max Planck Institute for Gravitational Physics
    (Albert Einstein Institute), Am M\"uhlenberg 1, Potsdam 14476, Germany}} %

\newcommand\MIT{\affiliation{LIGO Laboratory, Massachusetts Institute of
Technology, Cambridge, Massachusetts 02139, USA}}
\newcommand{\MKI}{\affiliation{Department of Physics and Kavli Institute for Astrophysics and Space Research, Massachusetts Institute of Technology, 77 Massachusetts Ave, Cambridge, MA 02139, USA}}
\newcommand{\CCA}{\affiliation{Center for Computational Astrophysics, Flatiron Institute, New York NY 10010, USA}}
\newcommand{\StonyBrook}{\affiliation{Department of Physics and Astronomy, Stony Brook University, Stony Brook NY 11794, USA}}

\begin{document}

\title{Measuring binary black hole orbital-plane spin orientations}

\author{Vijay Varma}
\email{vijay.varma@aei.mpg.de}
\thanks{Klarman fellow; Marie Curie fellow}
\CornellPhys
\Cornell
\AEI

\author{Maximiliano Isi}
\thanks{NHFP Einstein fellow}
\MIT
\MKI

\author{Sylvia Biscoveanu}
\MIT
\MKI

\author{Will M. Farr}
\StonyBrook{}
\CCA{}

\author{Salvatore Vitale}
\MIT
\MKI
\hypersetup{pdfauthor={Varma el al.}}

\date{\today}

\begin{abstract}
Binary black hole spins are among the key observables for gravitational wave
astronomy. Among the spin parameters, their orientations within the orbital
plane, $\phi_1$, $\phi_2$ and $\Delta \phi=\phi_1-\phi_2$, are critical for
understanding the prevalence of the spin-orbit resonances and merger recoils in
binary black holes. Unfortunately, these angles are particularly hard to
measure using current detectors, LIGO and Virgo. Because the spin directions
are not constant for precessing binaries, the traditional approach is to
measure the spin components at some reference stage in the waveform evolution,
typically the point at which the frequency of the detected signal reaches 20
Hz. However, we find that this is a poor choice for the orbital-plane spin
angle measurements. Instead, we propose measuring the spins at a fixed
\emph{dimensionless} time or frequency near the merger.  This leads to
significantly improved measurements for $\phi_1$ and $\phi_2$ for several
gravitational wave events. Furthermore, using numerical relativity injections,
we demonstrate that $\Delta \phi$ will also be better measured near the merger
for louder signals expected in the future. Finally, we show that numerical
relativity surrogate models are key for reliably measuring the orbital-plane
spin orientations, even at moderate signal-to-noise ratios like $\sim 30-45$.
\end{abstract}

\maketitle

\section{Introduction.}

\begin{figure}[thb]
\includegraphics[width=0.3\textwidth]{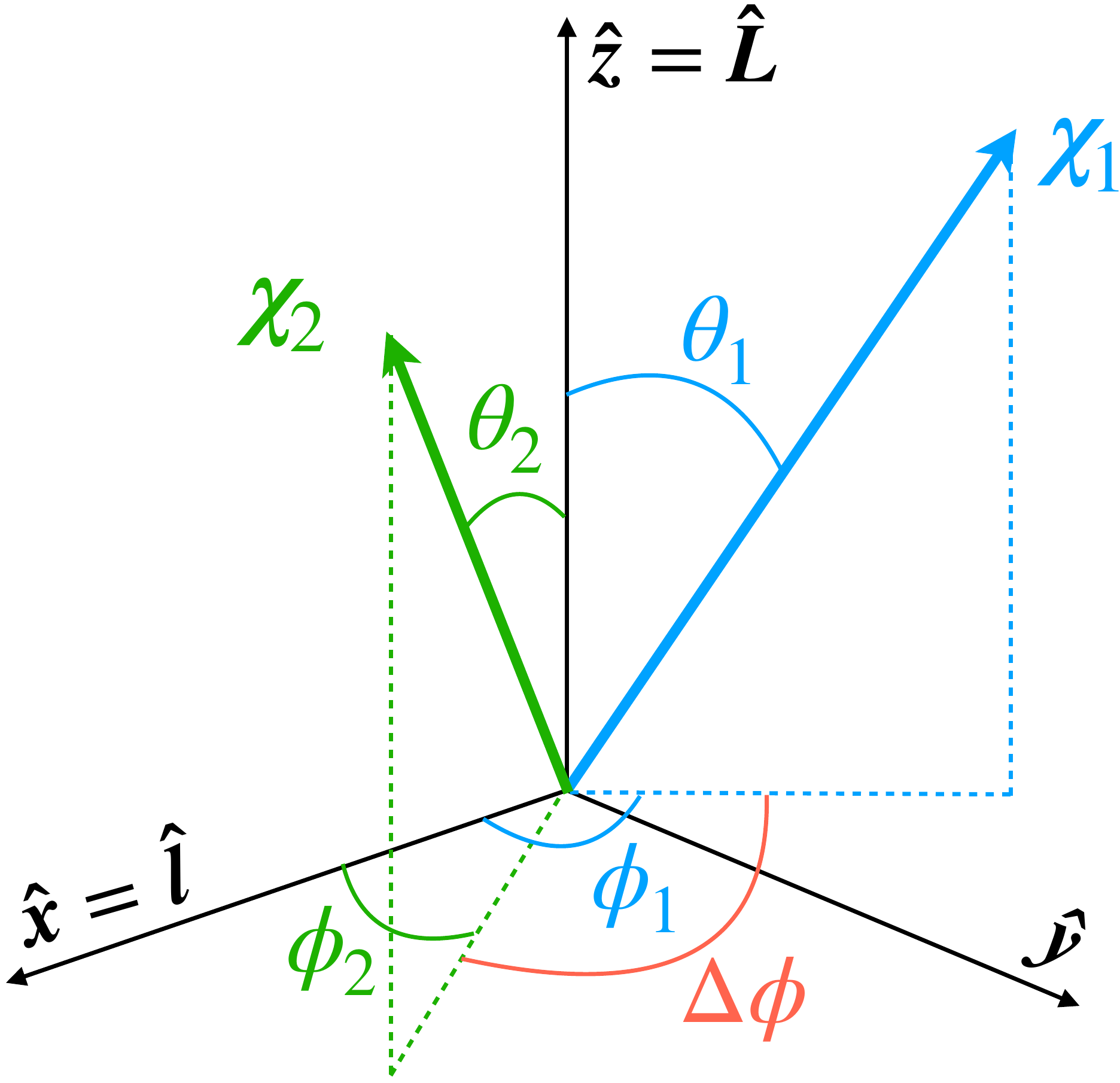}
\caption{
The binary BH spin parameters. The spins are represented by 3-vectors $\bchi_1$
and $\bchi_2$, with index 1 (2) denoting the heavier (lighter) BH.  It is
convenient to parameterize the spins by their dimensionless magnitudes
$\chi_1,\chi_2\leq1$, tilts $\theta_1,\theta_2$ w.r.t the orbital angular
momentum $\bL$~\cite{spinanglespaperfottnoteL}, and orbital-plane spin angles
$\phi_1,\phi_2$ w.r.t the line of separation $\bm{l}$ from the lighter to the
heavier BH. Finally, $\delphi=\phi_1-\phi_2$.
}
\label{fig:spin_frames}
\end{figure}

Binary black hole (BH) spins leave characteristic imprints on the
gravitational-wave (GW) signals observed by
LIGO~\cite{TheLIGOScientific:2014jea} and Virgo~\cite{TheVirgo:2014hva}.
Measuring the spin parameters (illustrated in Fig.~\ref{fig:spin_frames}) from
these signals will allow us to identify which astrophysical processes play a
role in the binary evolution. For example, if the spins are tilted with respect
to the orbital angular momentum, spin-orbit and spin-spin coupling cause both
the spins and the orbital plane to precess~\cite{Apostolatos:1994pre,
Kidder:1995zr}. On the other hand, the orbital-plane spin orientations,
$\phi_1$, $\phi_2$ and $\delphi=\phi_1 - \phi_2$, can be used to identify
spin-orbit resonances~\cite{Schnittman:2004vq} and infer merger kick
velocities~\cite{Varma:2020nbm}.

Unfortunately, measuring the individual spin degrees of freedom from GW events
is challenging at current detector sensitivities. This is particularly true for
the orbital-plane spin angles $\phi_1$, $\phi_2$ and
$\delphi$~\cite{Vitale:2014mka, Schmidt:2014iyl, Biscoveanu:2021nvg} (although
see Refs.~\cite{Gerosa:2014kta, Trifiro:2015zda, Afle:2018slw}). For instance,
these measurements are typically not shown in LIGO-Virgo Collaboration (LVC)
publications (e.g.  Ref.~\cite{Abbott:2020niy}) as they are very poorly
constrained. In this paper, we show that this can be significantly improved by
a simple change in the reference point at which the spins are measured.

Because the spin directions are not constant for precessing binaries, spin
measurements are inherently tied to a specific moment in the binary's
evolution. In practice, the spins are measured by varying the spin parameters
of a GW model at a given reference point in the inspiral and matching the
predicted signal to the observed data (cf.~Sec.~\ref{sec:pe_setup}).  The
traditional approach is to measure the spins at the point where the frequency
of the GW signal at the detector reaches a prespecified reference value,
typically $\frefTwentyHz$~\cite{Abbott:2020niy}. This is mainly motivated by
the fact that the sensitivity band of current detectors begins near this
value~\cite{TheLIGOScientific:2014jea, TheVirgo:2014hva}.

\begin{table*}[t]
\centering
\begin{tabular}{|c|}
\hline
GW events with $M\gtrsim 60 M_{\odot}$ \\
\hline
\hspace{0.2cm} \pbox{6.5in}{\raggedright
    \vspace{0.2cm}
    GW150914  \myspace  \colspace GW170729  \myspace \colspace GW170809  \myspace \colspace GW170818 \myspace  \colspace GW170823
    GW190413\und052954  \colspace GW190413\und134308 \colspace GW190421\und213856 \colspace GW190424\und180648 \colspace GW190503\und185404
    GW190513\und205428  \colspace GW190514\und065416 \colspace GW190517\und055101 \colspace GW190519\und153544 \colspace GW190521  \myspace
    GW190521\und074359  \colspace GW190527\und092055 \colspace GW190602\und175927 \colspace GW190620\und030421 \colspace GW190630\und185205
    GW190701\und203306  \colspace GW190706\und222641 \colspace GW190719\und215514 \colspace GW190727\und060333 \colspace GW190731\und140936
    GW190803\und022701  \colspace GW190828\und063405 \colspace GW190909\und114149 \colspace GW190910\und112807 \colspace GW190915\und235702
    GW190929\und012149
    \vspace{0.2cm} } \\
\hline
\end{tabular}
\caption{
The \numEv GW events for which we use the \NRSur model.
}
\label{tab:sur_events}
\end{table*}

We propose a different approach: instead of measuring the spins at a given
signal frequency, we can measure them at a reference point near the merger.
This can be achieved for all binaries by measuring spins at either (i) the
point where the GW frequency reaches a fixed \emph{dimensionless frequency},
$\frefISCOLong$, or (ii) a fixed \emph{dimensionless time}, $\trefmHundredM$
before the GW amplitude reaches its peak, as defined in Eq.~(5) of
Ref.~\cite{Varma:2019csw}. Throughout this paper, we use geometric units with
$G=c=1$, and masses refer to the redshifted, detector-frame values. Also, here
$M=m_1+m_2$ is the total mass of the binary with component masses $m_1 \geq
m_2$, and ISCO stands for the Schwarzschild innermost stable circular
orbit~\cite{Kaplan:1949isco}. Although the ISCO is only well-defined in the
point-particle limit, we follow previous literature (e.g.,
Ref.\cite{Buonanno:2009zt}) in \emph{defining} the binary's ISCO frequency to
be that of an isolated Schwarzschild BH with mass equal to the total binary
mass $M$.  This reference point typically occurs within $\sim 1-4$ GW cycles
before the peak amplitude. Similarly, the reference point of $\trefmHundredM$
typically falls within $\sim 2-4$ GW cycles before the peak amplitude.
Therefore, independent of the binary parameters, both of these choices allow us
to measure the spins near the merger.

Because the spins evolve deterministically, all choices of reference point lead
to the same waveform prediction if correctly specified (i.e., the GW model is
also evaluated using spins evolved to that reference point). Nevertheless, not
all reference points are equivalent for spin measurements in practice. There
are two considerations to take into account when choosing a reference point:
(1) if the reference point falls well outside the sensitive window of the
detector, parameter inference can become inefficient~\cite{Farr:2014qka};
additionally, (2) the waveform itself can be more (less) sensitive to
variations in the spin parameters at some reference point, leading to more
(less) precise constraints on the spins at those reference points. In
particular, the key finding of this paper is that choosing a reference point
near the merger leads to improved constraints on the orbital-plane spin angles.
We show that this is due to the waveform being more sensitive to
parameterspace
variations in these angles near the merger (cf.~Sec.~\ref{sec:varying_tref}).

While the traditional choice of $\frefTwentyHz$ accounts for consideration (1)
above, it is not optimal when it comes to consideration (2)---we find that this
makes it a poor choice for measuring the orbital-plane spin angles. On the
other hand, the reference points that we propose, $\trefmHundredM$ and
$\frefISCO$, satisfy both criteria, and so they provide improved spin
measurements. Furthermore, since binary BHs observed by LIGO-Virgo are expected
to always merge within the instruments' sensitivity band, $\trefmHundredM$ and
$\frefISCO$ tend to fall within the detector bandwidth and are just as
straightforward to interpret as spins measured at $\frefTwentyHz$.

An executive summary of this paper is as follows.  We find that measuring spins
near the merger (at either
$\frefISCO$
or $\trefmHundredM$) leads to a
marked improvement in the constraints of $\phi_1$ and $\phi_2$ (but not
$\delphi$) for several GW signals in the latest GWTC-2 catalog of
events~\cite{Abbott:2020niy, LIGOScientific:2018mvr, GWOSC_paper, GWOSC:GWTC,
GWOSC:GWTC-2} released by the LVC.  Furthermore, we use numerical relativity
(NR) injections to demonstrate that all three angles, including $\delphi$, will
be better measured near the merger for louder signals expected in the future.
Finally, we study how well different waveform models are able to recover the
orbital-plane spin angles from NR injections, and show that NR surrogate models
alone are accurate enough to reliably measure these angles, even for moderate
signal-to-noise ratios (SNRs) like $\sim 30-45$.

The improvement in the constraints obtained by measuring the spins near merger
does not reflect a gain of new information about the source, but rather that
the waveform is more sensitive to variations in the orbital-plane spin angles
at this point.  For instance, we find that evolving the spins measured at
$\frefTwentyHz$ to the reference point $\trefmHundredM$ ($\frefISCO$) gives
results consistent with measuring the spins directly at $\trefmHundredM$
($\frefISCO$). This also means that our method can be applied entirely in
post-processing. Even though no new information is extracted from the data, our
improved constraints can have important implications, providing a better
representation of the measurement. For example, in a companion
paper \citeauthor{Varma:2021xbh}~\cite{Varma:2021xbh}, we use the
GWTC-2 spin measurements form this work to constrain the astrophysical
distributions of the orbital-plane spin angles at $\trefmHundredM$. Some of the
features found in Ref.~\cite{Varma:2021xbh}, such as an unexpected peak in
the $\phi_1$ distribution, are only resolvable when the spins are measured near
the merger.

The rest of the paper is organized as follows. We describe our parameter
estimation setup in Sec.~\ref{sec:pe_setup}. In
Sec.~\ref{sec:spin_measurements}, we discuss the orbital-plane spin angle
measurements for GWTC-2 events. In Sec.~\ref{sec:nr_inj}, we describe our NR
injection study for louder signals as well as comparison of different waveform
models. Finally, in Sec.~\ref{sec:conclusion}, we provide concluding remarks.

\section{Parameter estimation setup}
\label{sec:pe_setup}

We obtain measurements of binary parameters from GW signals using Bayes'
theorem (see Ref.~\cite{Thrane:2019pe} for a review):
\begin{gather}
    p(\blambda|d) \propto \Li(d|\blambda) \, \pi(\blambda),
\label{eq:Bayes_single_orig_prior}
\end{gather}
where $p(\blambda|d)$ is the \emph{posterior} probability distribution of the
binary parameters $\blambda$ given the observed data $d$, $\Li(d|\blambda)$ is
the \emph{likelihood} of the data given $\blambda$, and $\pi(\blambda)$ is the
\emph{prior} probability distribution for $\blambda$. For quasicircular binary
BHs, the full set of binary parameters $\blambda$ is 15
dimensional~\cite{Abbott:2020niy}, and includes the masses and spins of the
component BHs as well as extrinsic properties such as the distance and sky
location. Under the assumption of Gaussian detector noise, the likelihood
$\Li(d|\blambda)$ can be evaluated for any $\blambda$ using a gravitational
waveform model and the observed data stream $d$~\cite{Thrane:2019pe}. A
stochastic sampling algorithm is then used to draw \emph{posterior samples} for
$\blambda$ from $p(\blambda|d)$.

Our main results are obtained using the time-domain NR surrogate waveform model
\NRSur~\cite{Varma:2019csw}. This model accurately reproduces precessing NR
simulations and is currently the most accurate model in its regime of
validity~\cite{Varma:2019csw}.
\NRSur is trained on generically precessing NR simulations with mass ratios
$q\leq4$ and spin magnitudes $\chi_{1},\chi_{2} \leq 0.8$, but can be
extrapolated to $q=6$ and $\chi_{1},\chi_{2}\leq1$~\cite{Varma:2019csw}.
Wherever comparison with NR is possible in the extrapolated region, \NRSur
performs better than alternate models~\cite{Varma:2019csw, Beth:2021_inprep}.

We use the \texttt{Parallel Bilby}~\cite{Smith:2019ucc} parameter estimation
package with the \texttt{dynesty}~\cite{Speagle:2019dynesty} sampler. Following
Ref.~\cite{Abbott:2020niy}, we choose a prior that is uniform in spin
magnitudes (with $0 \leq \chi_{1},\chi_{2}\leq 0.99$) and component masses, and
isotropic in spin orientations, sky location and binary orientation. Our
distance prior is flat-in-comoving-volume~\cite{Romero-Shaw:2020owr,
Abbott:2020niy}. In addition, we place the following constraints:
$12\leq\mathcal{M}\leq400$, $q \leq 6$, and $60\leq M \leq 400$, where
$\mathcal{M}=\frac{(m_1 m_2)^{3/5}}{(m_1+m_2)^{1/5}}$ is the chirp mass, and
$q=m_1/m_2\geq 1$ is the mass ratio. These choices are motivated by the regime
of validity of \NRSur.

In addition to predicting the waveform, \NRSur also predicts the spin and
orbital dynamics by numerically solving a set of ordinary differential
equations (ODEs)~\cite{Varma:2019csw}. The ODE integration can be initialized
at any reference point in the inspiral. The model then evolves the component
spins (and orbital dynamics) both forwards and backwards in time, and uses the
evolved spins for its internal fits. During the inspiral, the ODE is informed
by NR spins and dynamics. However, once the two BHs merge, the individual BH
spins are no longer available in NR~\cite{Boyle:2019kee}. Therefore, starting
at a time $\tmHundredM$ before the peak amplitude, \NRSur switches to
post-Newtonian-inspired equations to evolve the individual BH spins past the
merger-ringdown stage~\cite{Blackman:2017pcm, Varma:2019csw}.  Here, the choice
of $\tmHundredM$ is arbitrary, but once again, designed to be near the merger.
The spins extended past $\tmHundredM$ are not meant to be physical, but rather
a convenient parameterization for the \NRSur internal fits in the
merger-ringdown~\cite{Varma:2019csw}. For this reason, we choose to measure the
spins at $\trefmHundredM$, the closest point to the merger where the spins are
still guaranteed to be physical.

Besides $\trefmHundredM$, we measure the spins at $\frefISCOLong$ and
$\frefTwentyHz$ for comparison. Measuring the spins at $\frefISCO$ also has the
same benefits as $\trefmHundredM$, but $\frefISCO$ is more convenient for
frequency-domain models. While \NRSur also allows this, we find that for some
GW events, the ISCO
is reached
at a time after $\trefmHundredM$, which can result in unphysical spins.
Therefore, while we provide some results at $\frefISCO$ to demonstrate its
efficacy, we will use spin measurements at $\trefmHundredM$ for our main
results.

\section{Orbital-plane spin angle measurements}
\label{sec:spin_measurements}

The GWTC-2 catalog~\cite{Abbott:2020niy, LIGOScientific:2018mvr} includes a
total of 46 binary BH events. However, because \NRSur only includes ${\sim}20$
orbits before merger, it can only be applied to events with $M \gtrsim 60
\,M_{\odot}$~\cite{Varma:2019csw} (for a detector start frequency of 20Hz).
This reduces the set of events to \numEv; these are listed in
Tab.~\ref{tab:sur_events}.  All results in this section are obtained using
\NRSur for these events, which we will refer to as the ``\NRSur events'' for
convenience. We provide some results for all 46 events using the \PhenomT
model~\cite{Estelles:2021gvs} in App.~\ref{sec:app_PhenomT_posteriors}.

\subsection{Spin measurements for GW events}
\label{sec:all_phi_posteriors}

\begin{figure*}[thb]
\includegraphics[width=0.8\textwidth]{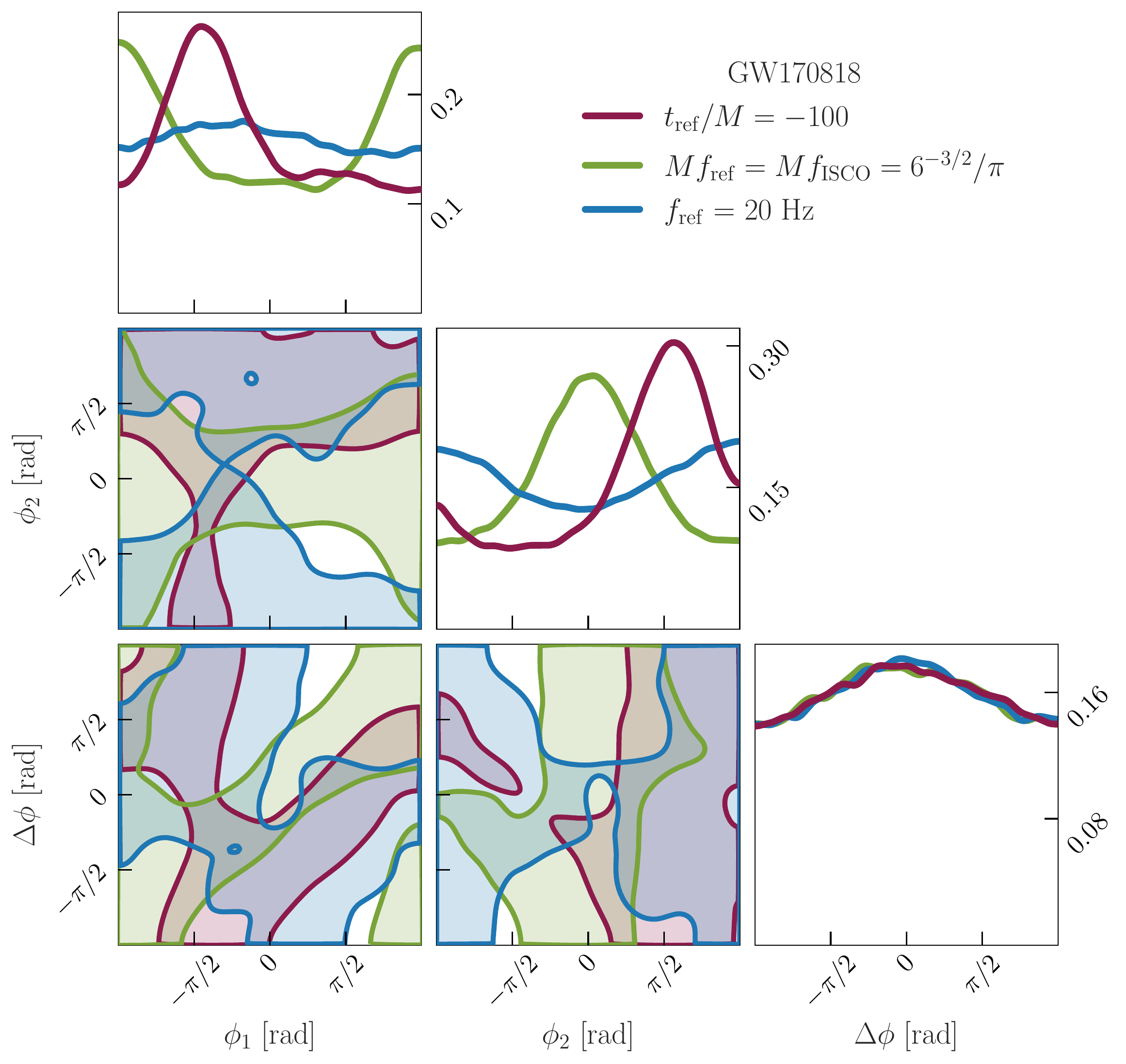}
\caption{
Spin angles $\phi_1$, $\phi_2$ and $\delphi$ for GW170818 using \NRSur at the
three different reference points. The lower-triangle subplots show central 70\%
credible regions of joint 2D posteriors, while the diagonal subplots show
marginalized 1D posteriors. The 1D $\phi_1$ and $\phi_2$ posteriors are more
sharply peaked when measured at $\trefmHundredM$ or $\frefISCO$. Similarly, the
2D posteriors are generally better constrained at $\trefmHundredM$ or
$\frefISCO$, compared to $\frefTwentyHz$.
}
\label{fig:phi_corner_GW170818}
\end{figure*}

\begin{figure*}[p]
\includegraphics[width=\textwidth]{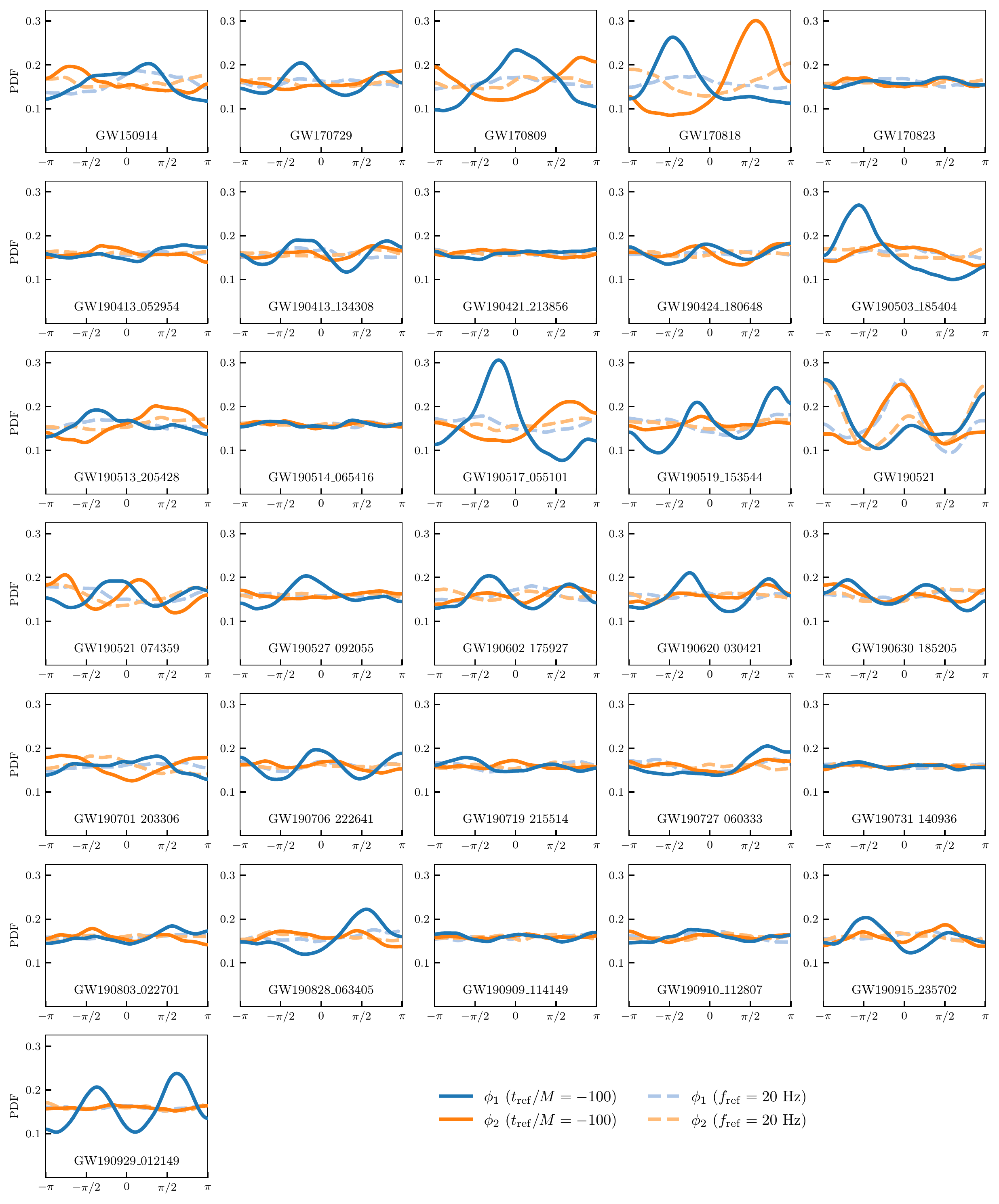}
\caption{
$\phi_1$ and $\phi_2$ posteriors at $\trefmHundredM$ (solid) and
$\frefTwentyHz$ (dashed) for \NRSur. The distributions at $\frefTwentyHz$ are
mostly flat (with the exception of GW190521, which is explained in
Sec.~\ref{sec:all_phi_posteriors}). By contrast, at $\trefmHundredM$, several
cases show a clear deviation from a flat distribution.
}
\label{fig:phi_all_comparison}
\end{figure*}

\begin{figure*}[p]
\includegraphics[width=\textwidth]{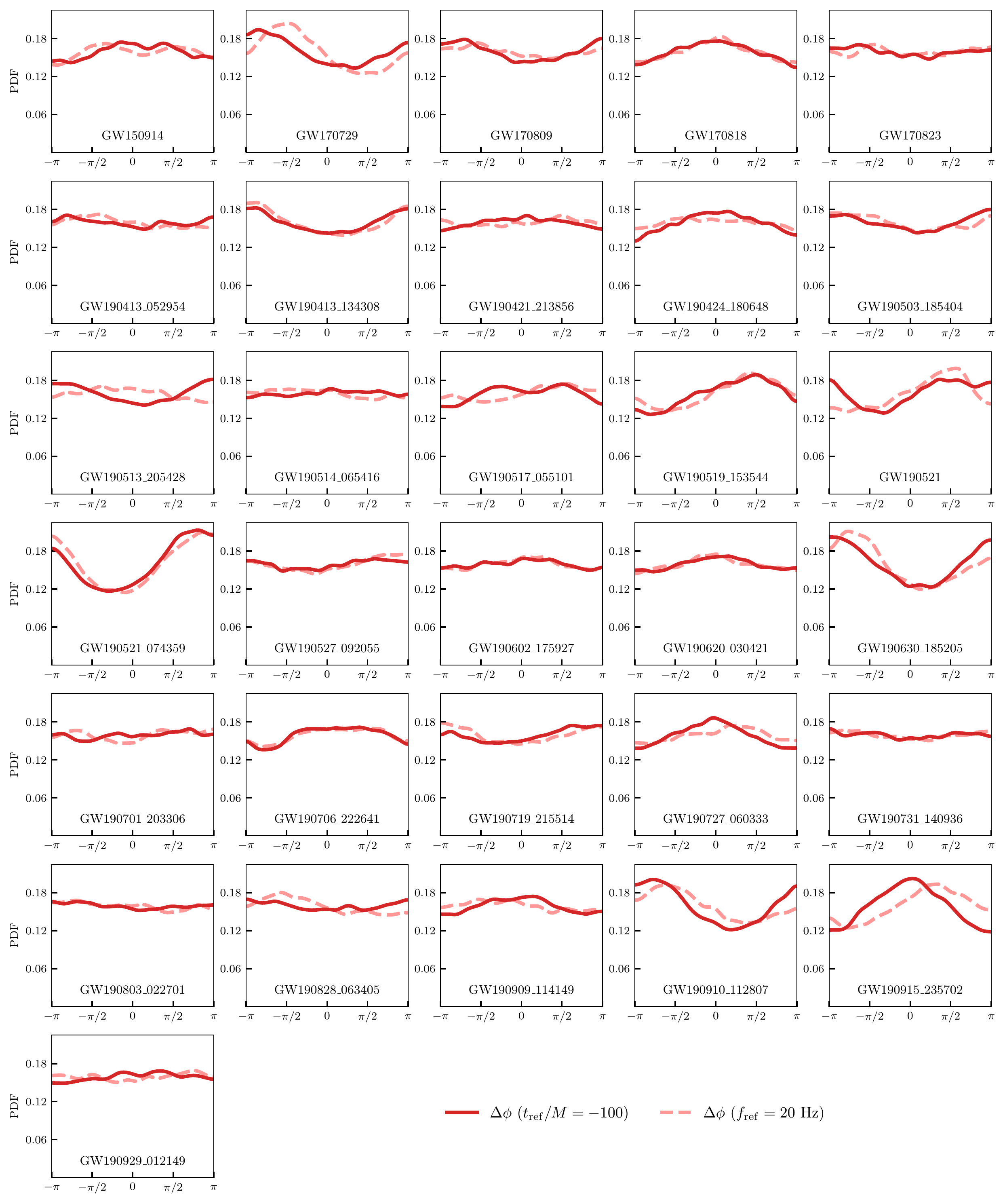}
\caption{
$\delphi$ posteriors at $\trefmHundredM$ (solid) and $\frefTwentyHz$ (dashed)
for \NRSur. Even at $\trefmHundredM$, $\delphi$ is less well-measured than
$\phi_1$ or $\phi_2$ (cf. Fig.~\ref{fig:phi_all_comparison}). In fact,
$\delphi$ measurements are comparable at $\trefmHundredM$ and $\frefTwentyHz$.
}
\label{fig:phi_all_comparison_delphi}
\end{figure*}

We first consider GW170818~\cite{LIGOScientific:2018mvr}, the event for which
we see the greatest improvement when the spins are measured near the merger.
Figure~\ref{fig:phi_corner_GW170818} shows the $\phi_1$, $\phi_2$ and $\delphi$
measurements for GW170818 at the three different reference points,
$\trefmHundredM$, $\frefISCO$ and $\frefTwentyHz$.  Here, for the joint 2D
posteriors, we show 70\% contours instead of the more commonly used 90\% and
50\% contours~\cite{Abbott:2020niy}, as we find the 70\% contours represent the
bulk of the probability mass while being more instructive to discuss the
correlations below.

First considering the marginalized 1D distributions in
Fig.~\ref{fig:phi_corner_GW170818}, we find that $\phi_1$ and $\phi_2$ measured
at $\trefmHundredM$ and $\frefISCO$ are significantly better constrained than
those at $\frefTwentyHz$. Note that $\phi_1$ and $\phi_2$ peak in different
regions for the three different reference points, while $\delphi$ is consistent
for each of them. This is because $\phi_1$ and $\phi_2$ change on the orbital
timescale, as they are defined with respect to the line-of-separation (cf.
Fig.~\ref{fig:spin_frames}). On the other hand, $\delphi$ changes only on the
precession timescale, which is longer.  Furthermore, while the peaks of
$\phi_1$ and $\phi_2$ are approximately $\pi$ apart, this does not result in a
$\delphi$ peak near $\pm\pi$. Instead, the $\delphi$ posterior is much broader,
with a mild peak near $0$. This suggests that even for spins measured near the
merger ($\trefmHundredM$ or $\frefISCO$), the data is only informative about
$\phi_1$ or $\phi_2$, but not necessarily both at the same time. In fact, we do
not see any significant improvement in the 1D $\delphi$ posterior near the
merger for this event.

However, examining the 2D distributions in Fig.~\ref{fig:phi_corner_GW170818},
we find that the posteriors for all three combinations of $\phi_1$, $\phi_2$
and $\delphi$ are generally better constrained at $\trefmHundredM$ or
$\frefISCO$ compared to $\frefTwentyHz$. The 2D posteriors also show a
significant amount of correlation between the three angles. The main feature
here is that $\phi_1$ and $\phi_2$ are better measured than $\delphi$, as noted
above. Therefore, the correlations are along vertical and horizontal
directions in the $\phi_1 - \phi_2$ posterior, while they are along diagonal
directions for the $\phi_1 - \delphi$ posterior (and to a lesser extent for the
$\phi_2 - \delphi$ posterior).  We note that similar correlations are absent
for the higher SNR injections shown in Sec.~\ref{sec:nr_inj}. This suggests
that the degeneracies we see in the 2D posteriors of
Fig.~\ref{fig:phi_corner_GW170818} are a function of the SNR, and can be broken
for louder signals.

In the rest of this section, we will focus on 1D marginalized posteriors at
$\trefmHundredM$ and $\frefTwentyHz$ for simplicity.  While GW170818 shows the
biggest improvement in $\phi_1$ and $\phi_2$ when measured at $\trefmHundredM$,
we find that several other events in GWTC-2 also show significant improvements.
Figure~\ref{fig:phi_all_comparison} compares marginalized 1D posteriors for
$\phi_1$ and $\phi_2$ measured at $\frefTwentyHz$ and $\trefmHundredM$ for all
\numEv \NRSur events.  While the measurements at $\frefTwentyHz$ are mostly
uninformative and consistent with a uniform distribution, the measurements at
$\trefmHundredM$ show clear deviations from uniformity for several events.
Interestingly, GW190521~\cite{Abbott:2020tfl} is the only event with a good
measurement at $\frefTwentyHz$. This is explained by the fact that this binary
merges at a low frequency due to its high mass ($M \sim 270 M_{\odot}$),
therefore its $f_{\rm ISCO} \sim 16$ Hz happens to be close to 20 Hz.

\begin{figure*}[p]
\includegraphics[width=\textwidth]{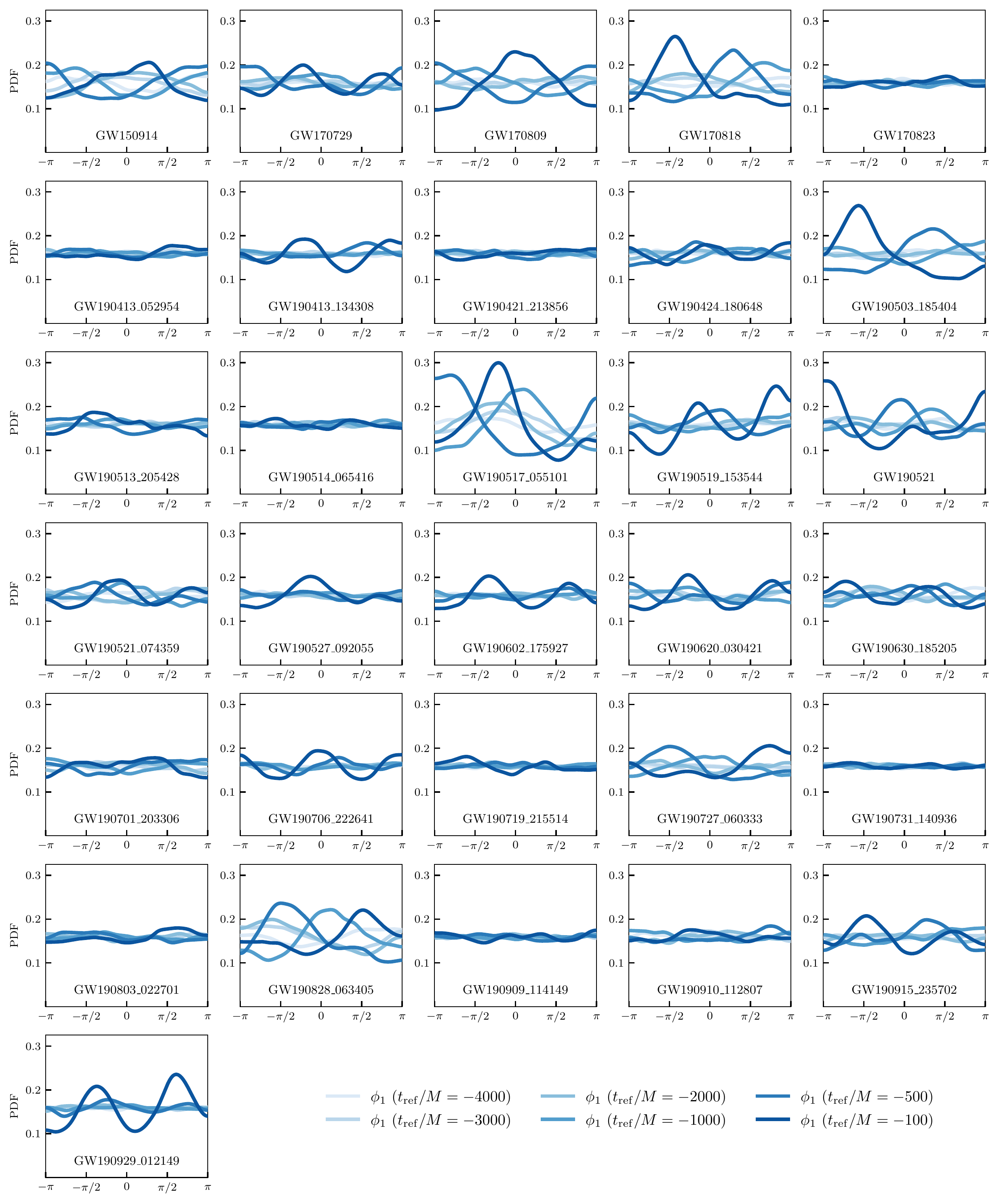}
\caption{
$\phi_1$ posteriors when measured at various different $\tref$ for \NRSur.
Going from $\trefmFourthousandM$ to $\trefmHundredM$, we see a clear
improvement in the $\phi_1$ measurement for several events.
}
\label{fig:phi_all_tref_comparison}
\end{figure*}

\begin{figure*}[p]
\includegraphics[width=\textwidth]{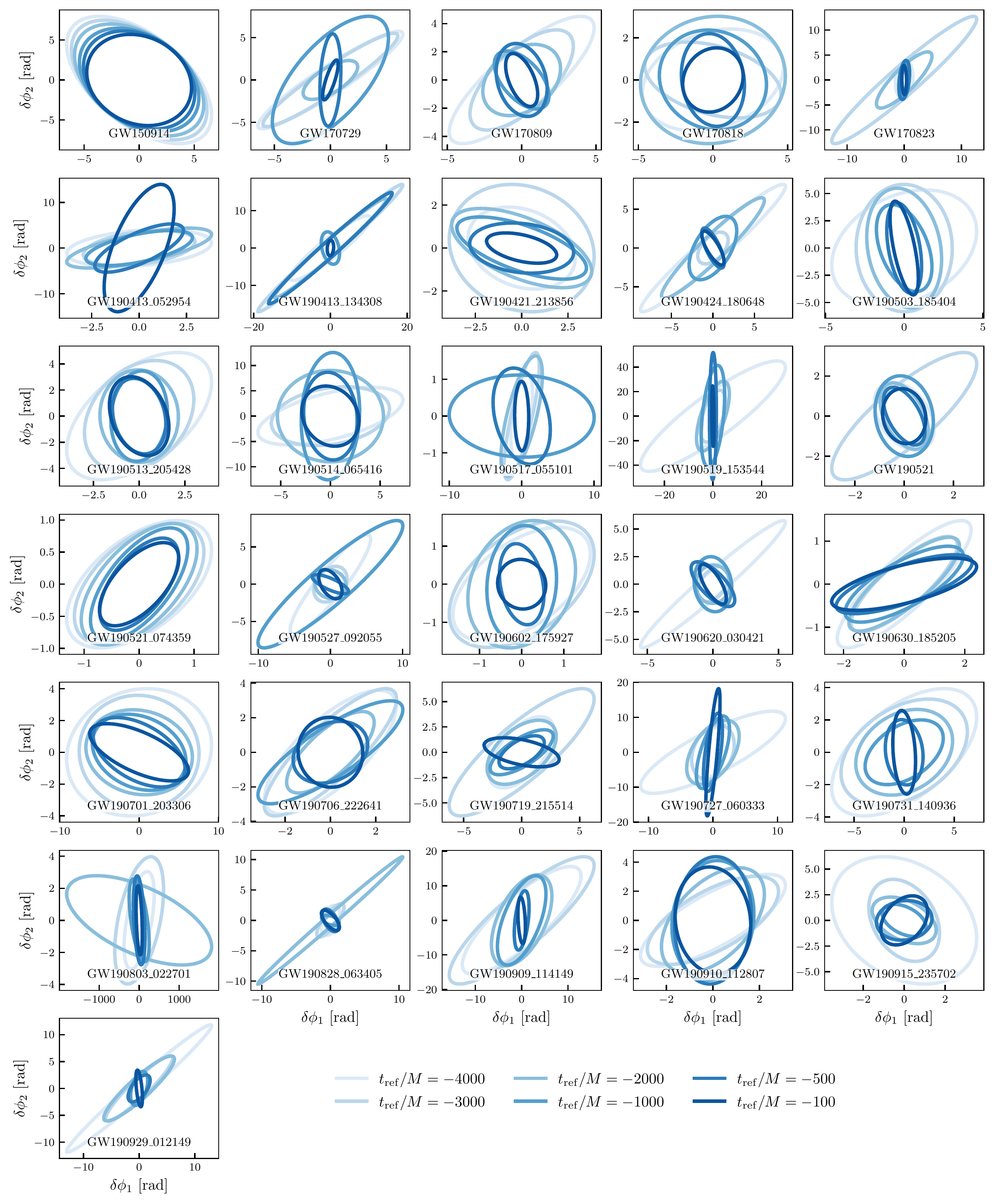}
\caption{
Estimated statistical uncertainty in $\phi_1$ and $\phi_2$ measurements for
\NRSur at the maximum likelihood parameters for the \NRSur events. The
uncertainties are estimated using the Fisher matrix by varying the spins at
various different $\tref$. In almost all cases, the expected uncertainty
decreases noticeably as one approaches $\trefmHundredM$, indicating that the
waveform is more sensitive to changes in $\phi_1$ and $\phi_2$ near merger.
Note that the Fisher matrix method does not place prior bounds on $\phi_1$ and
$\phi_2$ to be within $(-\pi, \pi)$, therefore the statistical biases are not
bound to be $\leq 2 \pi$.
}
\label{fig:fisher_errs}
\end{figure*}

Figure~\ref{fig:phi_all_comparison_delphi} compares 1D $\delphi$ posteriors
measured at $\frefTwentyHz$ and $\trefmHundredM$. Similar to
Fig.~\ref{fig:phi_corner_GW170818}, $\delphi$ is less well-measured than
$\phi_1$ and $\phi_2$, and there is no significant improvement when measuring
the spins at $\trefmHundredM$. However, as we will show in
Sec.~\ref{sec:nr_inj}, we expect this to change with louder signals.

An unambiguous measurement of the orbital-plane spin angles relies on being
able to constrain the spin magnitudes away from zero, and the tilt angles to be
neither 0 nor $\pi$. For the \NRSur events, our measurements of the spin
magnitudes and tilts are consistent with Refs.~\cite{Abbott:2020niy,
LIGOScientific:2018mvr},
and are shown in App.~\ref{sec:app_full_spins}.
Most of these events are consistent with having zero spin magnitudes for both
BHs~\cite{Abbott:2020niy}, but there is evidence for nonzero spin magnitude in
at least some of the events~\cite{Biscoveanu:2020are}. Secondly, even though
there is evidence of precession in the astrophysical binary BH
population~\cite{Abbott:2020gyp}, the individual events are not loud enough to
show clear evidence of precession on their own~\cite{Abbott:2020niy}.  Finally,
the posteriors for the tilt angles do not change significantly between
$\frefTwentyHz$ and $\trefmHundredM$ for these events. As a result, even with
the improvements at $\trefmHundredM$, the 1D posteriors for $\phi_1$, $\phi_2$
and $\delphi$ are still relatively broad (cf.
Figs.~\ref{fig:phi_all_comparison} and \ref{fig:phi_all_comparison_delphi}),
and do not exclude any of the allowed region between $(-\pi, \pi)$.
Nevertheless, these measurements still allow us to place interesting
constraints on the astrophysical distributions of the orbital-plane spin
angles. This is explored in a companion paper, Ref.~\cite{Varma:2021xbh}.

\subsection{Varying the reference point}
\label{sec:varying_tref}

Next, we systematically study the impact of the reference point at which the
spins are measured. Figure~\ref{fig:phi_all_tref_comparison} shows $\phi_1$
measurements at various different reference times for the \numEv \NRSur events.
Rather than repeat the parameter estimation at each $\tref$, we use the \NRSur
spin dynamics~\cite{Varma:2019csw} to evolve the spins backwards from
$\trefmHundredM$ to the earlier times. As noted in the introduction, we find
that this leads to results consistent with measuring spins directly at the new
reference time. The earliest reference time we consider is
$\trefmFourthousandM$, which is near the start of the \NRSur waveform's
validity~\cite{Varma:2019csw}.

In Fig.~\ref{fig:phi_all_tref_comparison}, as we move the reference point from
$\trefmFourthousandM$ to $\trefmHundredM$, we see a clear improvement in the
$\phi_1$ constraint for several events. A possible explanation for this
improvement is that the precession (and orbital) timescale decreases as the
merger approaches, making the waveform more sensitive to the orbital-plane spin
angles near the merger as a result. We provide further justification for this
with a Fisher matrix analysis in the following.

\begin{figure*}[thb]
\includegraphics[width=0.48\textwidth]{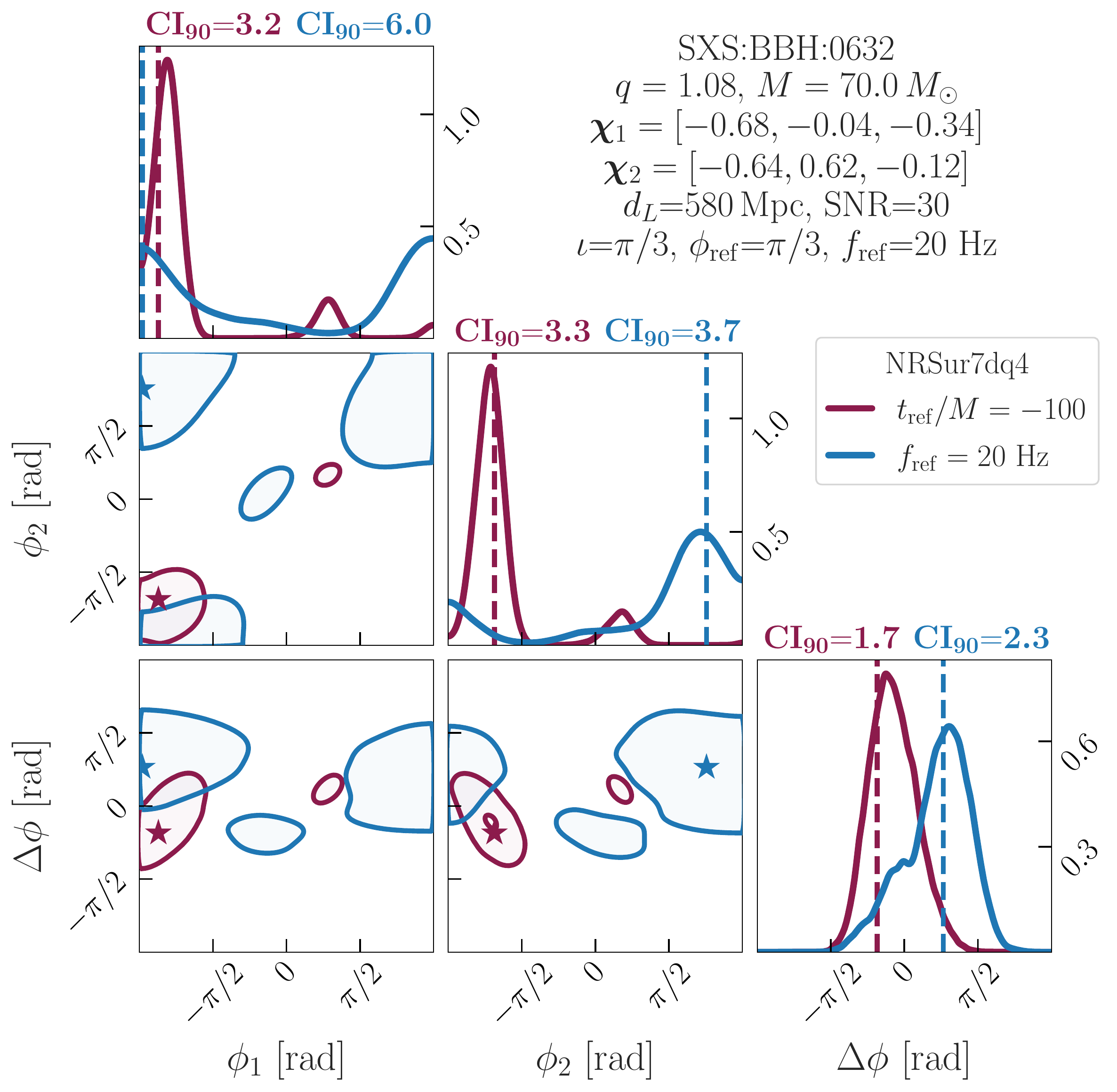} ~~~~~
\includegraphics[width=0.48\textwidth]{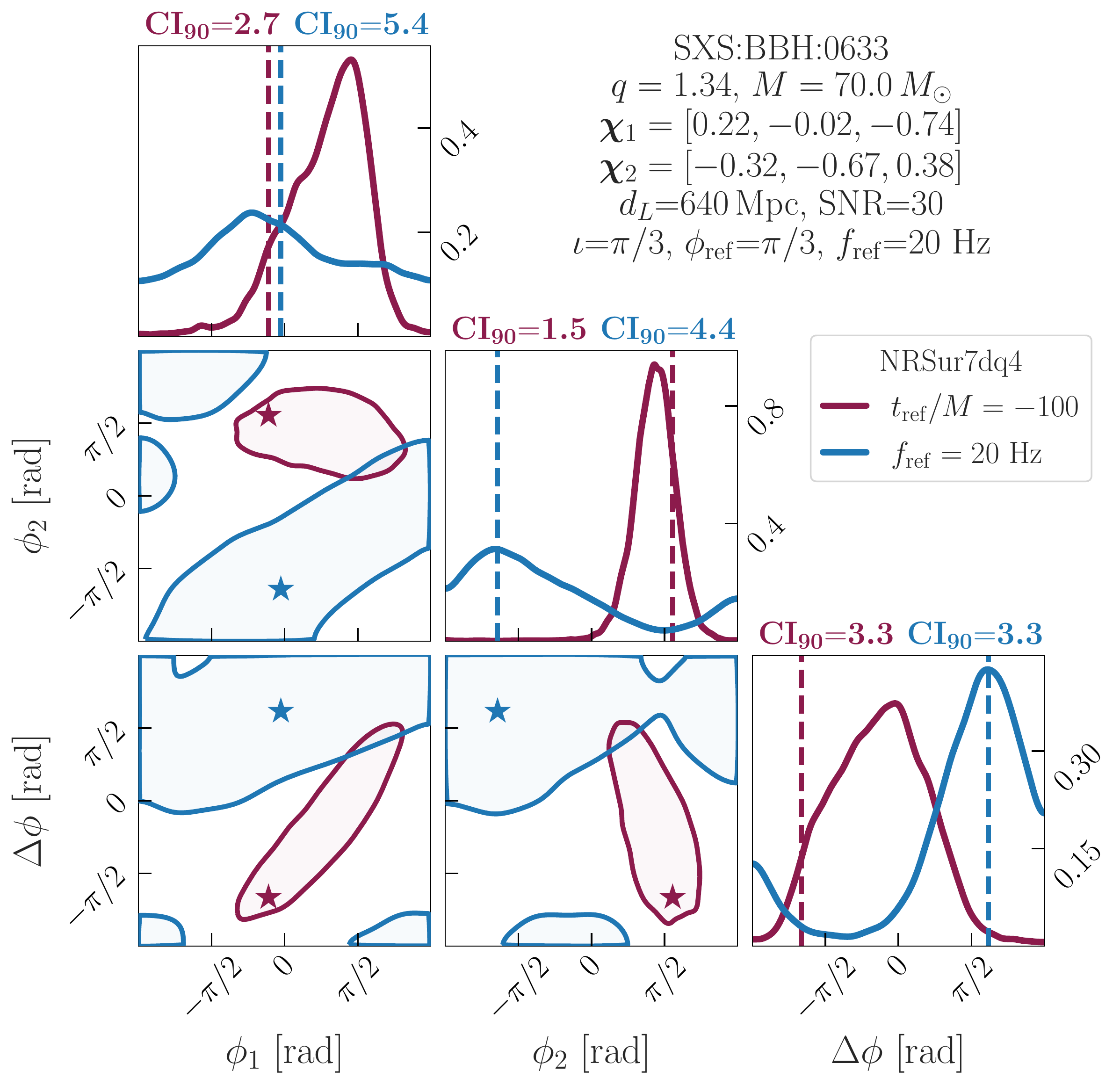}

\vspace{0.5cm}

\includegraphics[width=0.48\textwidth]{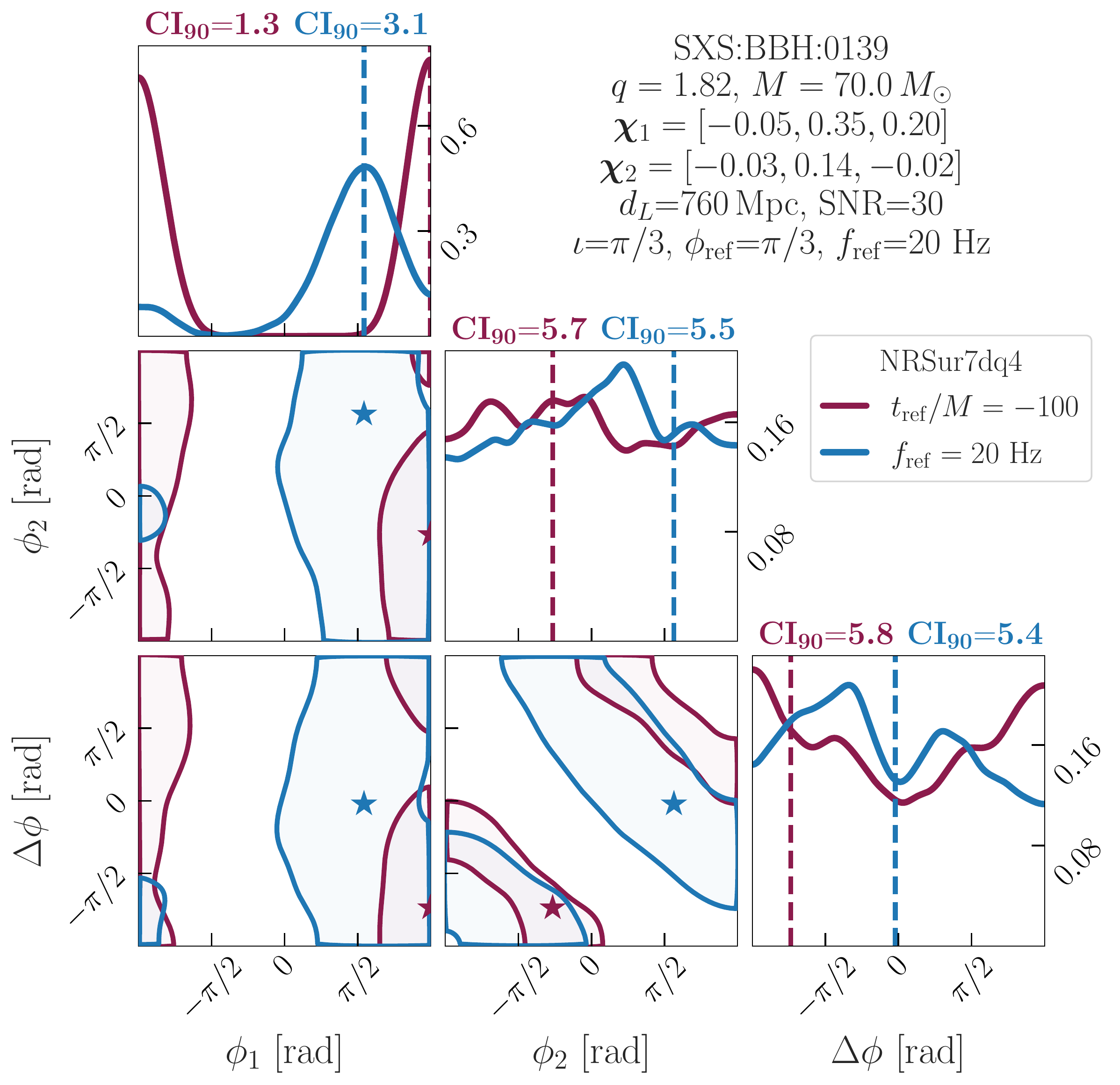} ~~~~~
\includegraphics[width=0.48\textwidth]{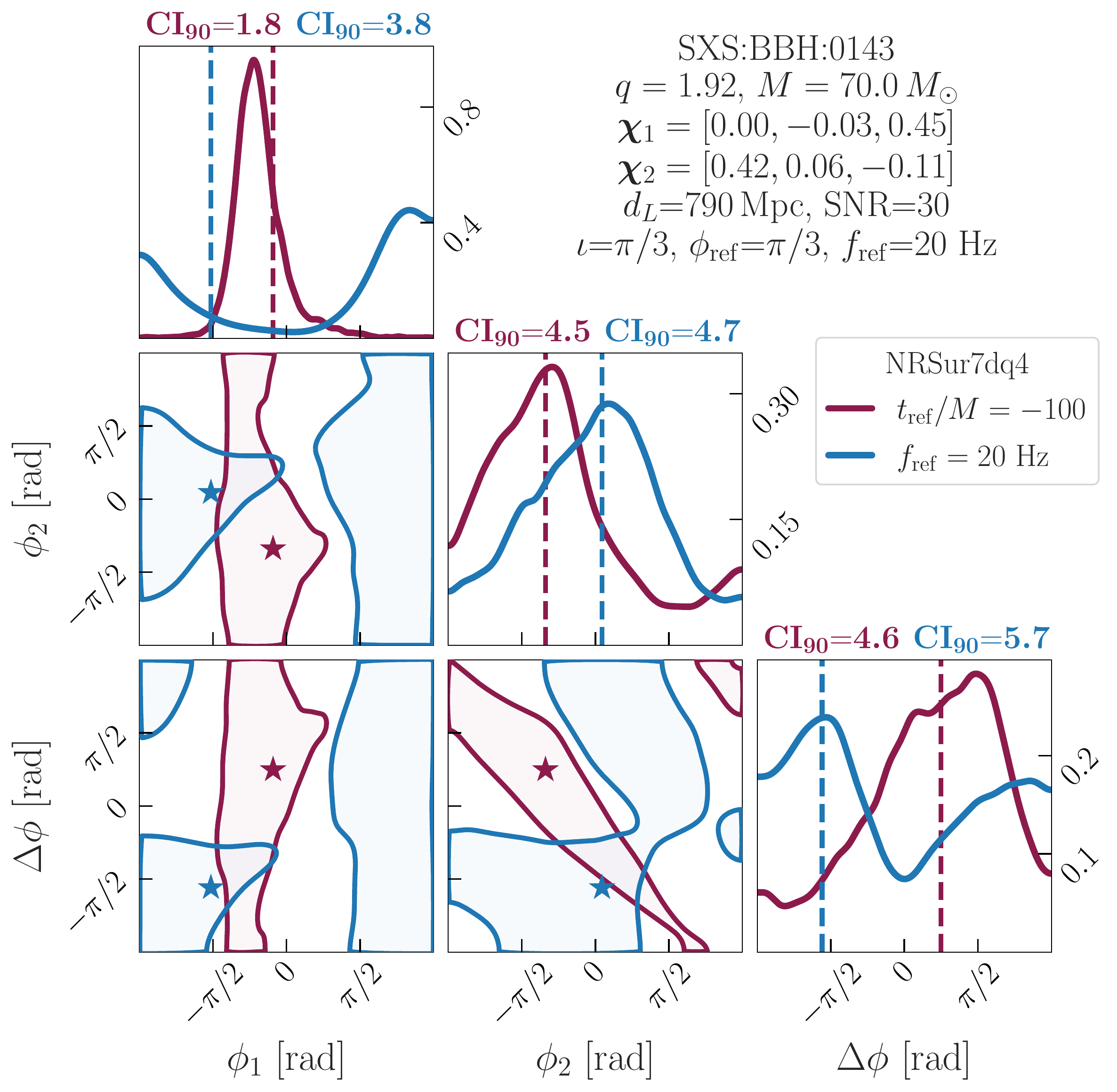}
\caption{
\NRSur posteriors for $\phi_1$, $\phi_2$ and $\delphi$ when measured at
$\trefmHundredM$ and $\frefTwentyHz$ for NR injections at SNR=$30$. In
each panel, the inset text shows the binary parameters for the injections, with
the spins, inclination angle and orbital phase defined at $\frefTwentyHz$. The
shaded regions in the lower-triangle subplots show the central 90\% credible
regions for joint 2D posteriors, with the true value indicated by star markers
(maroon for $\trefmHundredM$ and blue for $\frefTwentyHz$). The diagonal
subplots show marginalized 1D posteriors, with the true values indicated by
vertical dashed lines. The width of the central 90\% credible interval
($\mathrm{CI}_{90}$) for the 1D distributions are shown in text above the
diagonal subplots. All orbital-plane angles, including $\delphi$, are
significantly better measured at $\trefmHundredM$.
}
\label{fig:phi_corner_nr_inj_trefm100_snr30}
\end{figure*}

\subsubsection{Fisher matrix analysis}
As a proxy for the sensitivity of the waveform to the orbital-plane spin
angles, we can look to the Fisher information matrix. The Fisher matrix
provides a simple way to estimate the statistical uncertainty in measuring
binary BH parameters in the high-SNR limit~\cite{Finn:1992wt, Cutler:1994ys};
it is defined as
\begin{align}
\Gamma_{ij} = \left(\left.
    \frac{\partial h}{\partial \lambda^i}\right|
    \frac{\partial h}{\partial \lambda^j}\right),
\label{eq:fisher}
\end{align}
where $h(t)$ is the gravitational waveform with binary parameters
$\blambda=\{\lambda^i\}$ (cf. Sec~\ref{sec:pe_setup}), and the inner product
$(h|g)$ is defined as
\begin{align}
(h|g)=4{\rm Re} \int\frac{\tilde{h}^*(f)\tilde{g}(f)}{S_n(f)}df,
\end{align}
where $\tilde{h}(f)$ indicates the Fourier transform of $h(t)$, $*$ stands for
complex conjugation, and $S_n(f)$ is the one-sided power spectral density for
which we use the LIGO design sensitivity noise
curve~\cite{aLIGODesignNoiseCurve}. We use the \NRSur waveform model for
$h(t)$ and compute the derivatives in Eq.~(\ref{eq:fisher})
numerically~\cite{Ma:2021znq}. Then, using the Cramer-Rao
inequality~\cite{Cramer1999_Fisher, Rao1945_Fisher}, the measurement covariance
matrix ${\rm Var}(\lambda^i,\lambda^j)$ satisfies
\begin{align}
    {\rm Var}(\lambda^i,\lambda^j) \geq \left(\Gamma^{-1}\right)_{ij}\,.
\end{align}

Finally, taking the lower bound of the inequality, the statistical uncertainty
in $\lambda^j$ can be estimated as
\begin{align}
\delta\lambda^i=\sqrt{(\Gamma^{-1})_{ii}} \,.
\end{align}
and the correlation coefficient between $\lambda^i$ and $\lambda^j$ can be
estimated as,
\begin{align}
{\rm Corr}(\lambda^i,\lambda^j)=
    \frac{(\Gamma^{-1})_{ij}}{\sqrt{(\Gamma^{-1})_{ii}(\Gamma^{-1})_{jj}}} \,.
\end{align}

The Fisher matrix method reliably estimates the statistical uncertainty only in
the limit of high SNR (see, e.g., Ref.~\cite{Vallisneri:2007ev} for caveats).
Regardless, here we are not interested in the statistical uncertainty itself
but in quantifying the sensitivity of the waveform to variations in the binary
parameters. The Fisher matrix method is well-suited for this purpose, as we use
its bound on statistical uncertainties merely as a proxy for waveform
sensitivity: smaller $\delta \lambda^j$ indicates that the waveform is more
sensitive to $\lambda^j$.

In particular, we are interested in how sensitive the waveform is to changes in
the orbital-plane spin angles at various values of $\tref$. For this purpose,
we generate the waveform corresponding to the maximum likelihood parameters for
each of the \numEv \NRSur events. Then, we compute ${\rm Corr}(\phi_1, \phi_2)$
by varying the spins for the same waveform at different $\tref$. We initially
compute the statistical uncertainties at a fixed distance of 100 Mpc, but then
rescale them following $\delta \lambda^j \propto 1/\mathrm{SNR}$ to correspond
to an SNR (defined as $\sqrt{(h|h)}$\,) matching that of the observed event.

Figure~\ref{fig:fisher_errs} shows the statistical uncertainties in $\phi_1$
and $\phi_2$ as we move from $\trefmFourthousandM$ to $\trefmHundredM$. In
almost all cases, we see that the statistical uncertainty decreases as we
approach the merger, meaning that the waveform is generally more sensitive to
variations in $\phi_1$ and $\phi_2$ near the merger. This explains the improved
measurement at $\trefmHundredM$ in Fig.~\ref{fig:phi_corner_GW170818} and
Fig.~\ref{fig:phi_all_comparison}, as well as the systematic improvement as we
approach $\trefmHundredM$ in Fig.~\ref{fig:phi_all_tref_comparison}.

To summarize, since the observed waveform is most sensitive to the
orbital-plane spin angles near merger, the data can successfully constrain
these angles at that point. However, the precision of that measurement is not
preserved as we extrapolate the spins back in time because, even though the
dynamics are deterministic, this detail gets smeared out during the inspiral
cycles.

Finally, we note that the direction of the $\delta \phi_1-\delta \phi_2$
correlations in Fig.~\ref{fig:fisher_errs} depend on where in the evolution
they are evaluated. This is in agreement with Ref.~\cite{Ma:2021znq}, which
found that the inspiral and ringdown regions of the waveform carry
complimentary information.

\begin{figure*}[thb]
\includegraphics[width=0.48\textwidth]{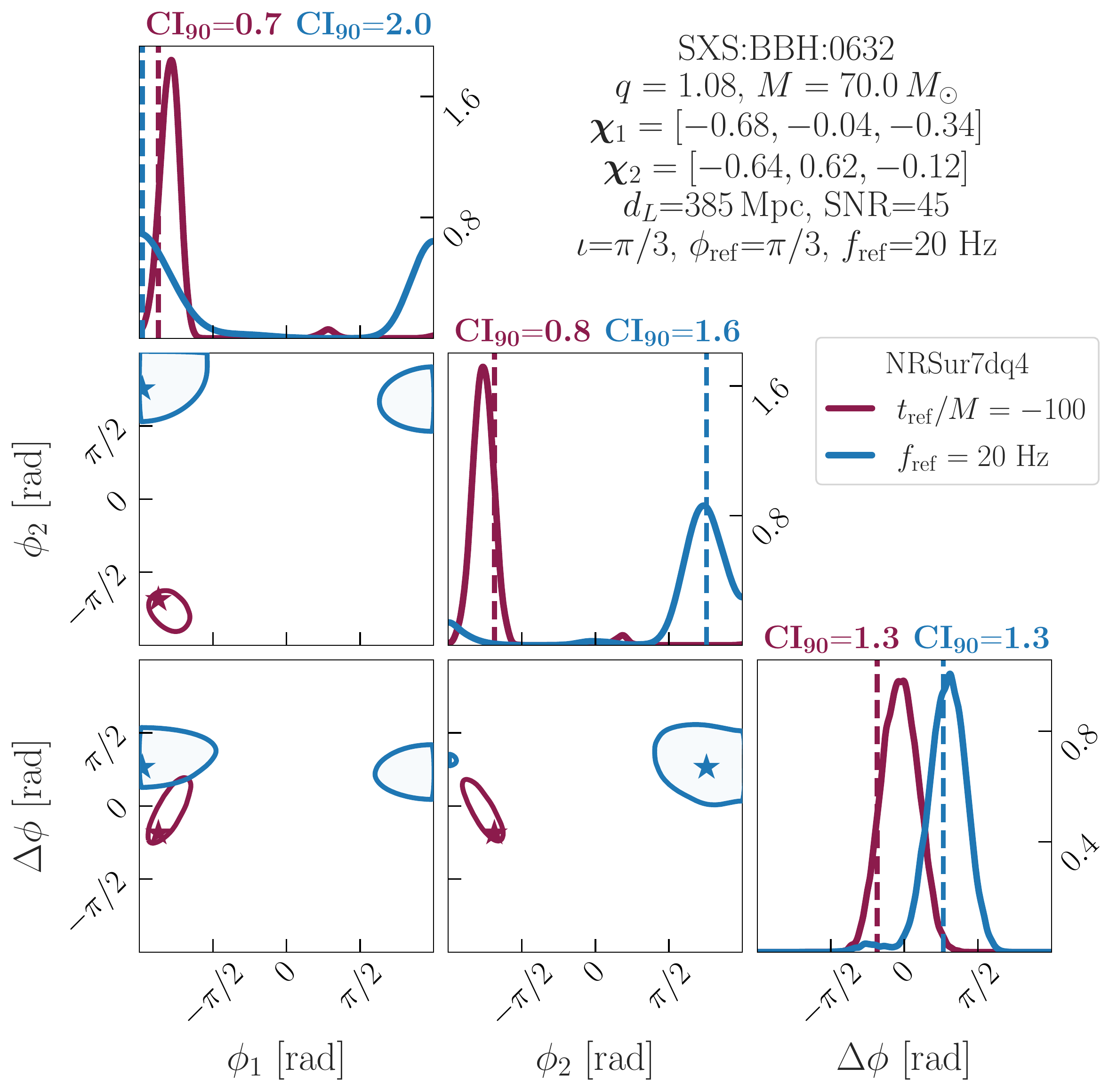} ~~~~~
\includegraphics[width=0.48\textwidth]{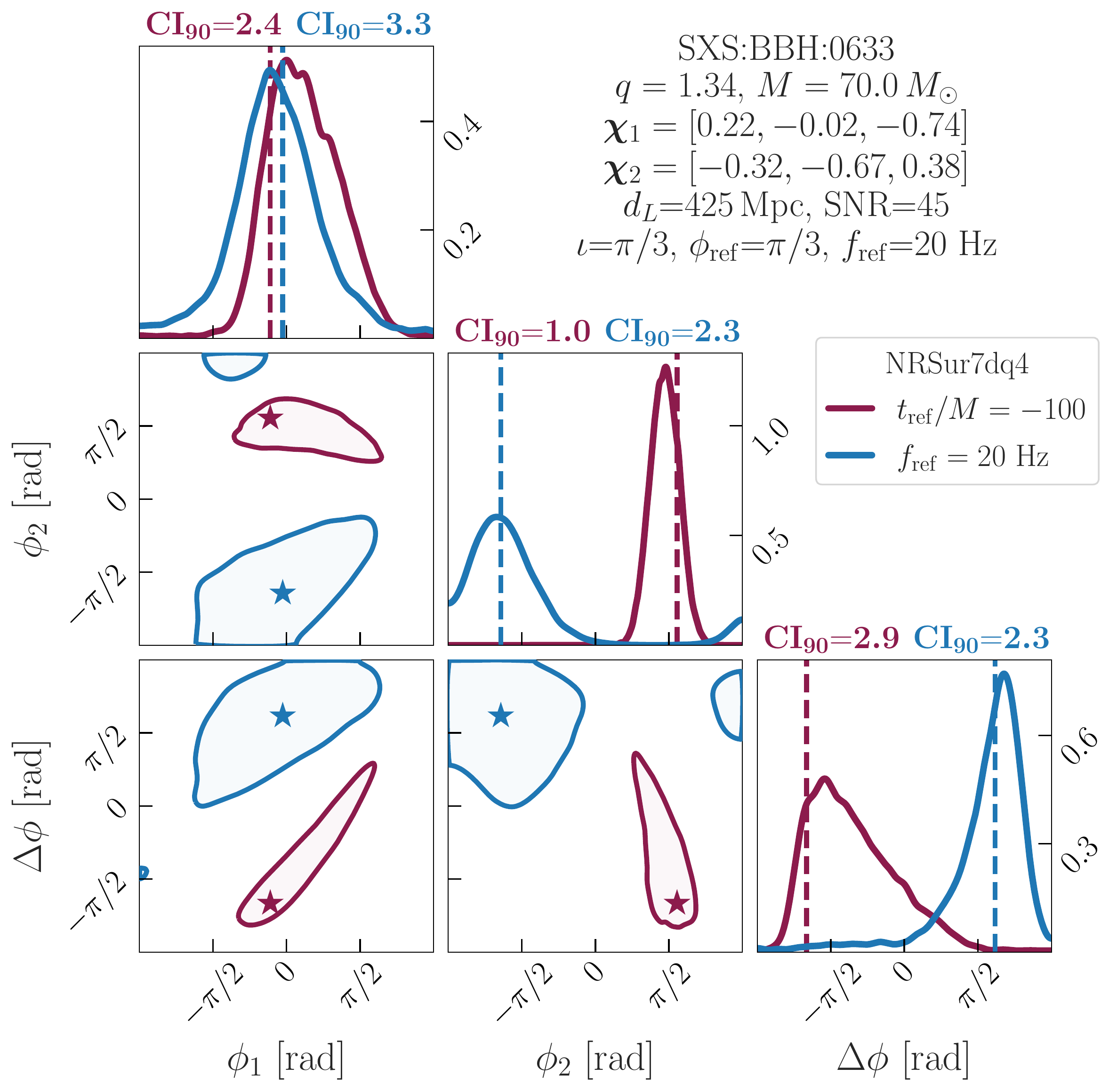}

\vspace{0.5cm}

\includegraphics[width=0.48\textwidth]{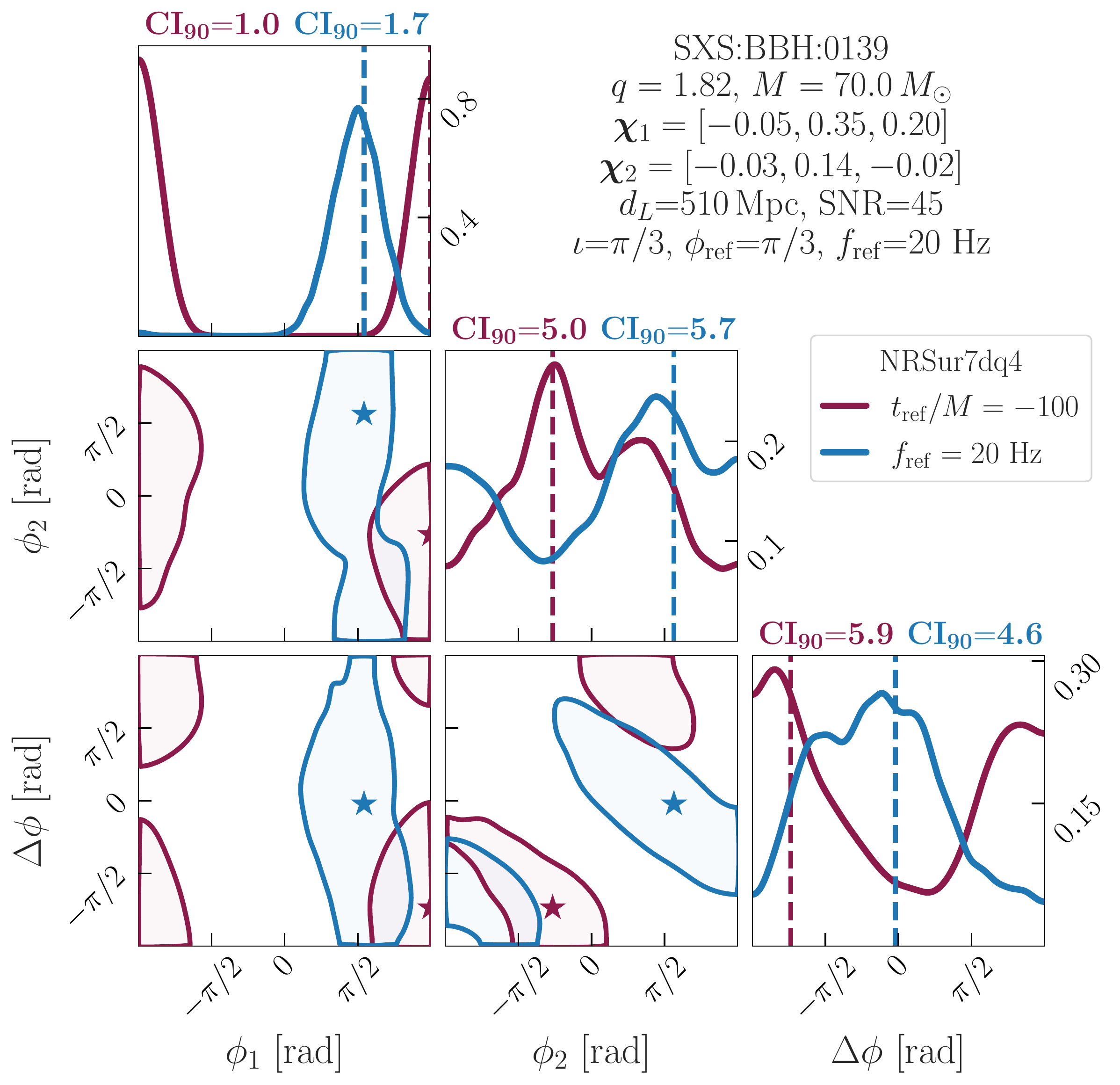} ~~~~~
\includegraphics[width=0.48\textwidth]{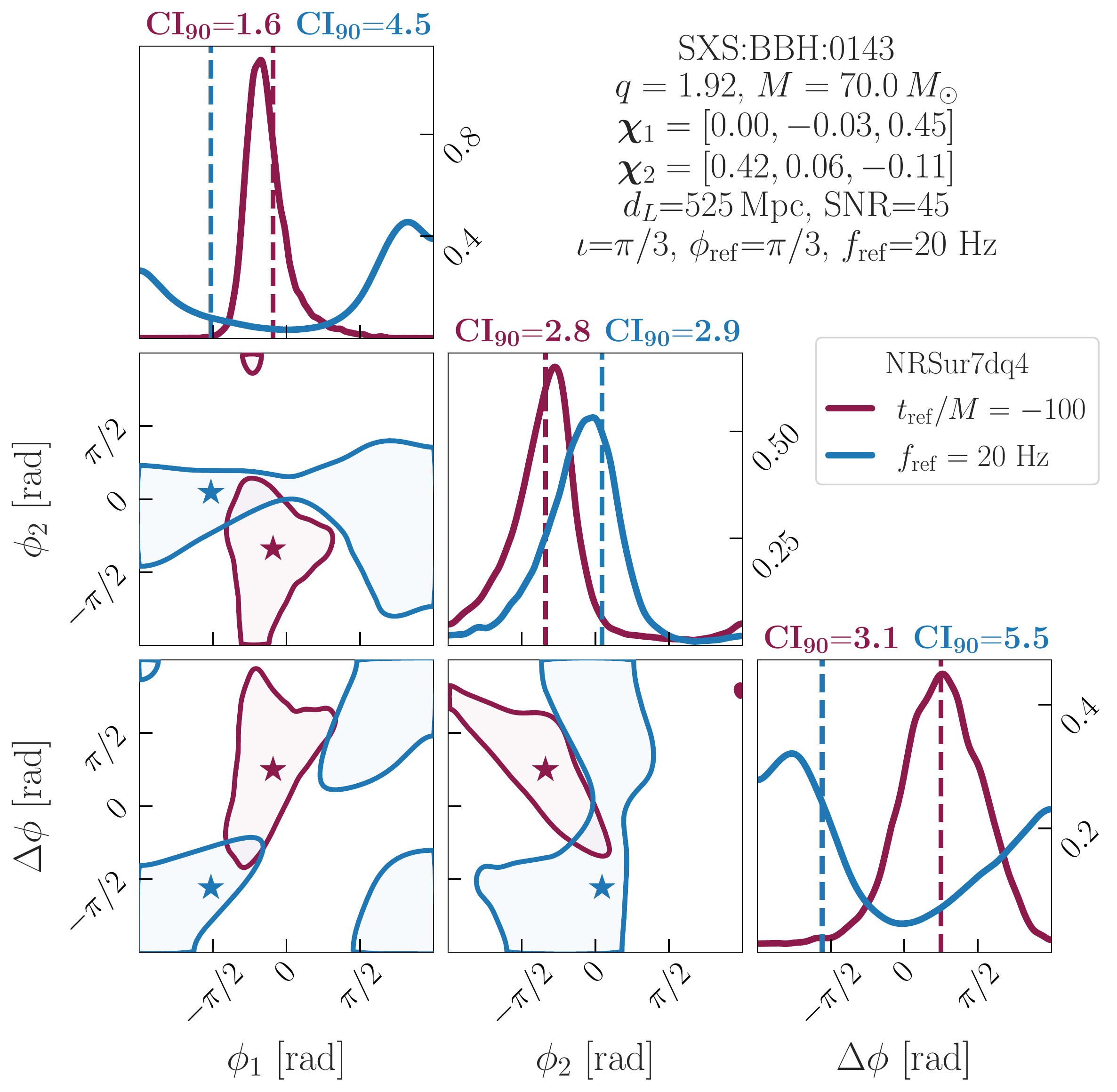}
\caption{
Same as Fig.~\ref{fig:phi_corner_nr_inj_trefm100_snr30}, but now the SNR is
increased to $45$.
}
\label{fig:phi_corner_nr_inj_trefm100_snr45}
\end{figure*}

\begin{figure*}[thb]
\includegraphics[width=0.48\textwidth]{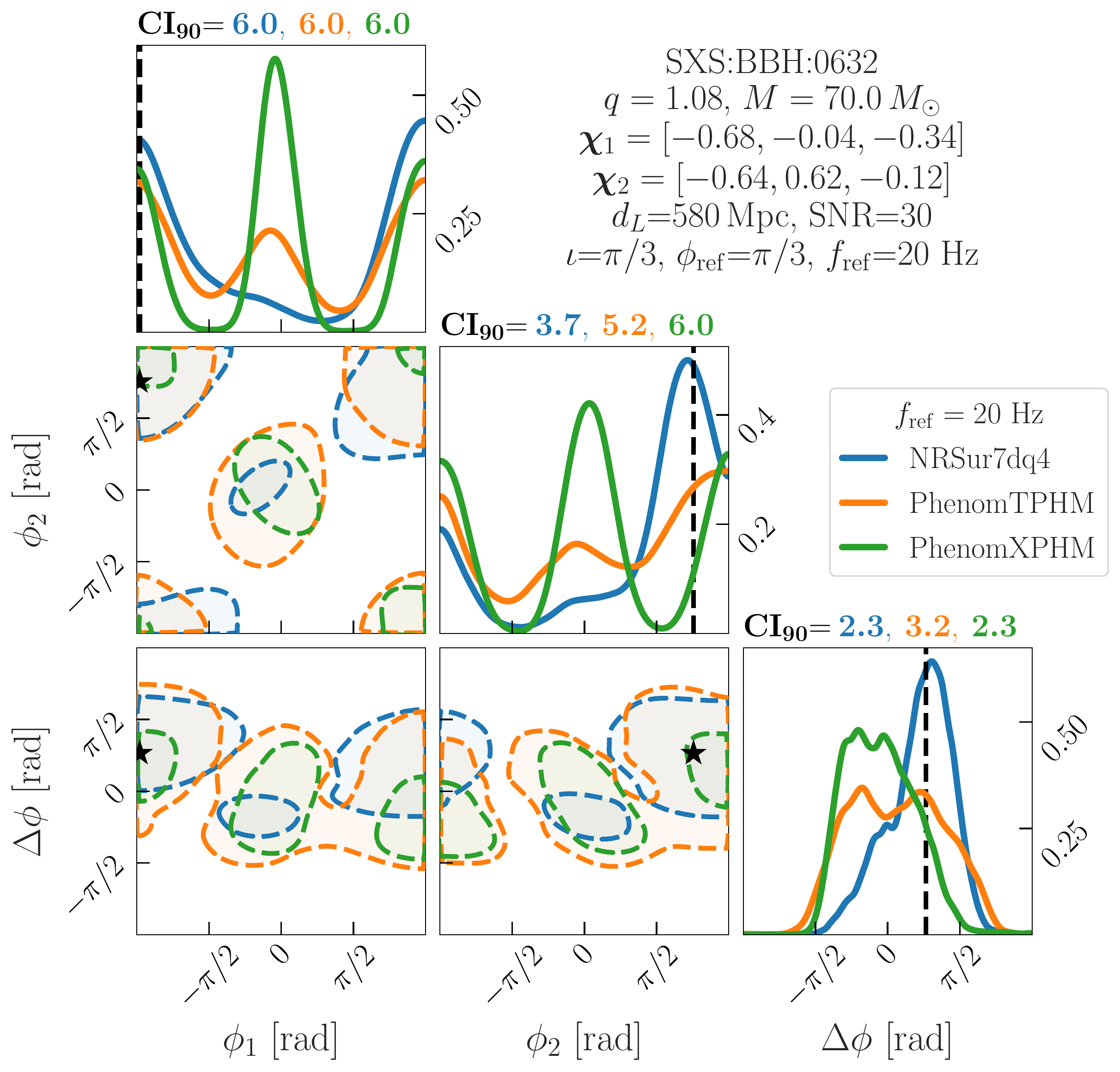} ~~~~~
\includegraphics[width=0.48\textwidth]{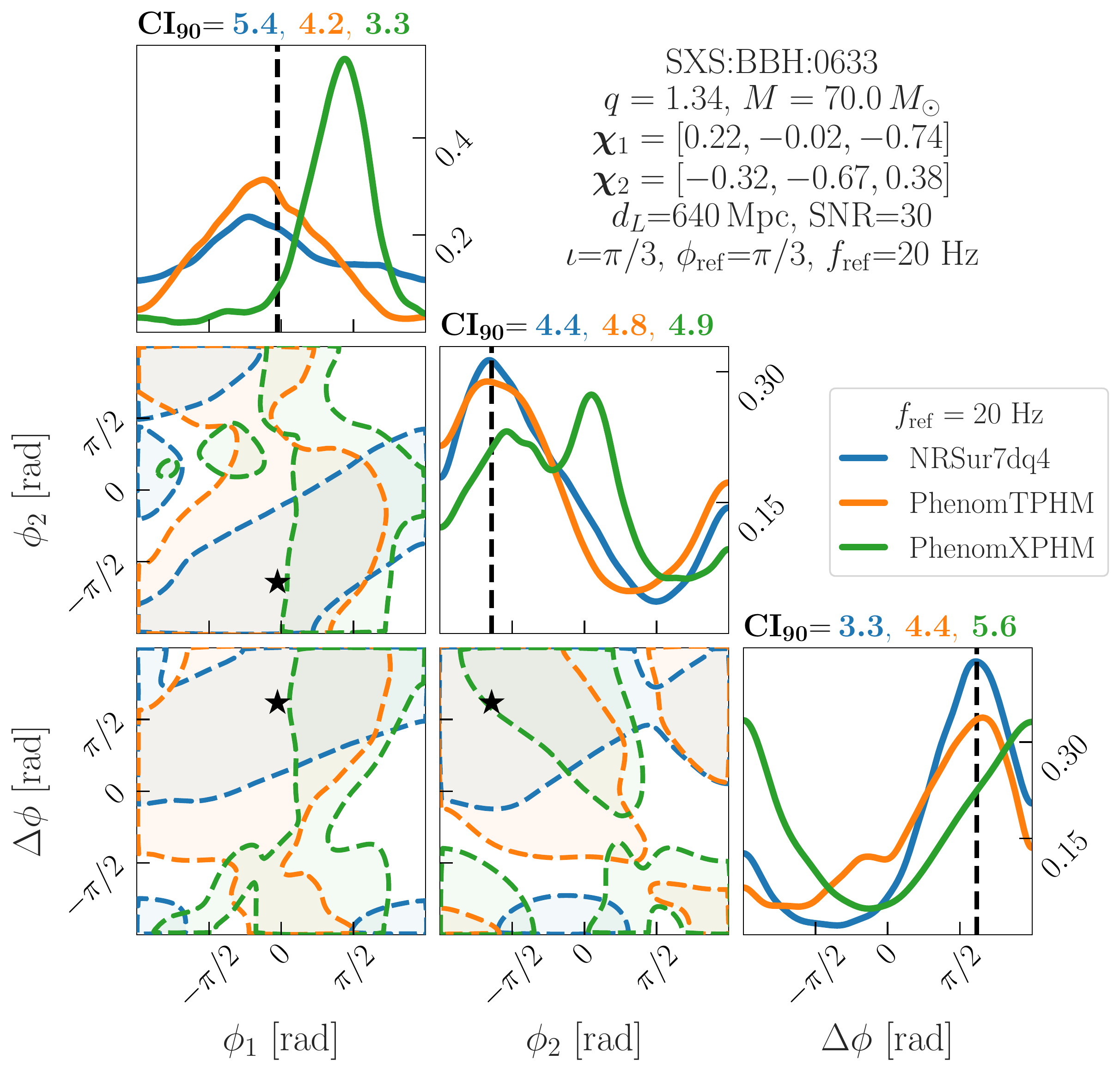}

\vspace{0.5cm}

\includegraphics[width=0.48\textwidth]{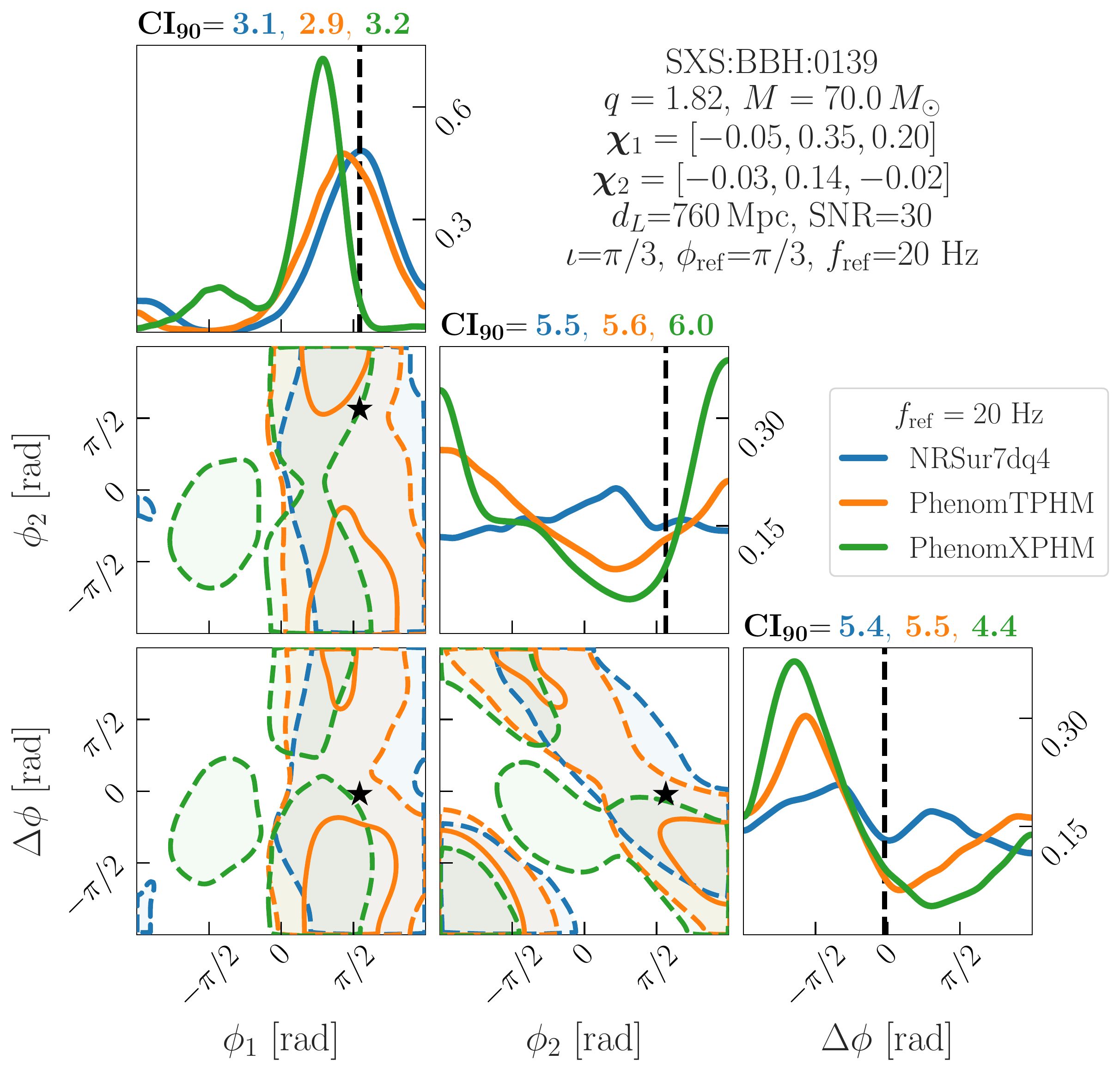} ~~~~~
\includegraphics[width=0.48\textwidth]{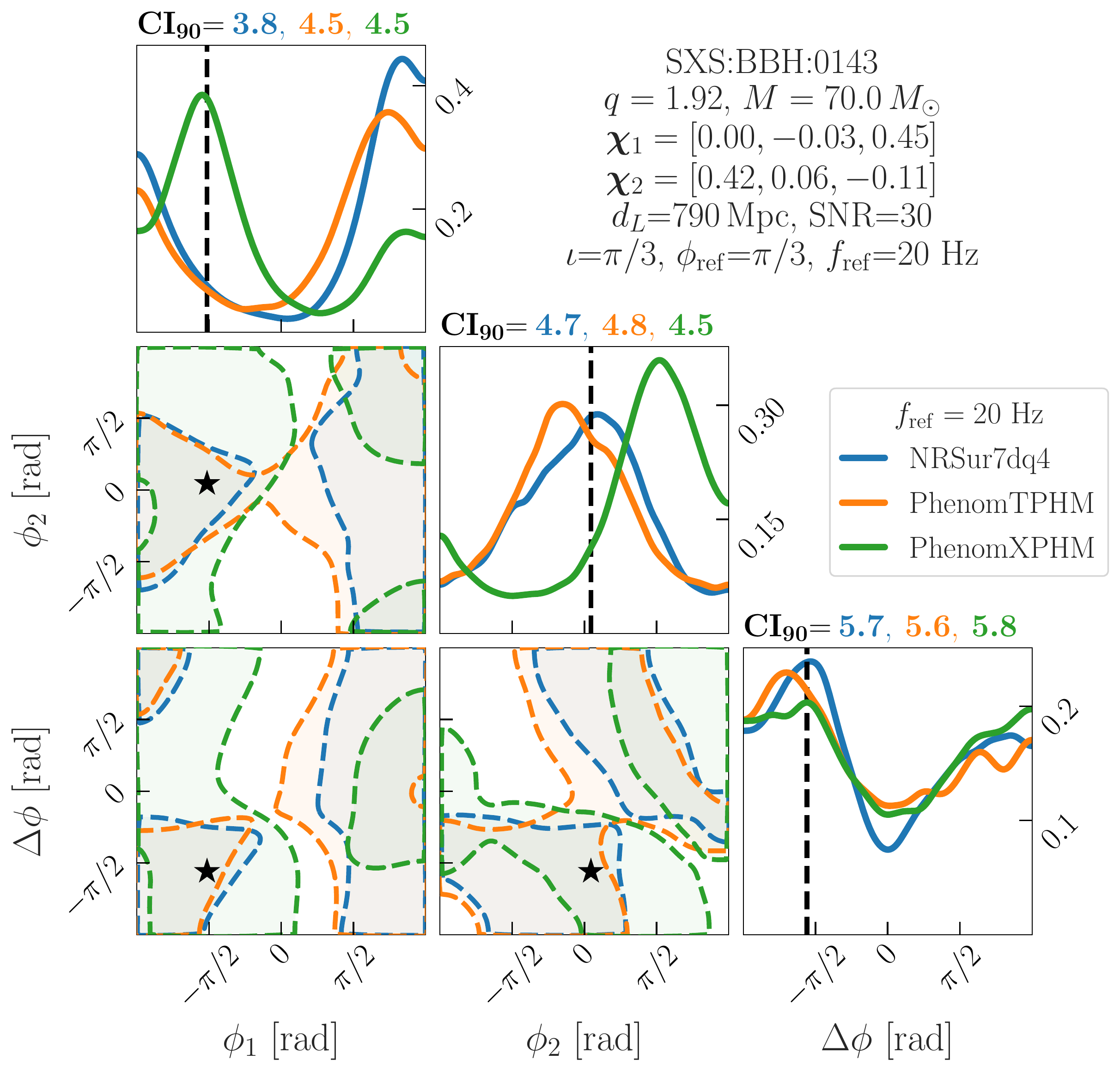}
\caption{
Same NR injections as Fig.~\ref{fig:phi_corner_nr_inj_trefm100_snr30}, but we
show posteriors at $\frefTwentyHz$ for \PhenomX and \PhenomT along with \NRSur.
The lower-triangle subplots show joint 2D posteriors for $\phi_1$, $\phi_2$ and
$\delphi$, with the true value indicated by a black star. The central 90\%
credible regions are shown as dashed contours. Only for the bottom-left panel
and \PhenomT, we also show the 50\% credible regions as solid contours to
demonstrate the systematic bias. The diagonal subplots show marginalized 1D
posteriors, with the true value indicated by a black vertical dashed line. All
injections are done at an SNR of $30$.
}
\label{fig:phi_corner_nr_inj_snr30}
\end{figure*}

\begin{figure*}[thb]
\includegraphics[width=0.48\textwidth]{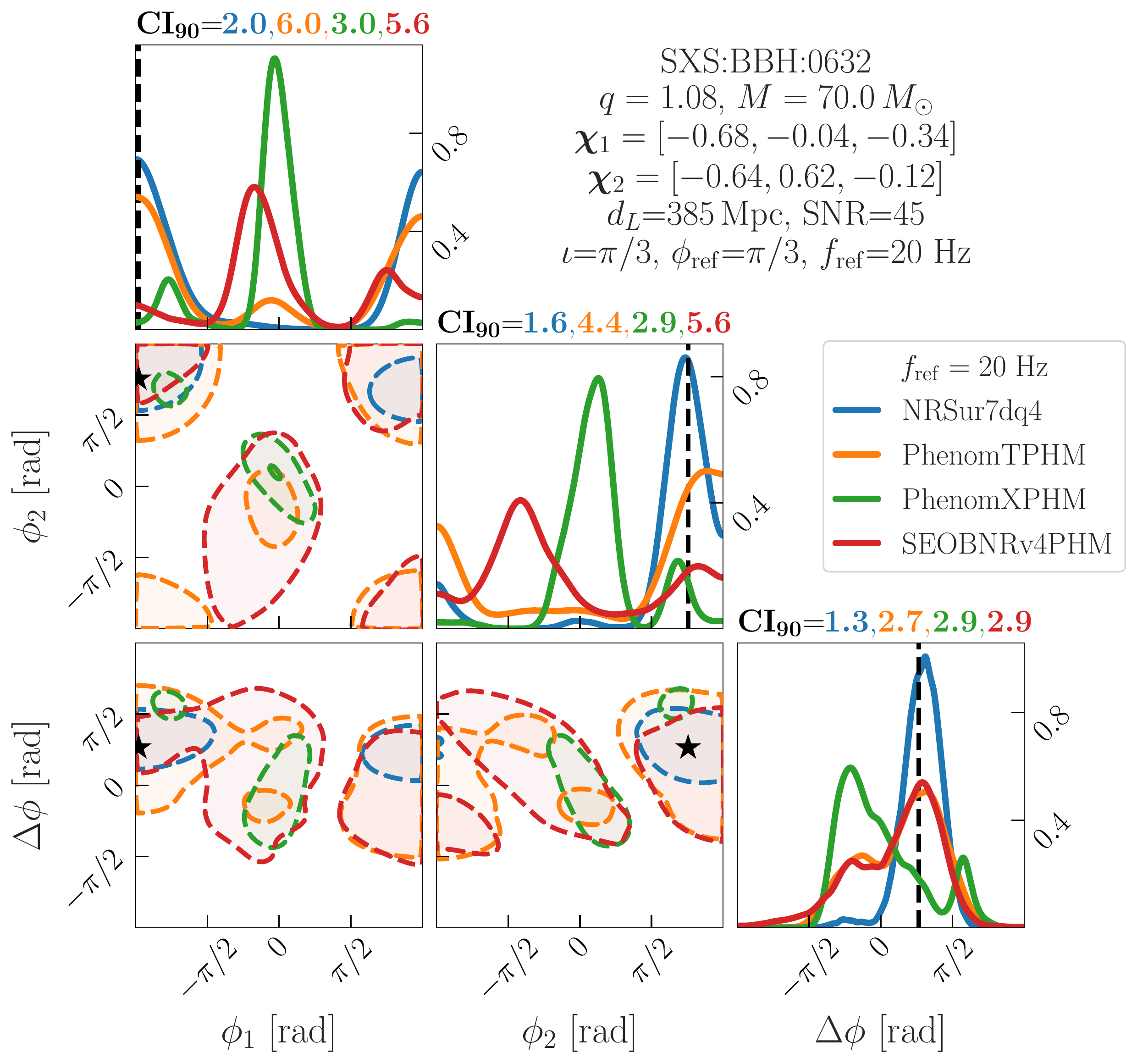} ~~~~~
\includegraphics[width=0.48\textwidth]{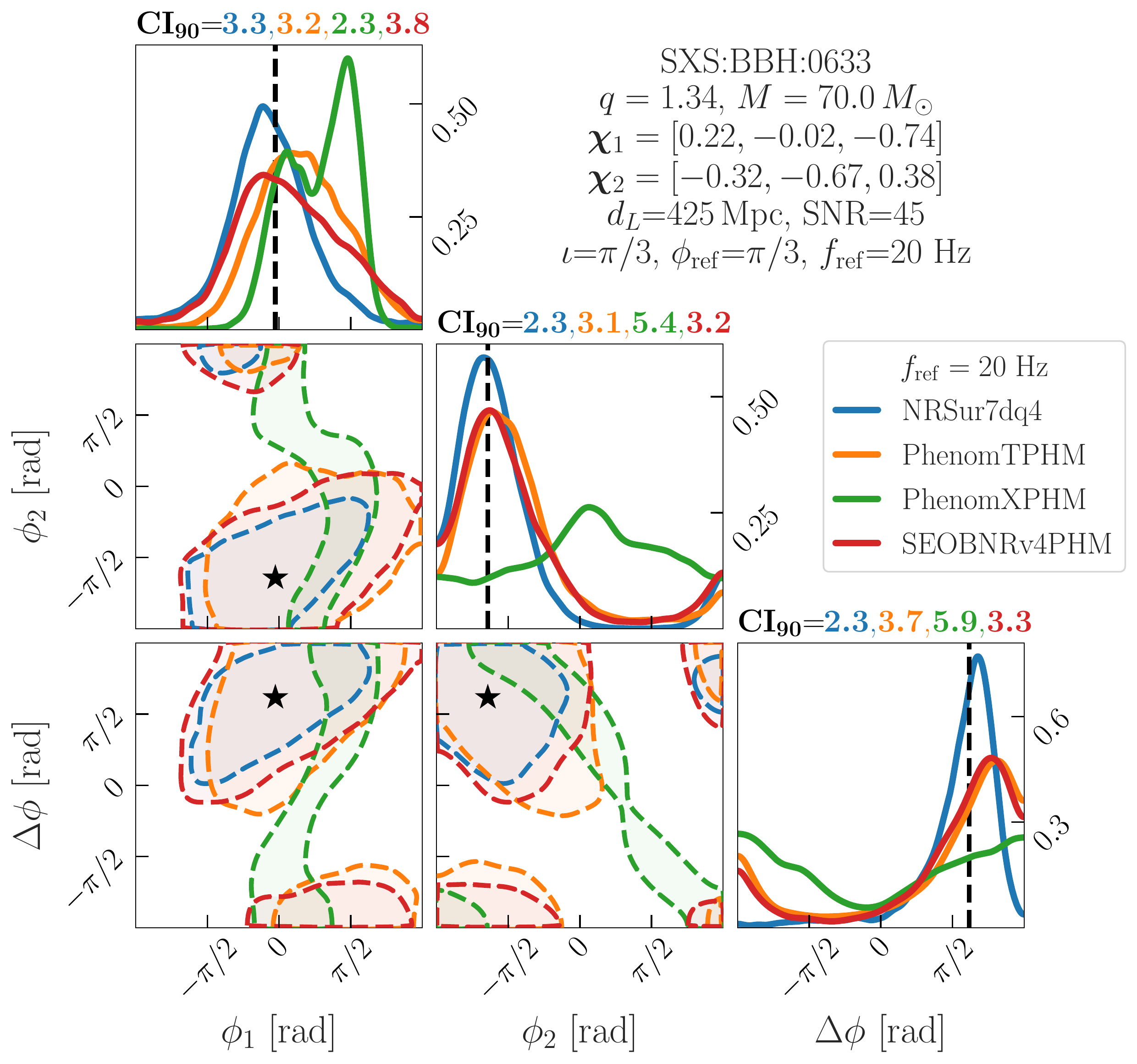}

\vspace{0.5cm}

\includegraphics[width=0.48\textwidth]{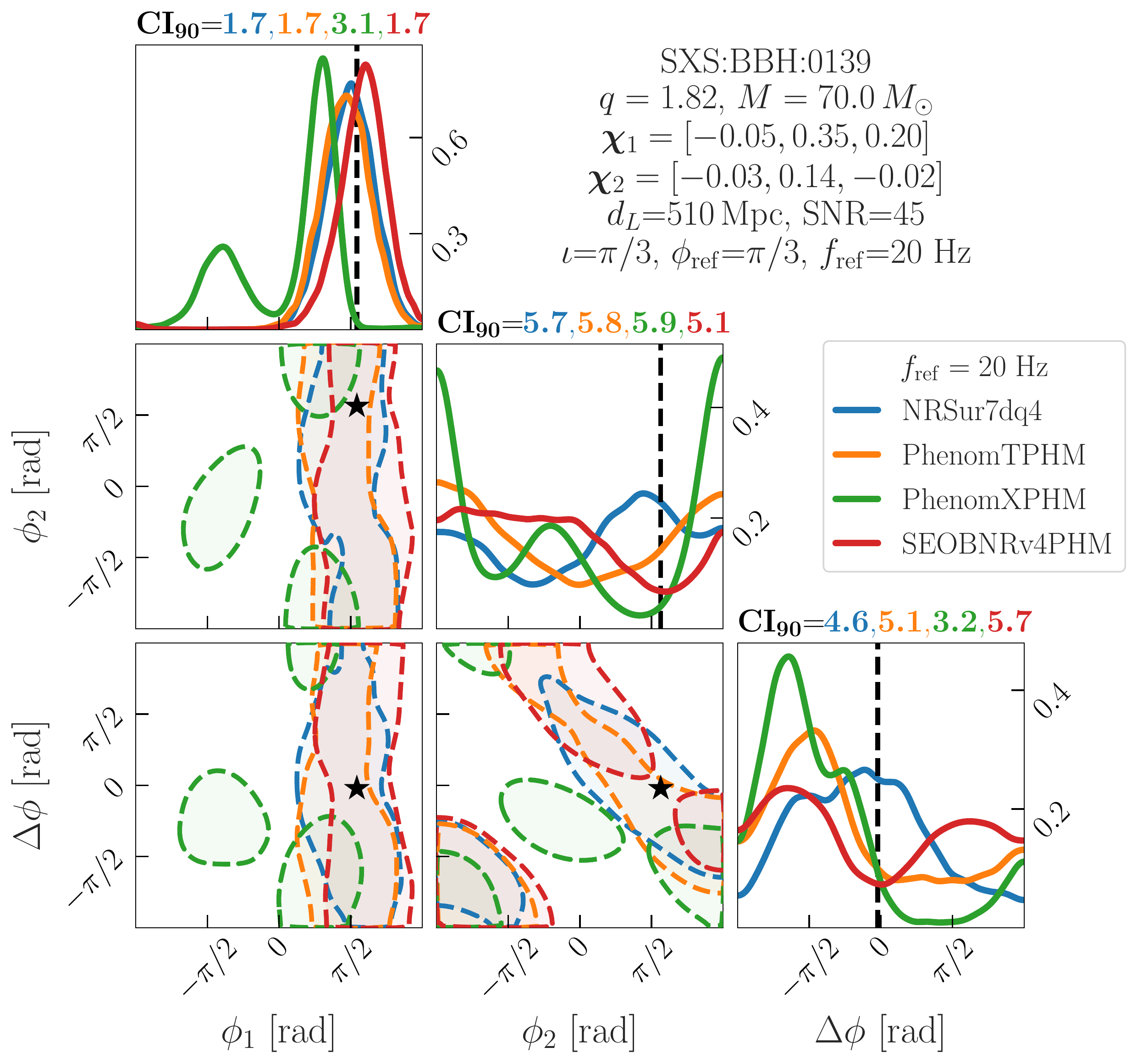} ~~~~~
\includegraphics[width=0.48\textwidth]{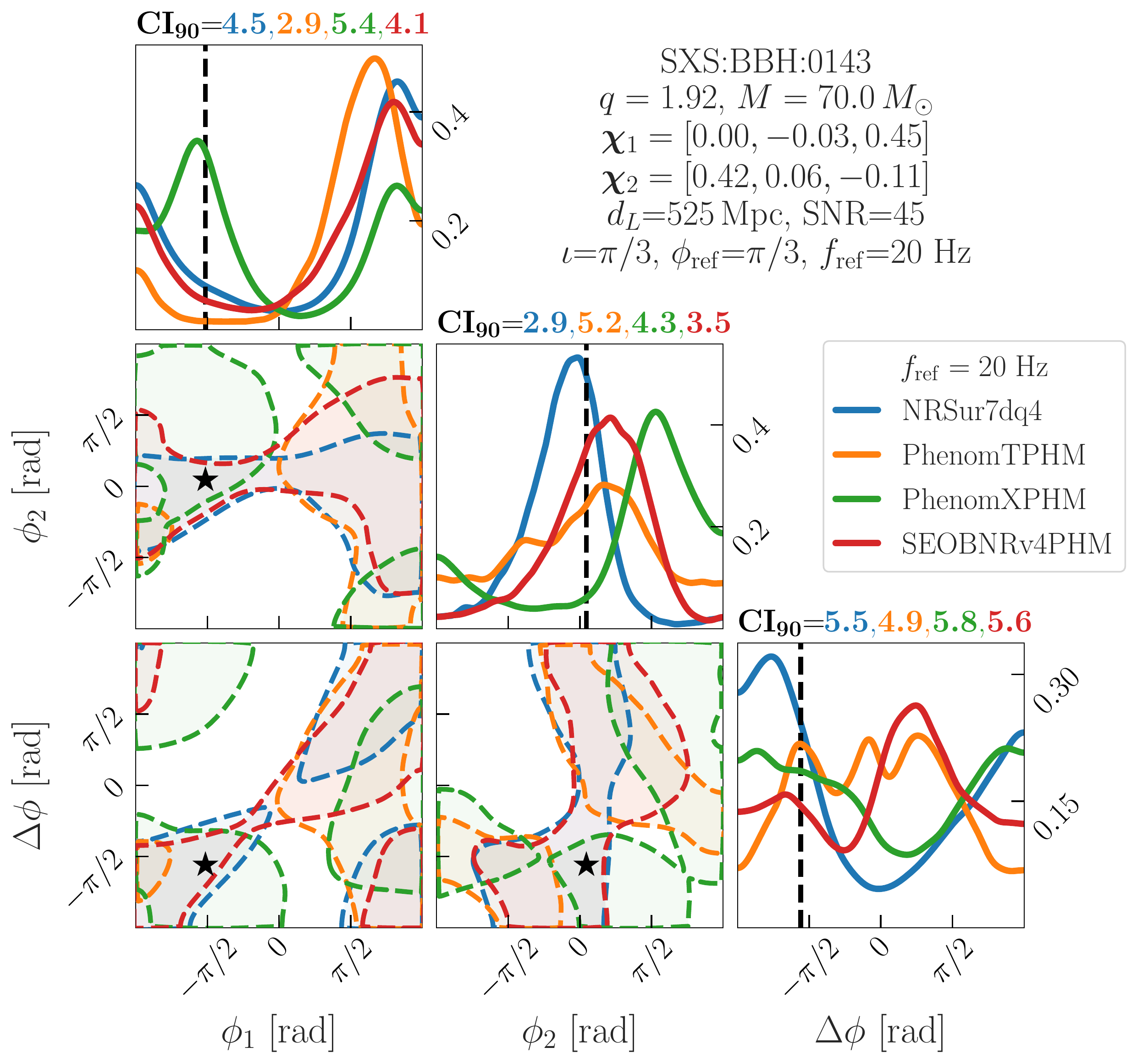}
\caption{
Same as Fig.~\ref{fig:phi_corner_nr_inj_snr30}, but now the SNR is increased to
$45$, and only 90\% contours are shown for all joint posteriors.
}
\label{fig:phi_corner_nr_inj_snr45}
\end{figure*}

\section{NR injection study}
\label{sec:nr_inj}

To further investigate the measurability of the orbital-plane spin angles, we
consider four NR waveforms, SXS:BBH:0139, SXS:BBH:0143, SXS:BBH:0632, and
SXS:BBH:0633, from the public SXS catalog~\cite{SXSCatalog, Mroue:2013xna,
Boyle:2019kee}. These waveforms correspond to systems with mass ratios $q<2$
and substantial orbital-plane spins. Note that none of these waveforms were
used to train \NRSur. We choose a total mass $M=70 M_{\odot}$, an inclination
angle $\iota=\pi/3$ between $\bL$ and the line-of-sight, and a
reference orbital phase $\phi_{\mathrm{ref}}=\pi/3$. Note that $\iota$ and
$\phi_{\mathrm{ref}}$ are defined at $\frefTwentyHz$. The luminosity distance
is chosen such that the network matched-filter SNR is either $30$ or $45$. The
rest of the binary parameters will be shown in figure insets below. We inject
these NR waveforms (in zero-noise) into a simulated LIGO-Virgo network
operating at design sensitivity~\cite{aLIGODesignNoiseCurve}, and recover them
using different waveform models. The injection and parameter inference are done
using the \texttt{Parallel Bilby}~\cite{Smith:2019ucc} package.

\subsection{$\trefmHundredM$ vs $\frefTwentyHz$ for \NRSur}

In Sec.~\ref{sec:spin_measurements}, we showed that the constraints on the
orbital-plane spin angles become tighter when measured near the merger. It is
important to verify that this tighter constraint does not lead to biased
estimates for these angles. We verify this in
Fig.~\ref{fig:phi_corner_nr_inj_trefm100_snr30}, where we show $\phi_1$,
$\phi_2$ and $\delphi$ measured at $\frefTwentyHz$ and $\trefmHundredM$, using
\NRSur against NR injections at SNR=$30$. We first note that the \NRSur model
indeed recovers the true values at both $\frefTwentyHz$ and $\trefmHundredM$.
Next, all orbital-plane angles, including $\delphi$, are significantly better
measured at $\trefmHundredM$. While this is not always clear from the 1D
marginalized distributions for $\delphi$, note that the 2D joint posteriors for
all three combinations of $\phi_1$, $\phi_2$ and $\delphi$ are always better
constrained at $\trefmHundredM$.

In Fig.~\ref{fig:phi_corner_nr_inj_trefm100_snr30}, for SXS:BBH:0633 (top-right
panel), the true value falls near the edge of the 2D 90\% credible region for
$\trefmHundredM$. To check whether this is indicative of a systematic bias in
\NRSur when spins are measured at $\trefmHundredM$, we repeat our injections at
at SNR=$45$ in Fig.~\ref{fig:phi_corner_nr_inj_trefm100_snr45}. As expected,
increasing the SNR leads to better constraints on the orbital-plane spins
angles for both $\frefTwentyHz$ and $\trefmHundredM$. For SXS:BBH:0633, the 2D
90\% credible regions in Fig.~\ref{fig:phi_corner_nr_inj_trefm100_snr45} still
include the true value at $\trefmHundredM$, suggesting that there are no
significant biases. Once again, we find that some 1D posteriors can be more
sharply peaked at $\frefTwentyHz$ (e.g. $\delphi$ in the top-right panel of
Fig.~\ref{fig:phi_corner_nr_inj_trefm100_snr45}), but the 2D posteriors are
always better constrained at $\trefmHundredM$ for all three combinations of
$\phi_1$, $\phi_2$ and $\delphi$.

In Figs.~\ref{fig:phi_all_comparison} and \ref{fig:phi_all_comparison_delphi},
we noted that while $\phi_1$ and $\phi_2$ measurements are improved at
$\trefmHundredM$ for current GW events, $\delphi$ measurements are not
significantly impacted.  Figures~\ref{fig:phi_corner_nr_inj_trefm100_snr30} and
\ref{fig:phi_corner_nr_inj_trefm100_snr45} show that as detector sensitivity
improves and GW signals are observed at higher SNR, our method will generally
lead to improved measurements in $\delphi$ as well.
Even when the 1D posterior for $\delphi$ is more sharply peaked at
$\frefTwentyHz$, the overall constraints on the orbital-plane spin angles are
better at $\trefmHundredM$. Measuring the full orbital-plane spin degrees of
freedom is necessary to constrain the kick population as done in
Ref.~\cite{Varma:2021xbh}.

\subsection{Waveform model comparison}
In this section, we study the performance of different waveform models in
recovering the orbital-plane spins angles. Apart from \NRSur, we also consider
the phenomenological models \PhenomT~\cite{Estelles:2021gvs} and
\PhenomX~\cite{Pratten:2020ceb}, as well as the effective-one-body model
\EOB~\cite{Ossokine:2020kjp}. While these models also include some effects of
precession, they are not calibrated on precessing NR simulations.  Note that
\PhenomT and \EOB are time-domain models, while \PhenomX is a frequency-domain
model. We only consider spin measurements at $\frefTwentyHz$ for these models
as specifying spins at a dimensionless time/frequency would require careful
modifications to how these models are implemented.
However, because binary BH spin evolution is deterministic, any biases seen at
$\frefTwentyHz$ should translate to biases at $\trefmHundredM$ as well.
We repeat our NR injections
at SNRs of $30$ and $45$, but because \EOB is significantly more expensive than
the other models, we only apply it to the injections at SNR=$45$.

Figure~\ref{fig:phi_corner_nr_inj_snr30} shows $\phi_1$, $\phi_2$ and $\delphi$
posteriors obtained using \NRSur, \PhenomT and \PhenomX for our NR injections
at SNR=$30$.  For two out of the four cases (SXS:BBH:0633 and SXS:BBH:0139),
the true value falls on the edge of the 90\% credible region of the 2D
posteriors for \PhenomX. The 1D marginalized posteriors are also biased in
several cases for \PhenomX;
in particular, there are cases (e.g. $\phi_1$ in top-right panel and $\delphi$
in the bottom-left panel of Fig.~\ref{fig:phi_corner_nr_inj_snr30}) where this
model has the strongest peaks in the 1D distributions, but prefers the wrong
value.
By contrast, for both \NRSur and \PhenomT, the true value is always within the
90\% credible region of the 2D posteriors.

However, for SXS:BBH:0139 (bottom-left of
Fig.~\ref{fig:phi_corner_nr_inj_snr30}), the 1D $\delphi$
posterior for \PhenomT is peaked away from the true value, even though the true
value is included in the 90\% credible region of the 2D posteriors. For
\NRSur, the peak in the 1D $\delphi$ posterior is much broader in this case and
includes the true value. To check whether this is indicative of a systematic
bias in \PhenomT, we consider the 50\% credible region of the 2D posteriors for
this case, which is shown only for \PhenomT for simplicity. The 50\% credible
region clearly excludes the true value for \PhenomT, meaning that the bulk of
the probability density for \PhenomT is concentrated in a region away from the
true value. This suggests that the deviation in the 1D $\delphi$ posterior for
$\PhenomT$ is indeed due to a systematic bias.
This serves as an another example where a waveform model has a stronger peak
than \NRSur in the 1D posterior, but is peaked at the wrong value.

Similarly, for SXS:BBH:0143 (bottom-right of
Fig.~\ref{fig:phi_corner_nr_inj_snr30}), the true value is included in the 2D
posteriors but the 1D $\phi_1$ posteriors appear biased for both \PhenomT and
\NRSur. However, in this case the primary BH has negligible spin in the
orbital-plane, therefore $\phi_1$ is not a meaningful parameter and hence the
offset from the true value is not of concern.

We repeat these NR injections at SNR=$45$ in
Fig.~\ref{fig:phi_corner_nr_inj_snr45}, now also including the \EOB model. We
now find that the true value is fully excluded from the 90\% credible region of
the 2D posteriors for \PhenomX for three out of our four injections. While
\PhenomT still performs better than \PhenomX, \PhenomT also excludes the true
value from the 90\% credible region of the 2D posteriors for SXS:BBH:0143
(bottom-right of Fig.~\ref{fig:phi_corner_nr_inj_snr45}). For SXS:BBH:0139
(bottom-left of Fig.~\ref{fig:phi_corner_nr_inj_snr45}), the 1D $\delphi$
posterior is still biased for $\PhenomT$, while \NRSur now has a clear peak
around the true value. This suggests that \NRSur is not prone to the systematic
biases present in \PhenomT (and \PhenomX) as noted above.

In Fig.~\ref{fig:phi_corner_nr_inj_snr45}, \EOB is comparable to \PhenomT,
including the true value in the 90\% region of the 2D posteriors for three out
of four cases, with the exception being SXS:BBH:0139 (bottom-left of
Fig.~\ref{fig:phi_corner_nr_inj_snr45}).
For this case, \EOB also shows similar biases in the $\delphi$ 1D posterior as
\PhenomT (and \PhenomX), meaning that this model is also prone to the
systematic biases noted above.
Furthermore, for SXS:BBH:0632 (top-left of
Fig.~\ref{fig:phi_corner_nr_inj_snr45}), the 1D posteriors for $\phi_1$ and
$\phi_2$ for \EOB are biased (although they are somewhat included in the
smaller secondary modes).

Finally, we note that \NRSur generally leads to the best constraints in
Figs.~\ref{fig:phi_corner_nr_inj_snr30} and \ref{fig:phi_corner_nr_inj_snr45},
which is expected as this is the only model trained on precessing NR
simulations (but not the ones injected here).  Instead, the \PhenomT, \PhenomX
and \EOB models approximate orbital precession by
``twisting''~\cite{Estelles:2021gvs, Pratten:2020ceb, Ossokine:2020kjp} a
corresponding nonprecessing waveform. While this captures the leading effect of
precession, it does not account for effects such as asymmetries between pairs
of $(\ell, m)$ and $(\ell, -m)$ spin-weighted spherical harmonic waveform
modes~\cite{Varma:2019csw}. Missing physics like this can lead to the
systematic biases we see in Figs.~\ref{fig:phi_corner_nr_inj_snr30} and
\ref{fig:phi_corner_nr_inj_snr45}.
Among the phenomenological models, \PhenomX is known to have a less accurate
precession treatment than \PhenomT~\cite{Estelles:2021gvs}, which could be
responsible for \PhenomX having the largest biases in our tests.

\section{Conclusions}
\label{sec:conclusion}

We propose that binary BH spins be measured at a reference point close to the
merger, at either a dimensionless GW frequency $\frefISCOLong$ or a
dimensionless time $\trefmHundredM$ before the peak of the GW amplitude. We
demonstrate that this leads to significant improvements in the measurement of
orbital-plane spin orientations $\phi_1$ and $\phi_2$ for various events in the
GWTC-2 catalog, while $\delphi$ is not significantly impacted. However, using
NR injections, we show that $\delphi$ will also be better measured near the
merger for louder signals expected in the future.

Using the same NR injections, we compare the performance of the waveform models
\NRSur, \PhenomX, \PhenomT, and \EOB, in recovering $\phi_1$, $\phi_2$ and
$\delphi$
at $\frefTwentyHz$.
As expected, \NRSur provides the most accurate constraints for
these angles, as this is the only model informed by precessing NR simulations.
Among the other models, in general, we find that \PhenomT and \EOB perform
better than \PhenomX. However, even at moderate SNRs ($\sim 30-45$), we find
examples where these models have biased estimates. This highlights the need to
train waveform models on precessing NR simulations in order to reliably extract
the full spin information from binary BH signals.
\PhenomX, \PhenomT, and \EOB do not currently allow specifying the spins at
$\trefmHundredM$ or $\frefISCO$, but we expect these biases will persist at
those reference points.

In a companion paper, Ref.~\cite{Varma:2021xbh}, we use the improved spin
measurements obtained here to constrain the astrophysical distribution of the
orbital-plane spin angles as well as merger kicks for the binary BH population.
Notably, we find a preference for $\delphi\sim\pm \pi$ in the population, which
can be a signature of spin-orbit resonances~\cite{Schnittman:2004vq}.

\section*{Acknowledgments.}
We thank Rory Smith and Avi Vajpeyi for support with the  \texttt{Parallel
Bilby}~\cite{Smith:2019ucc} package, and Hector Estelles, Sascha Husa, Geraint
Pratten, and Marta Colleoni for support with the phenomenological waveforms. We
thank Sizheng Ma for sharing his Fisher matrix code. We thank Davide Gerosa and
Katerina Chatziioannou for useful discussions.
V.V.\ was supported by a Klarman Fellowship at Cornell. This project has
received funding from the European Union’s Horizon 2020 research and innovation
programme under the Marie Skłodowska-Curie grant agreement No.~896869.
S.B.,  M.I. and S.V. acknowledge support of the National Science Foundation
and the LIGO Laboratory.
S.B. is also supported by the NSF Graduate Research Fellowship under Grant No. DGE-1122374.
M.I.\ is supported by NASA through the NASA Hubble Fellowship
grant No.\ HST-HF2-51410.001-A awarded by the Space Telescope
Science Institute, which is operated by the Association of Universities
for Research in Astronomy, Inc., for NASA, under contract NAS5-26555.
This research made use of data, software and/or web tools obtained from the
Gravitational Wave Open Science Center~\cite{GW_open_science_center}, a service
of the LIGO Laboratory, the LIGO Scientific Collaboration and the Virgo
Collaboration.
LIGO was constructed by the California Institute of
Technology and Massachusetts Institute of Technology with funding from the
National Science Foundation and operates under cooperative agreement PHY-1764464.
Computations were performed on the Wheeler cluster at Caltech, which is
supported by the Sherman Fairchild Foundation and by Caltech; and the High
Performance Cluster at Caltech.
\\

\appendix

\begin{figure*}[p]
\includegraphics[width=0.4\textwidth]{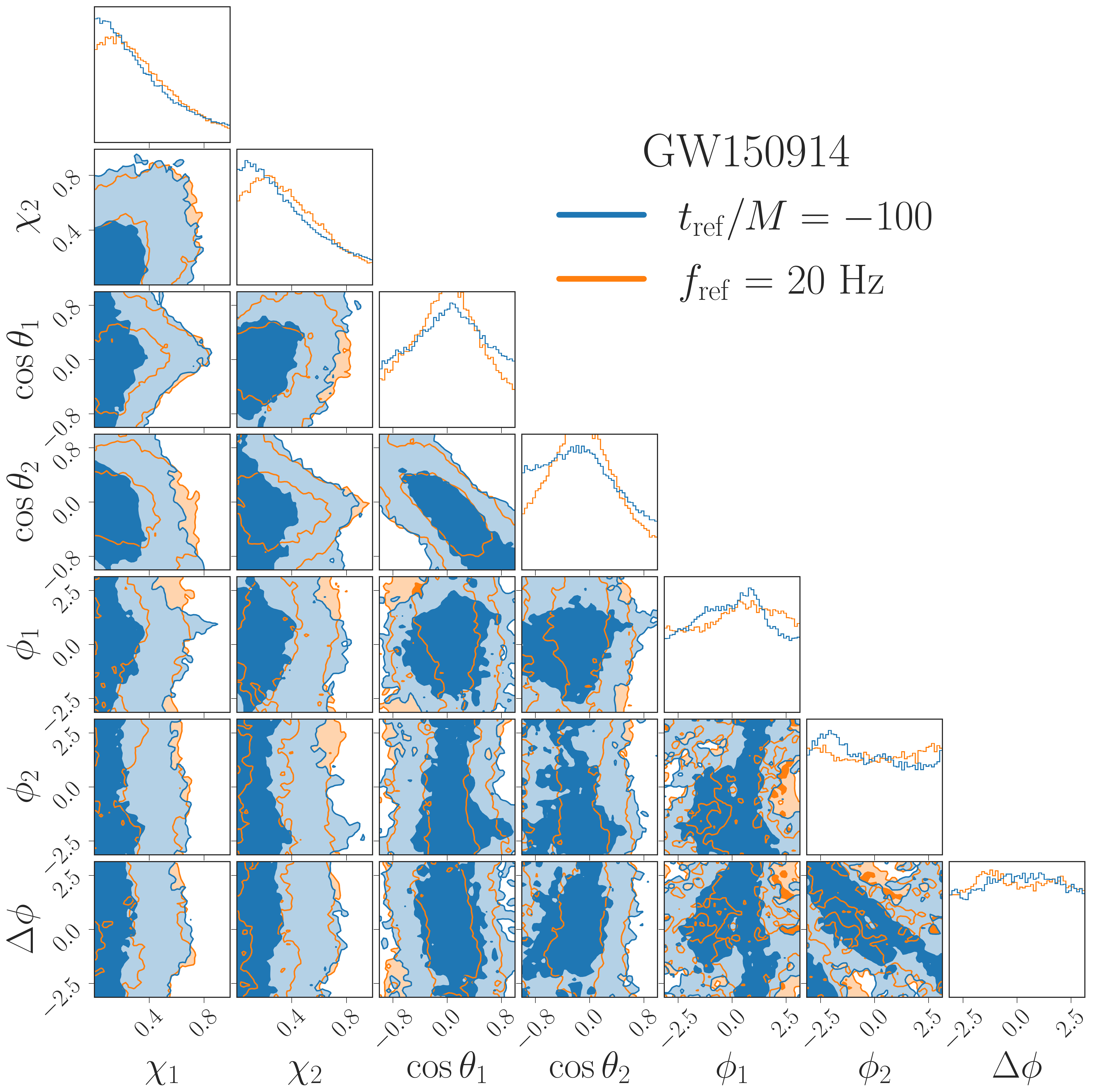} \hspace{1.5cm}
\includegraphics[width=0.4\textwidth]{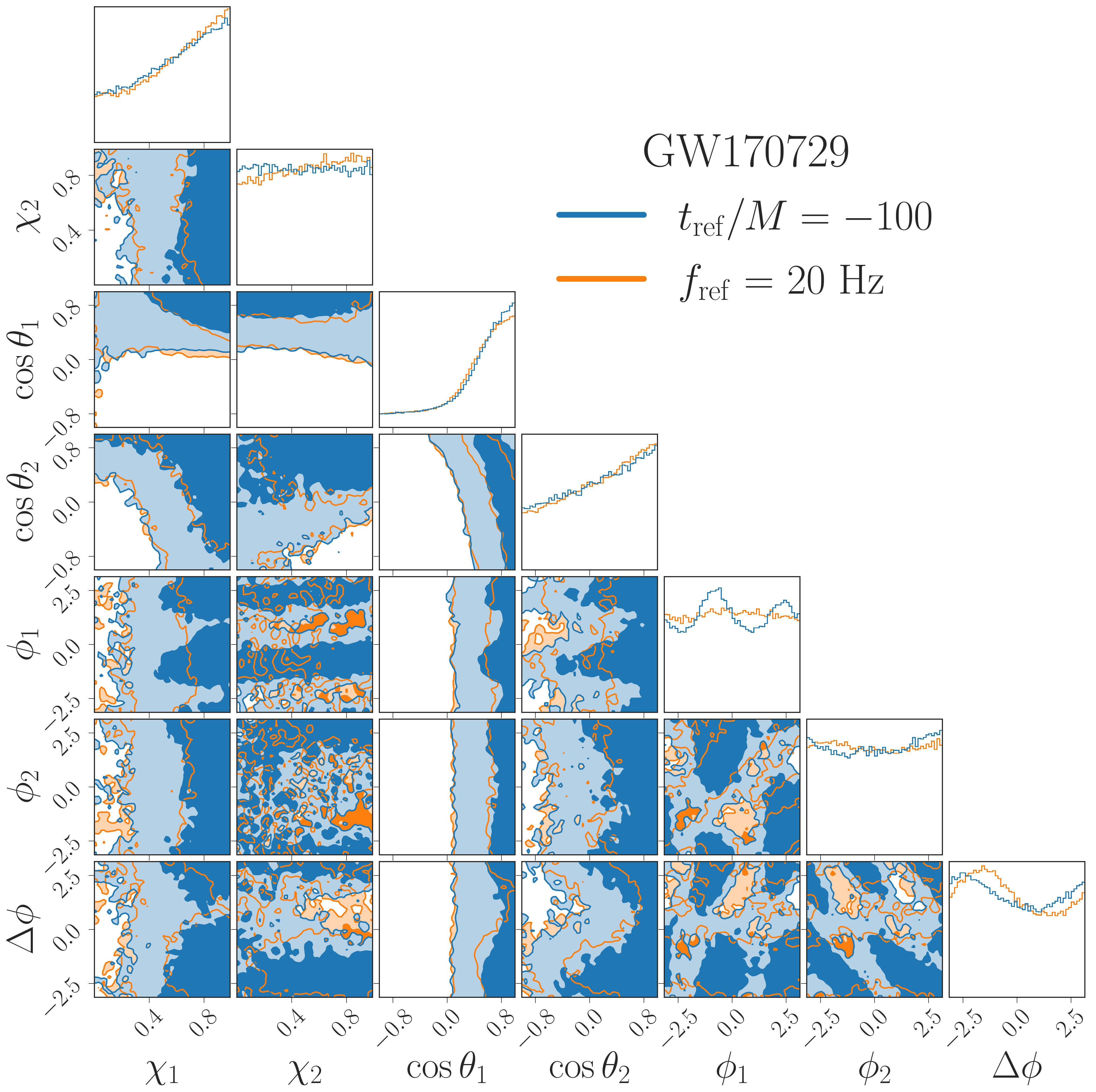} \\ \vspace{0.2cm}
\includegraphics[width=0.4\textwidth]{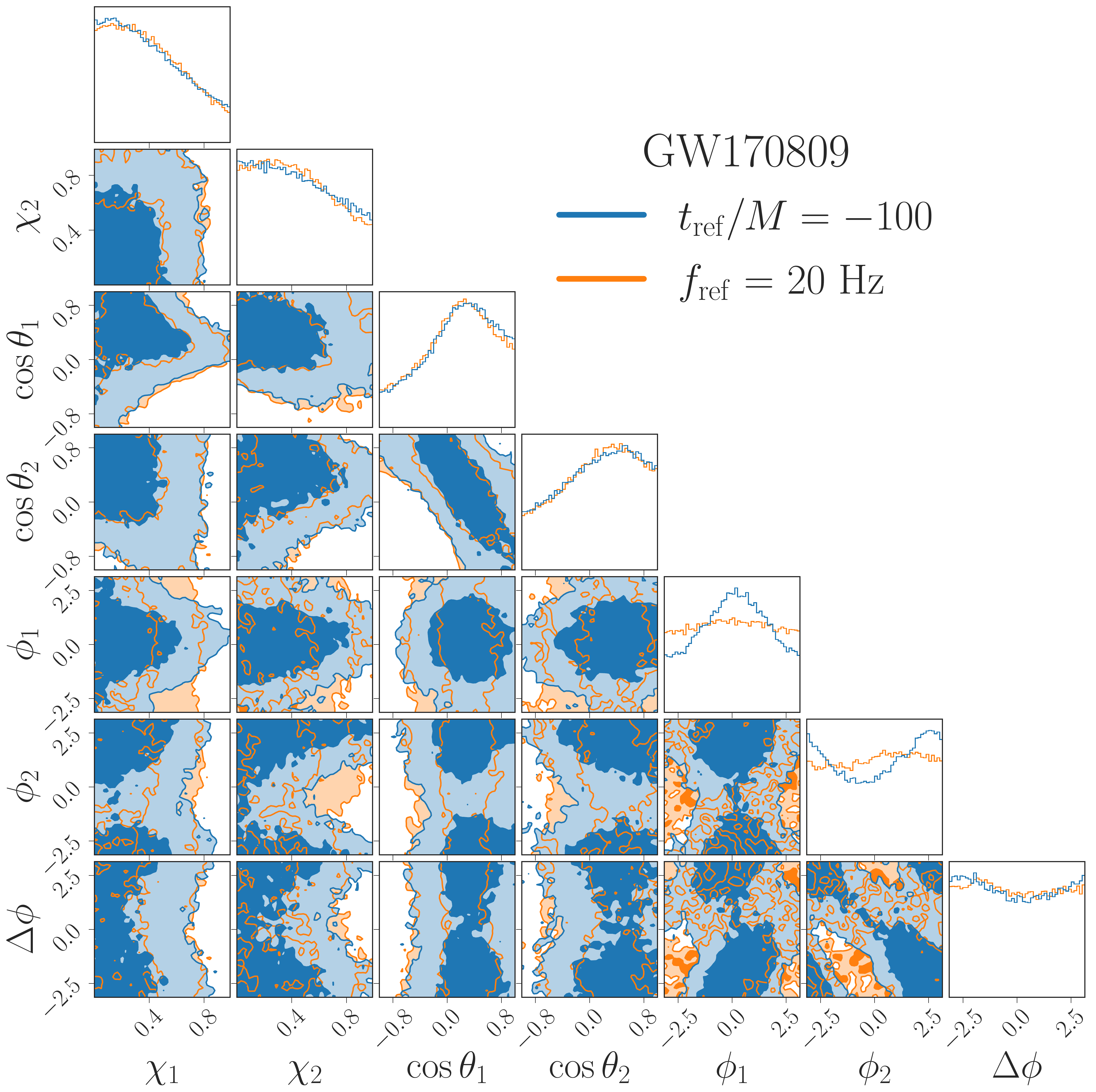} \hspace{1.5cm}
\includegraphics[width=0.4\textwidth]{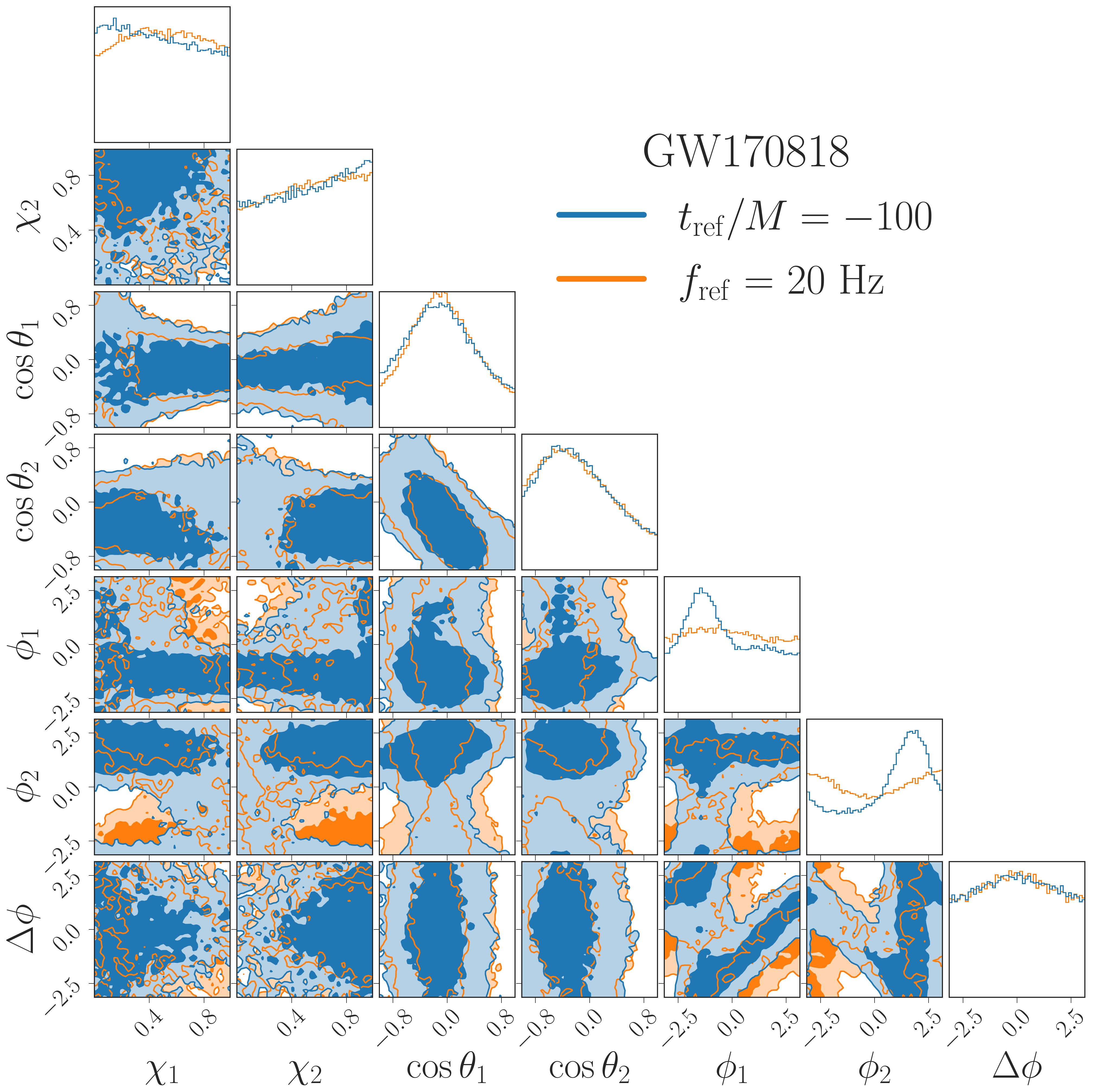} \\ \vspace{0.2cm}
\includegraphics[width=0.4\textwidth]{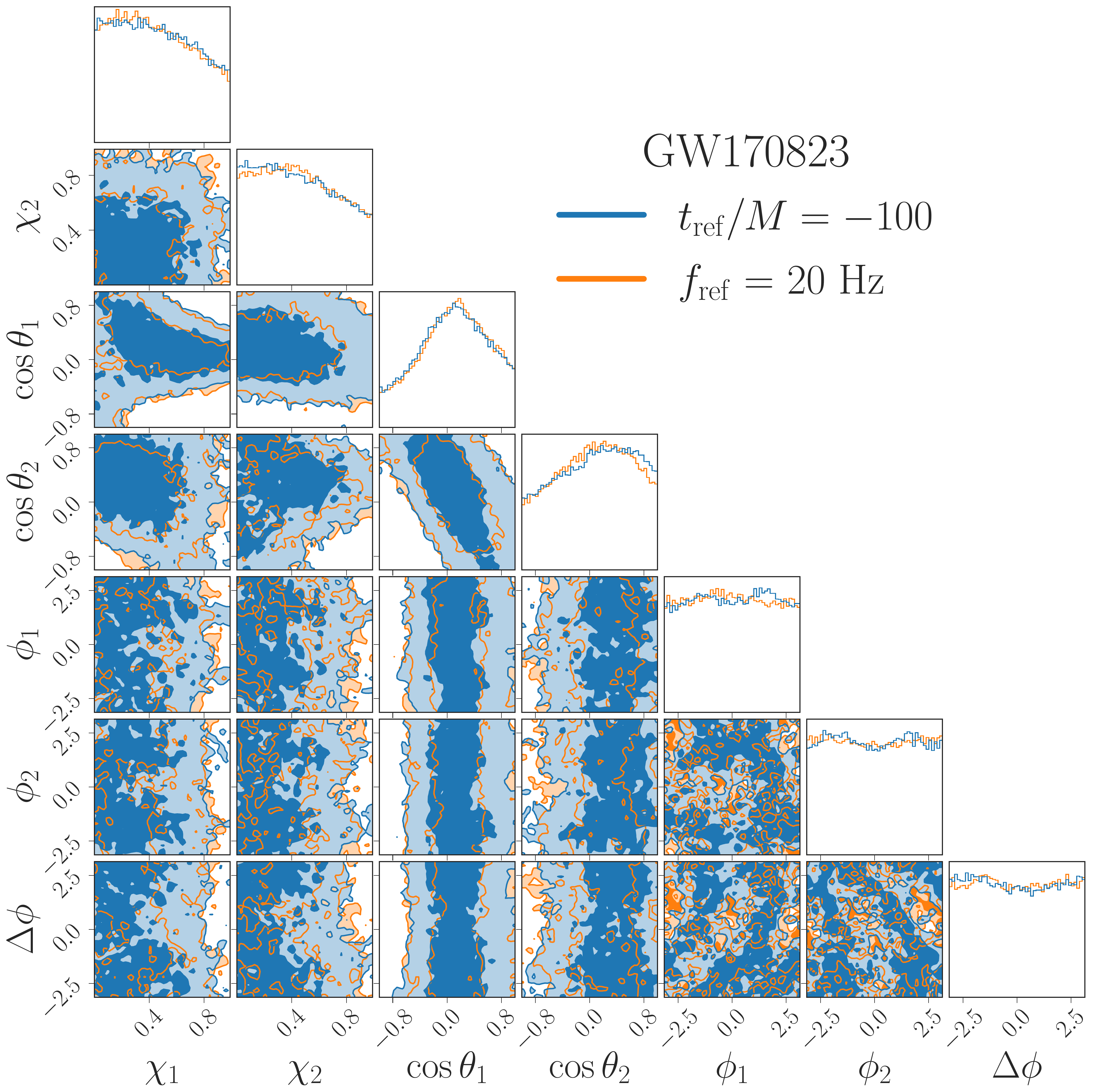}  \hspace{1.5cm}
\includegraphics[width=0.4\textwidth]{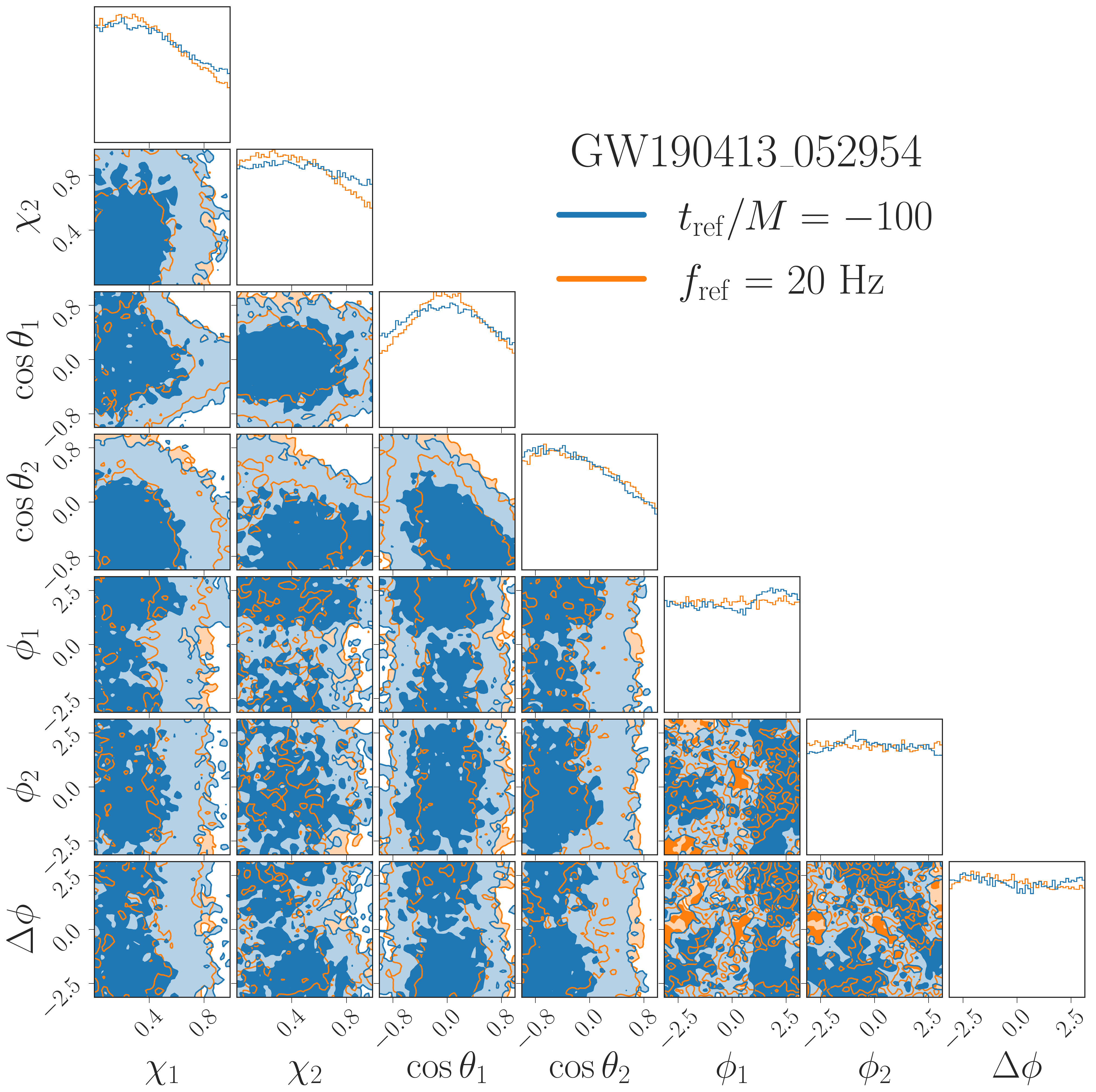}

\caption{
\fullspincaption{1}
}
\label{fig:full_spins_1}
\end{figure*}

\begin{figure*}[p]
\includegraphics[width=0.4\textwidth]{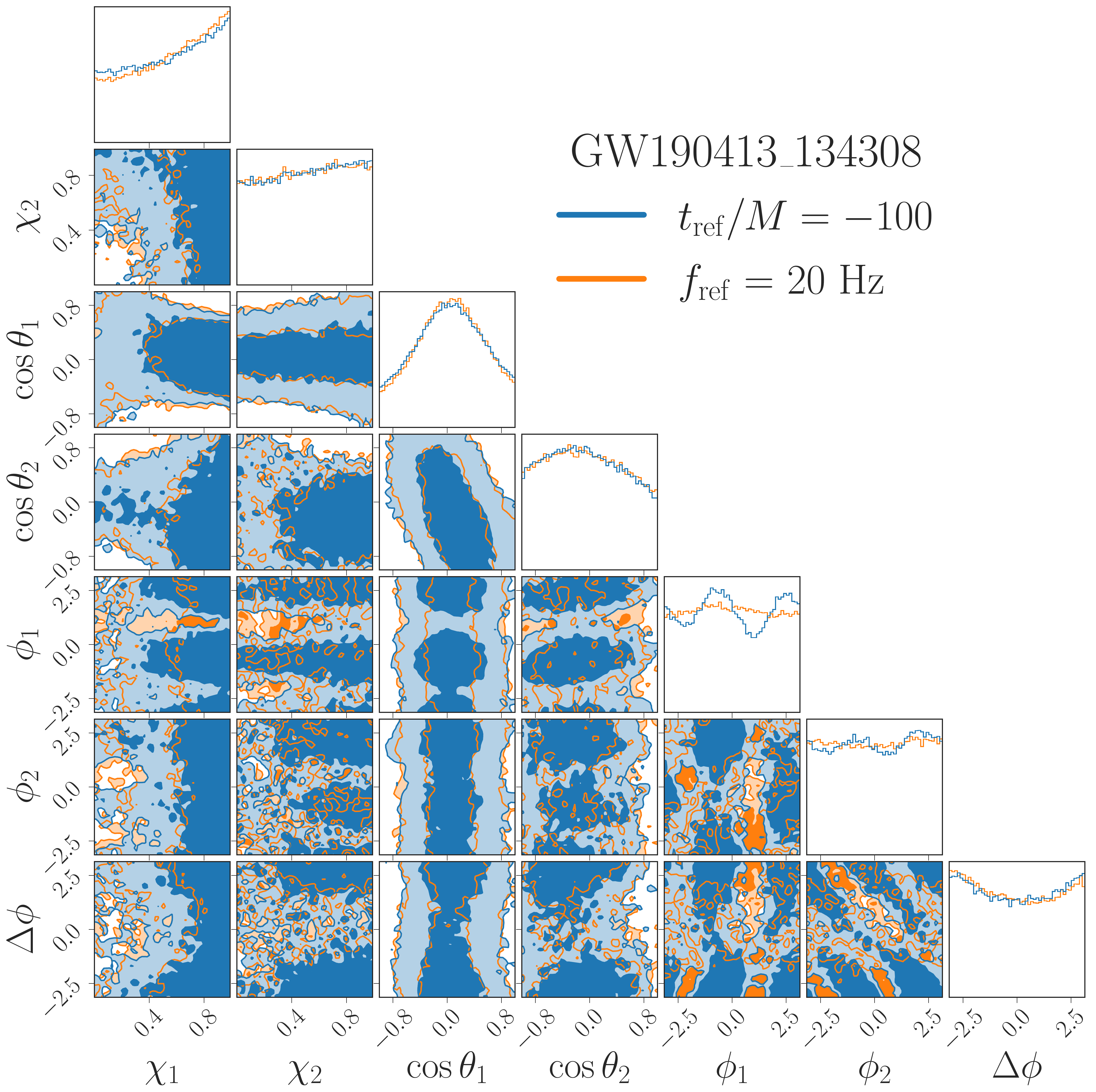} \hspace{1.5cm}
\includegraphics[width=0.4\textwidth]{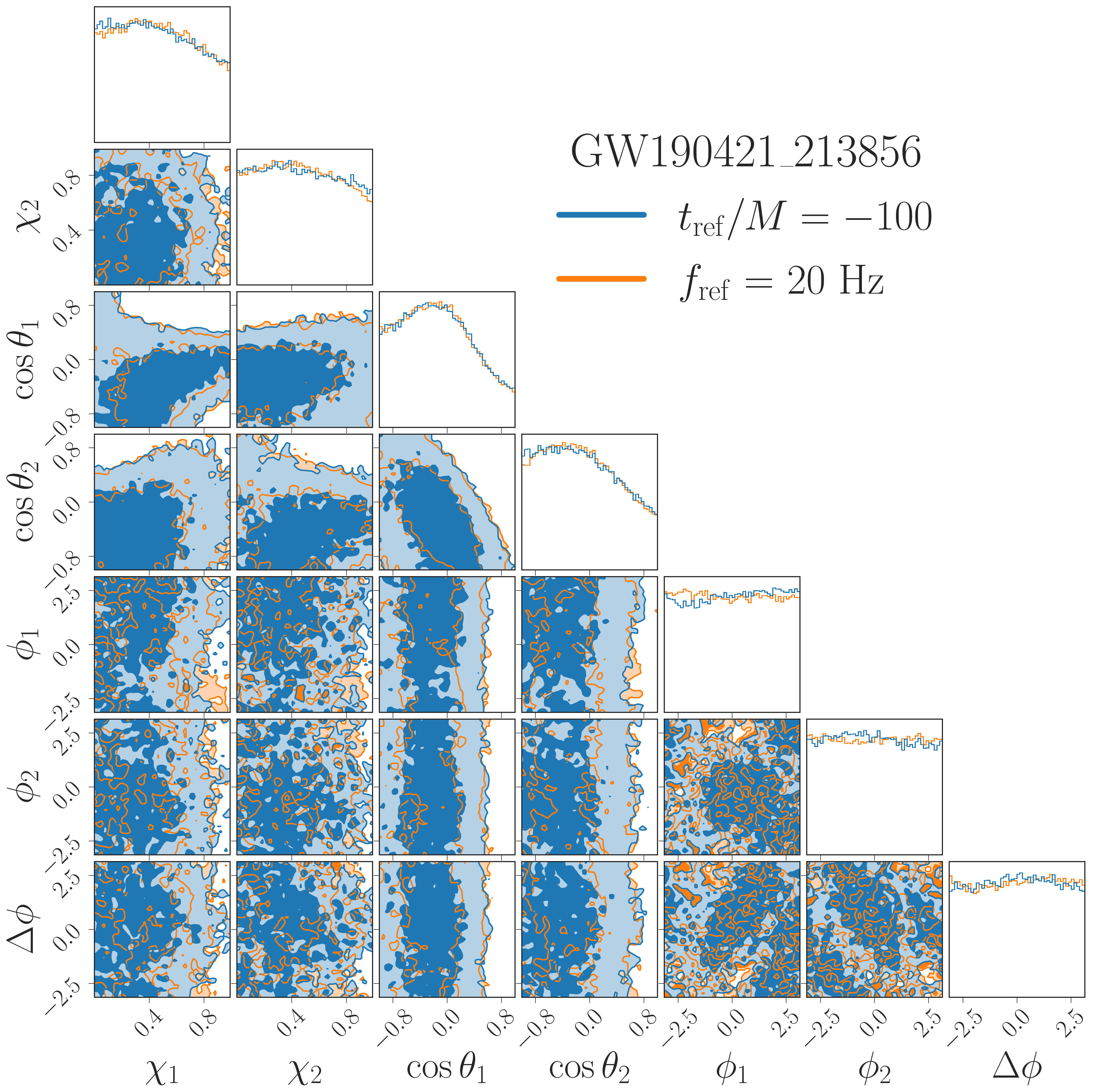} \\ \vspace{0.2cm}
\includegraphics[width=0.4\textwidth]{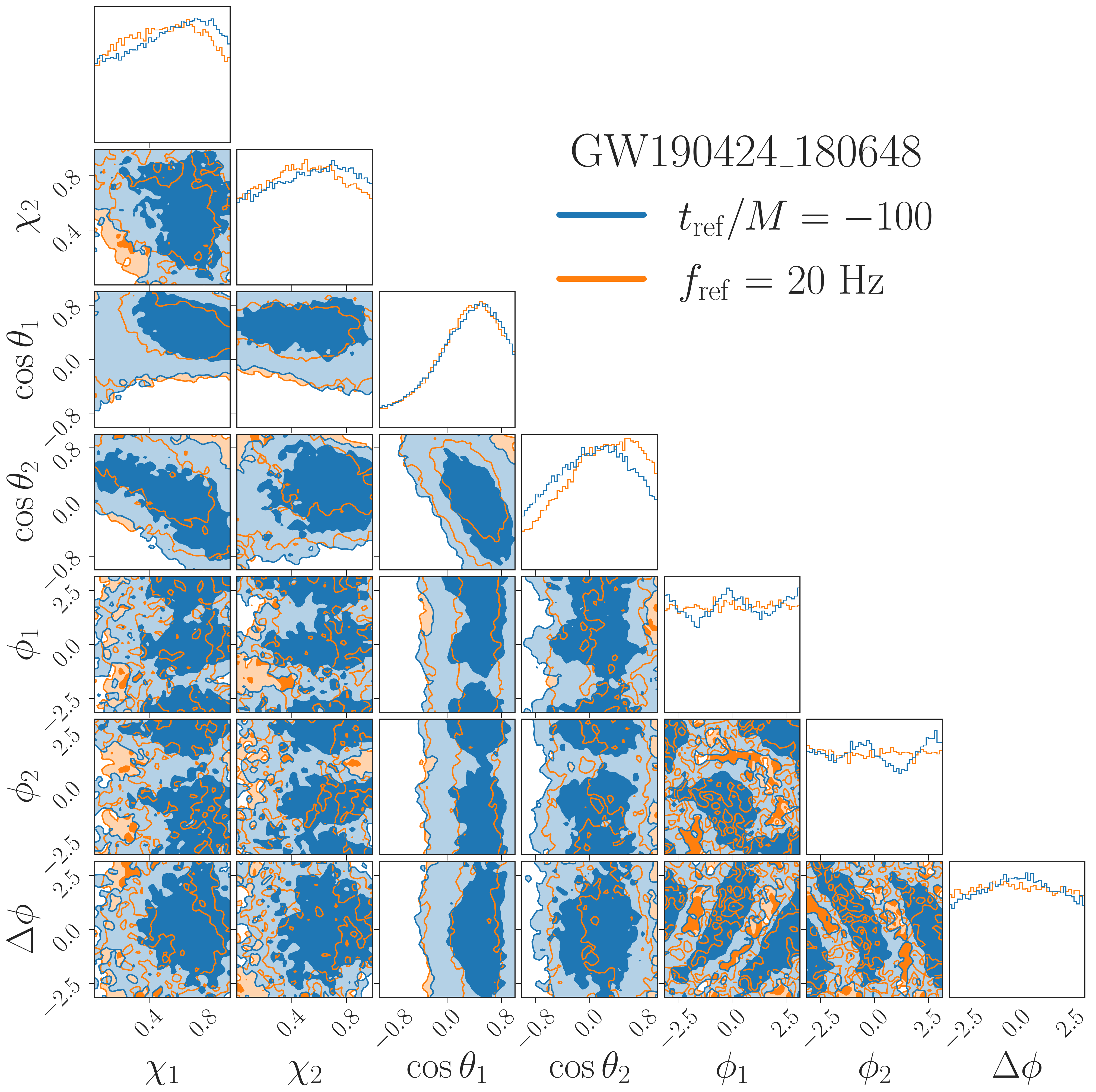} \hspace{1.5cm}
\includegraphics[width=0.4\textwidth]{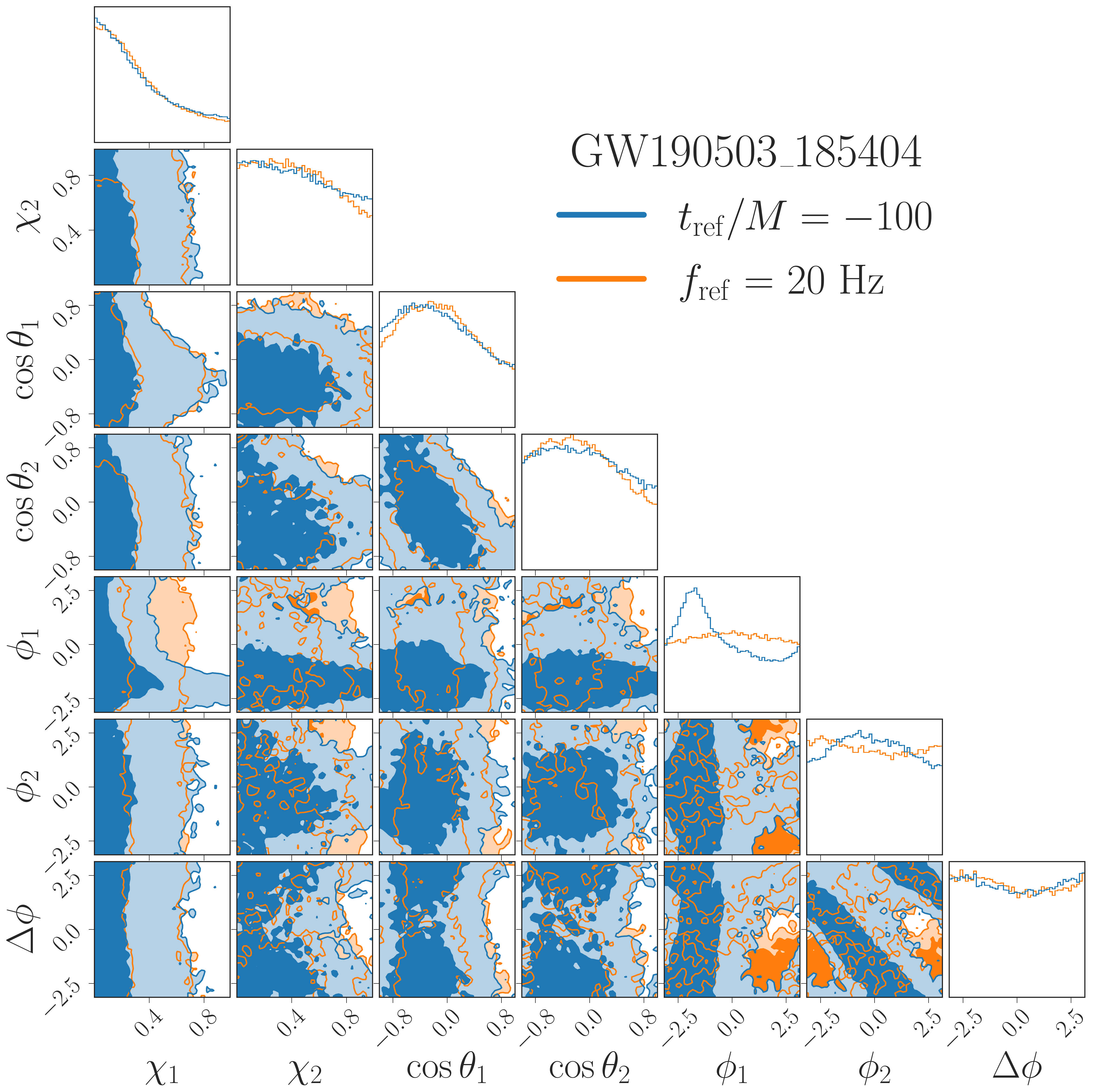} \\ \vspace{0.2cm}
\includegraphics[width=0.4\textwidth]{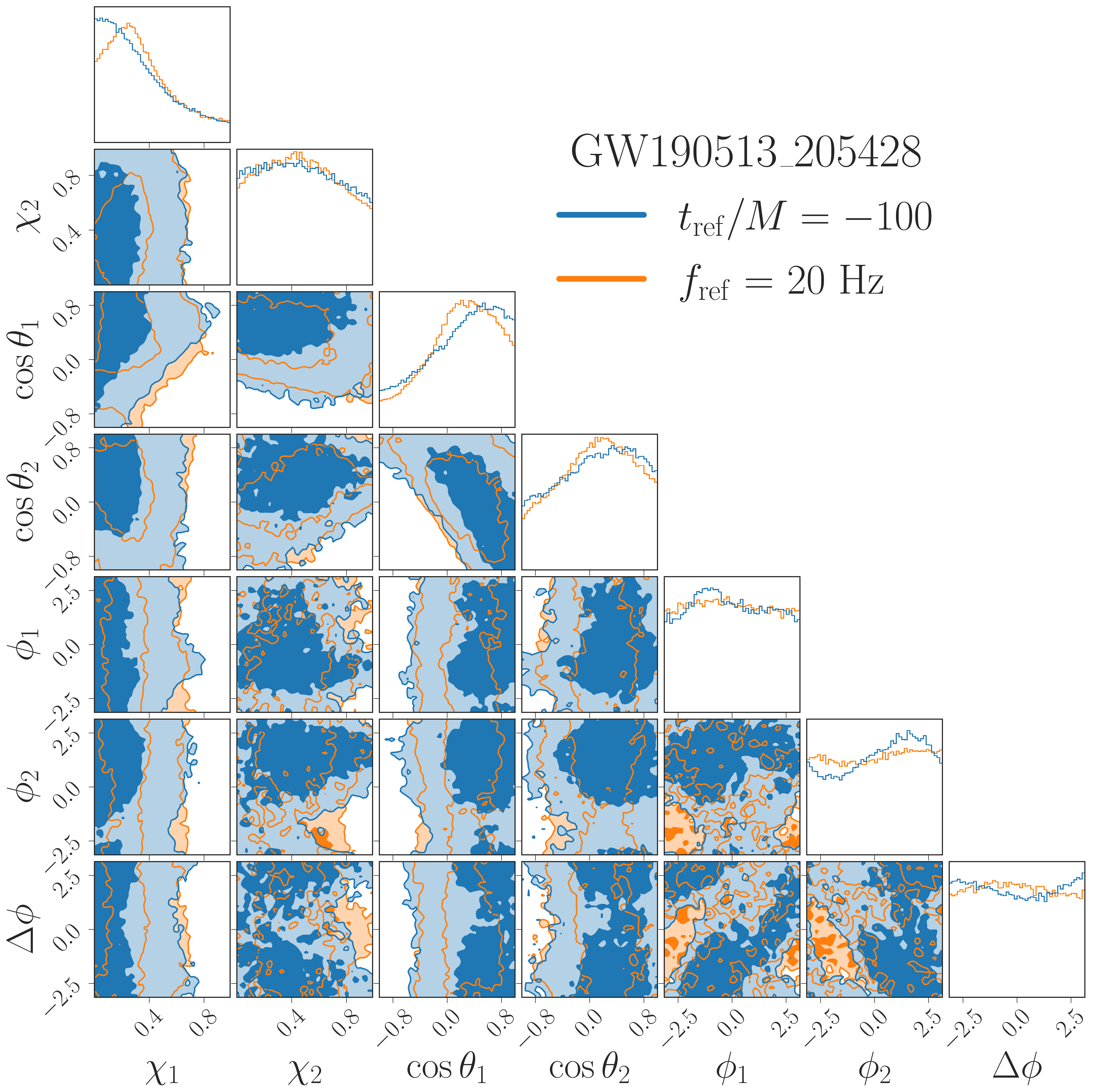} \hspace{1.5cm}
\includegraphics[width=0.4\textwidth]{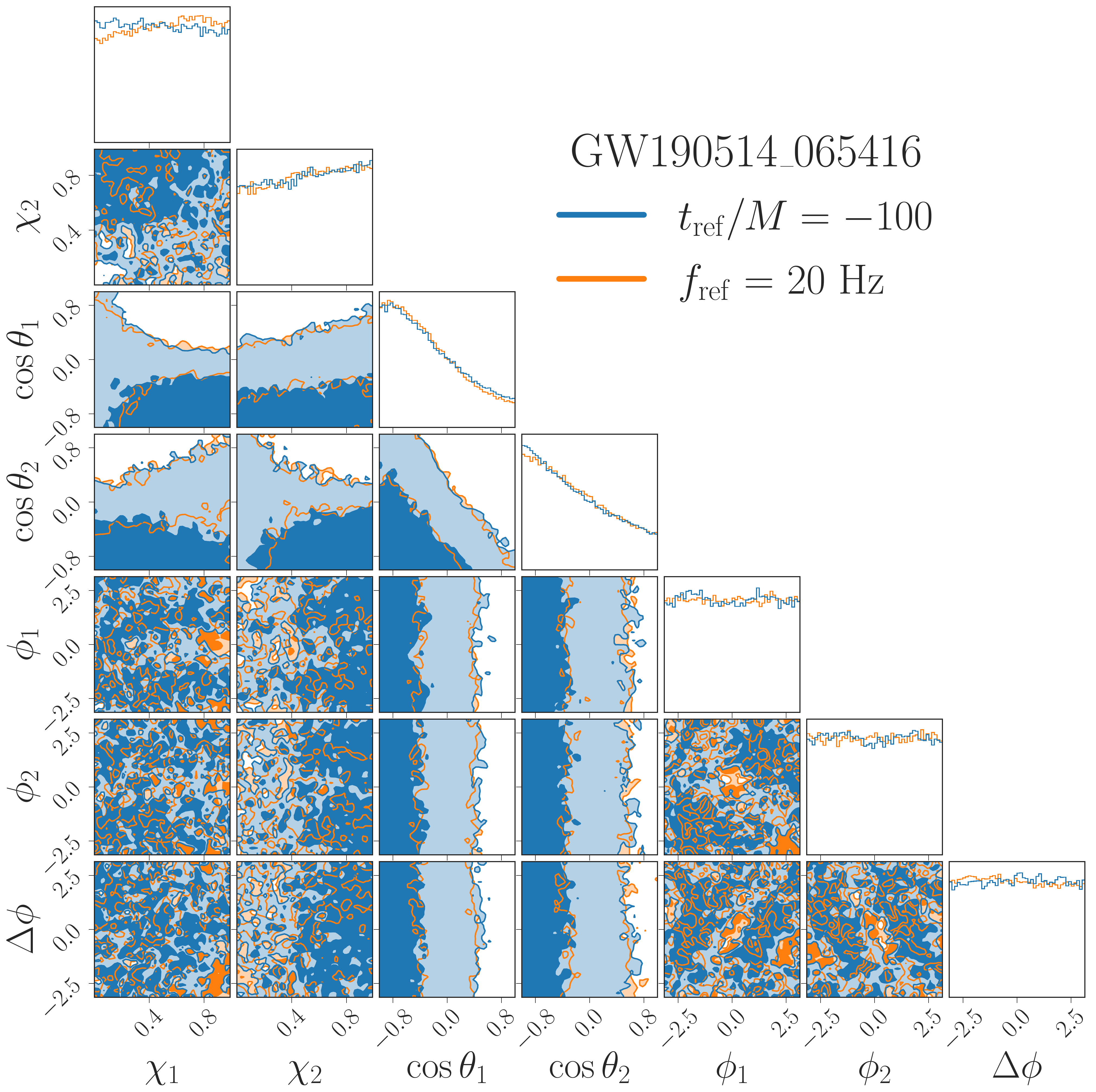}
\caption{
\fullspincaption{2}
}
\label{fig:full_spins_2}
\end{figure*}

\begin{figure*}[p]
\includegraphics[width=0.4\textwidth]{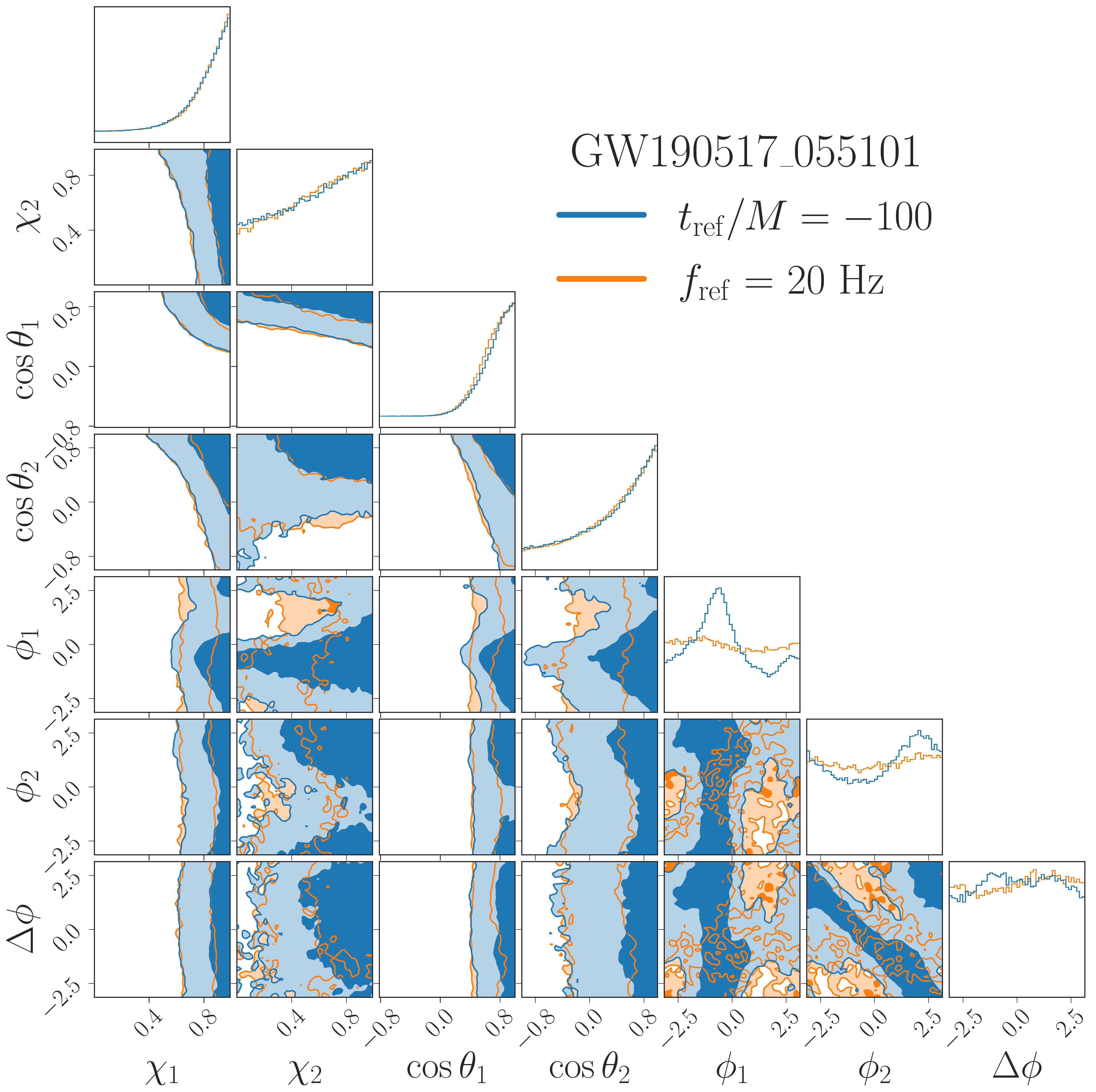}  \hspace{1.5cm}
\includegraphics[width=0.4\textwidth]{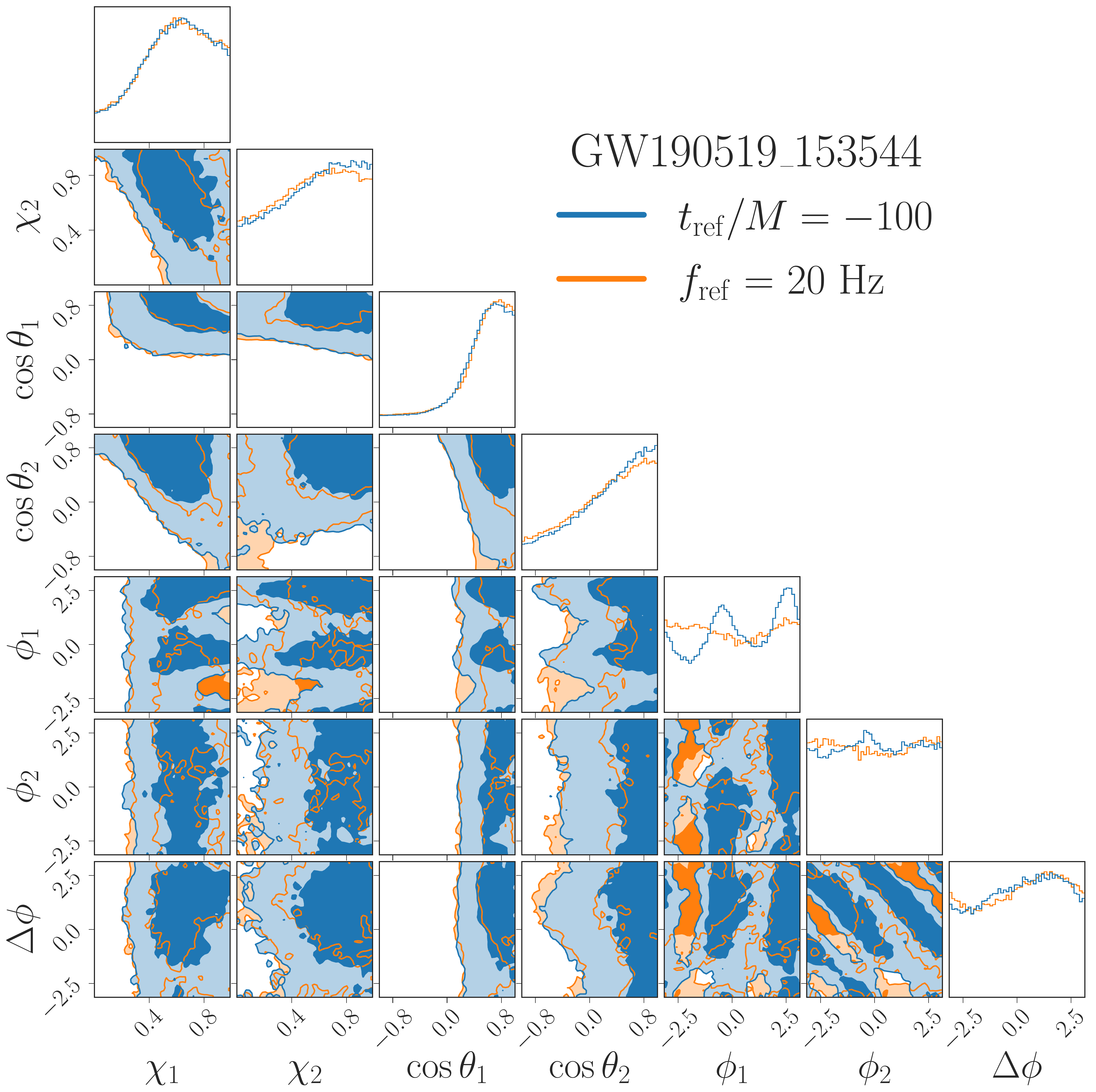}  \\ \vspace{0.2cm}
\includegraphics[width=0.4\textwidth]{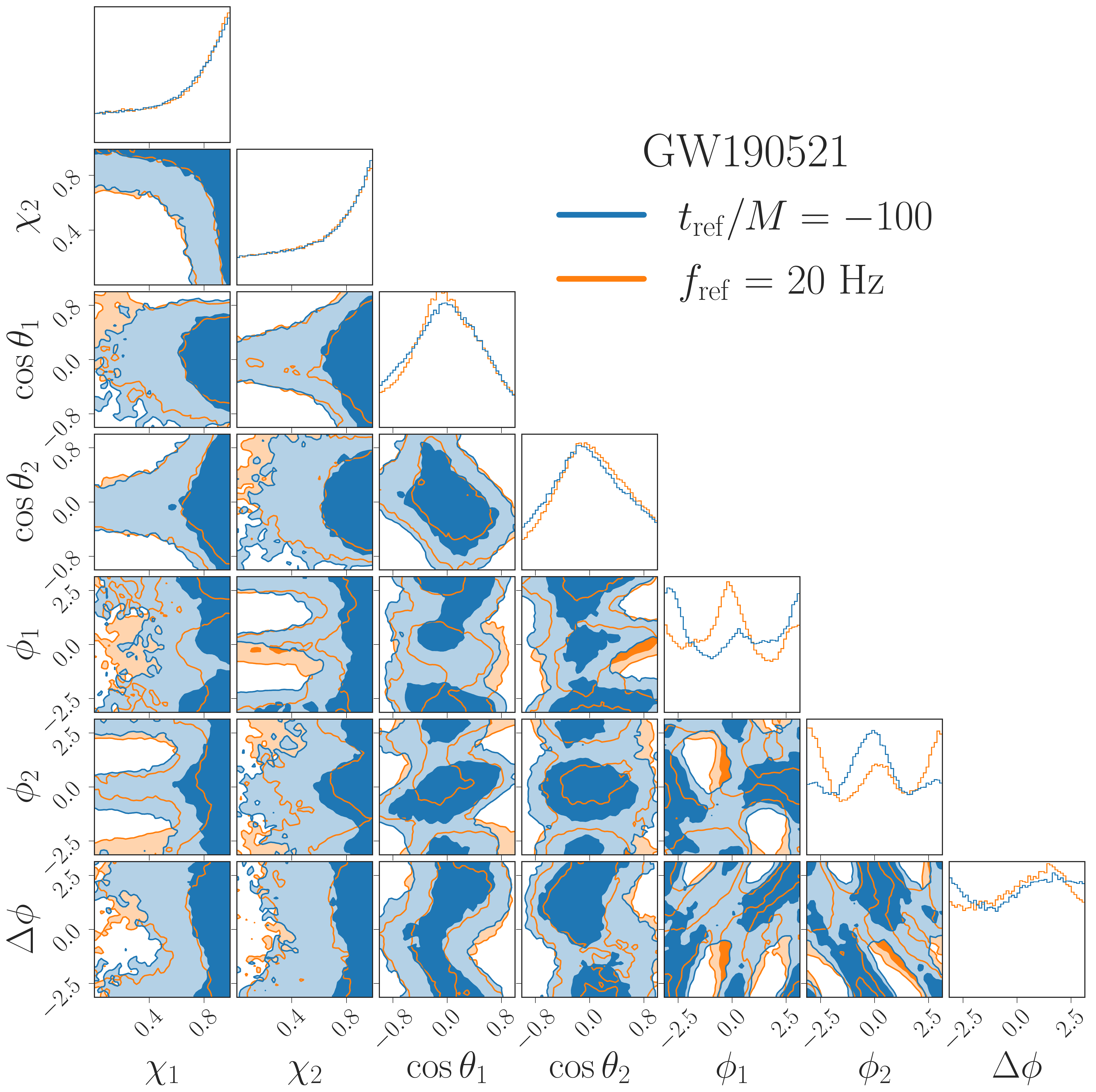}         \hspace{1.5cm}
\includegraphics[width=0.4\textwidth]{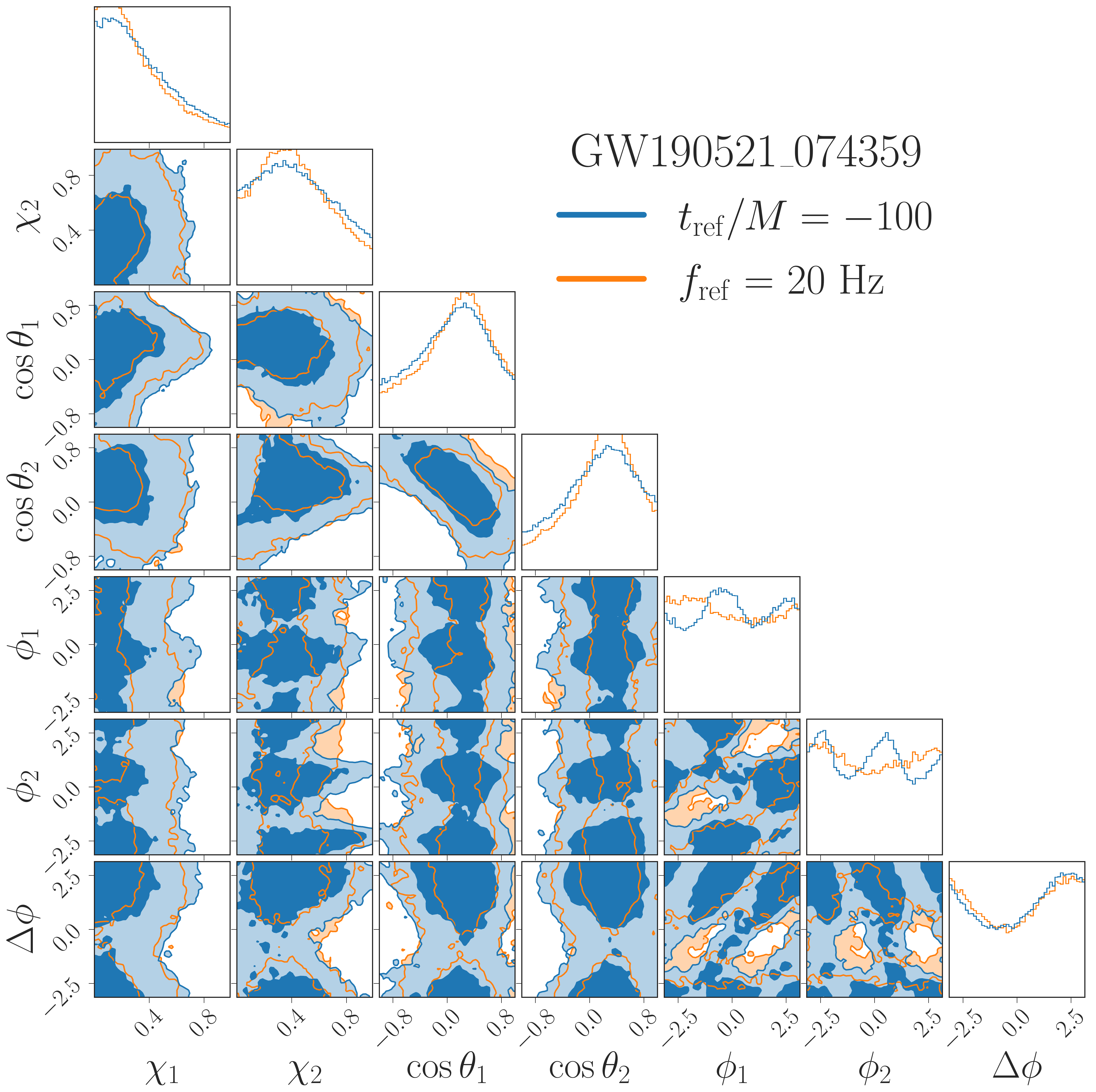}  \\ \vspace{0.2cm}
\includegraphics[width=0.4\textwidth]{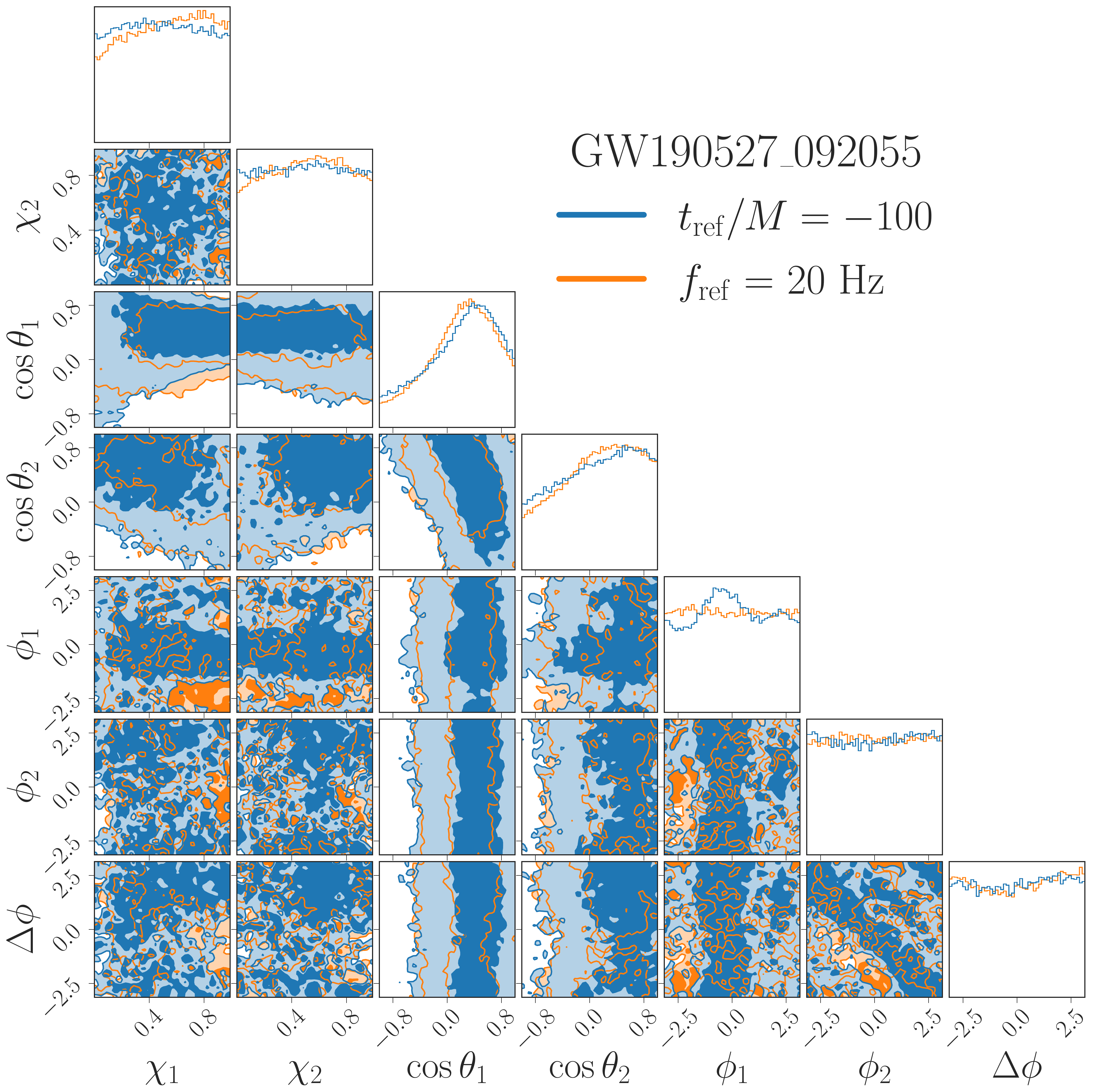}  \hspace{1.5cm}
\includegraphics[width=0.4\textwidth]{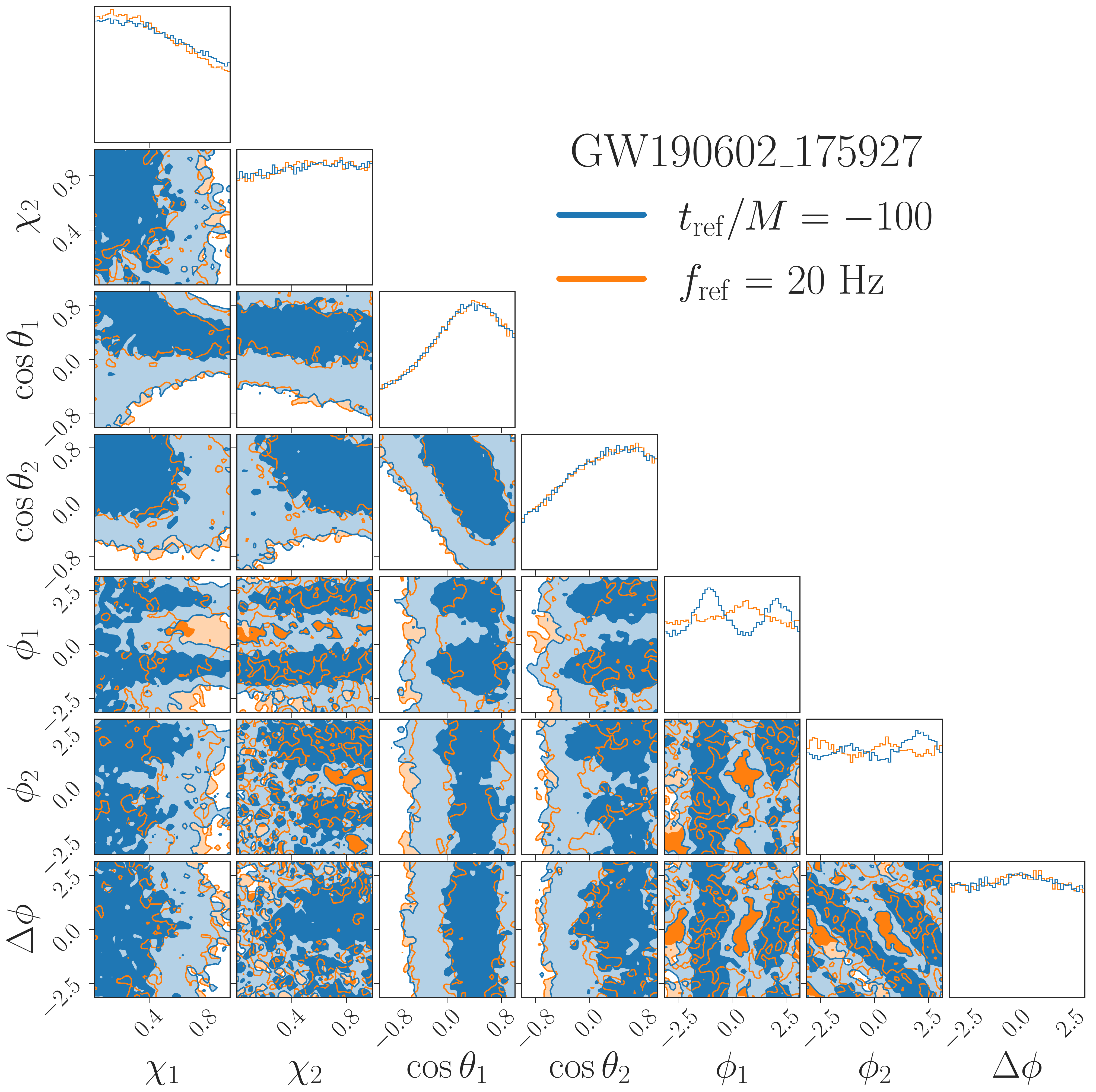}
\caption{
\fullspincaption{3}
}
\label{fig:full_spins_3}
\end{figure*}

\begin{figure*}[p]
\includegraphics[width=0.4\textwidth]{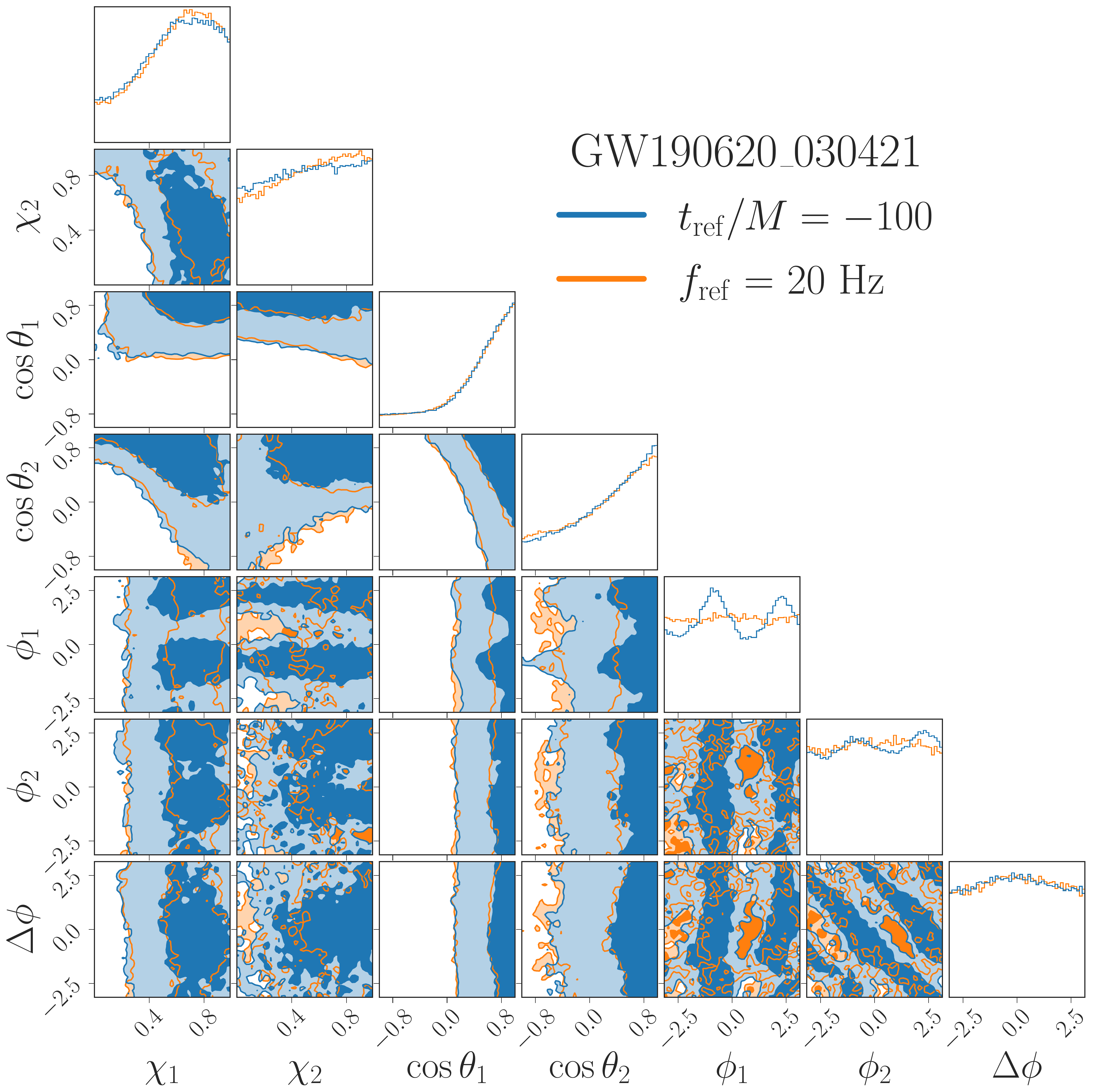}  \hspace{1.5cm}
\includegraphics[width=0.4\textwidth]{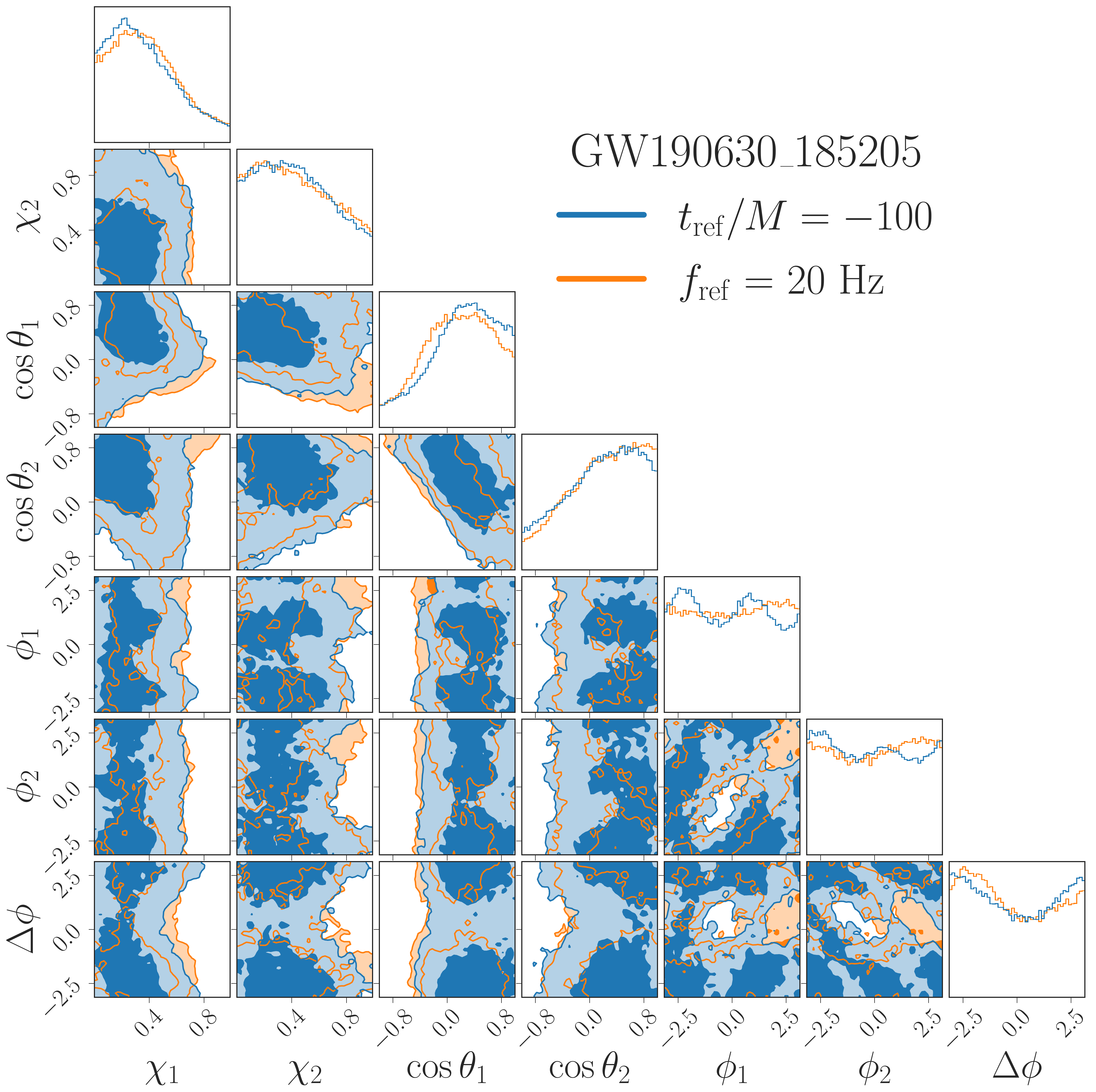}  \\ \vspace{0.2cm}
\includegraphics[width=0.4\textwidth]{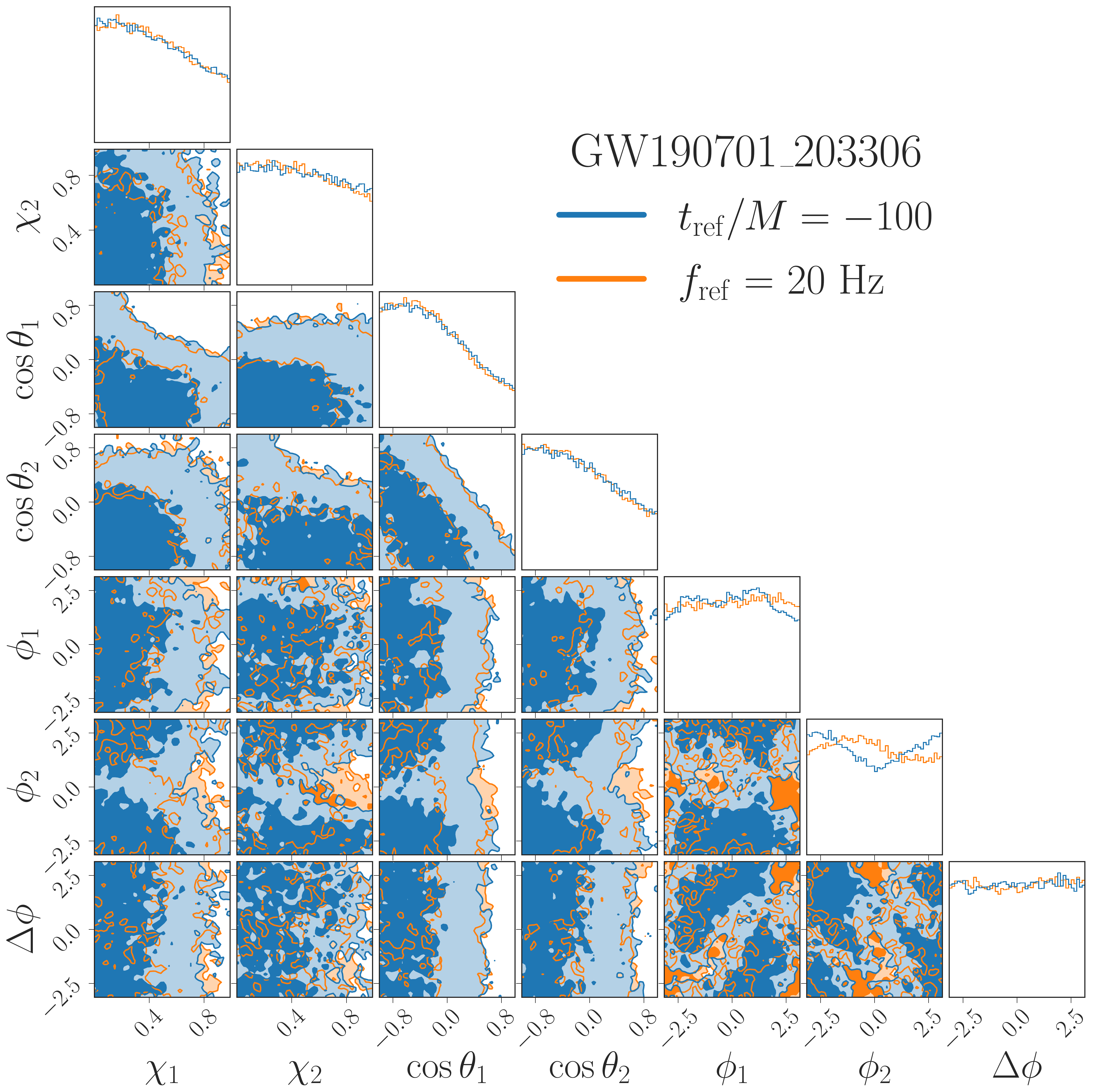}  \hspace{1.5cm}
\includegraphics[width=0.4\textwidth]{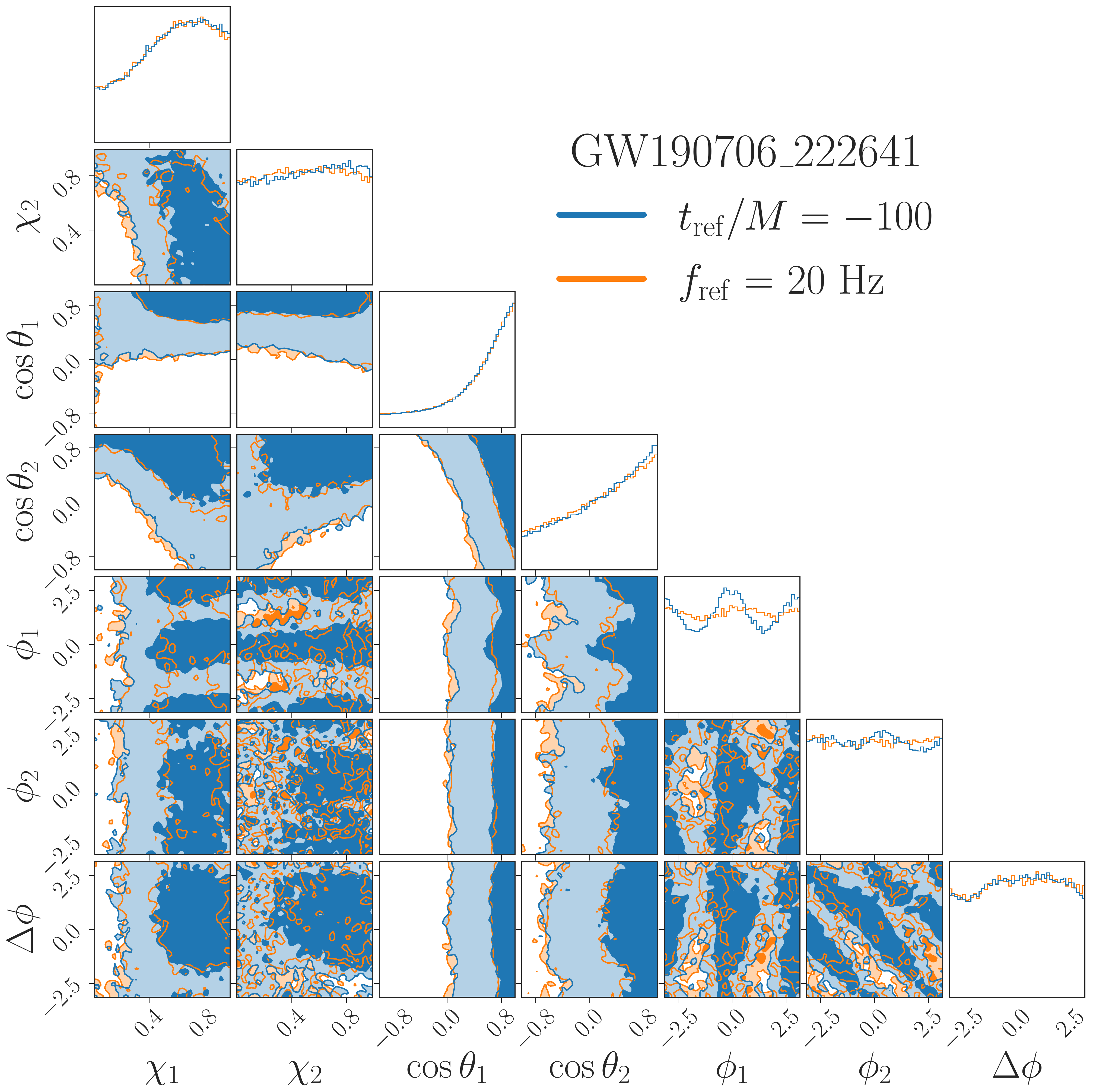}  \\ \vspace{0.2cm}
\includegraphics[width=0.4\textwidth]{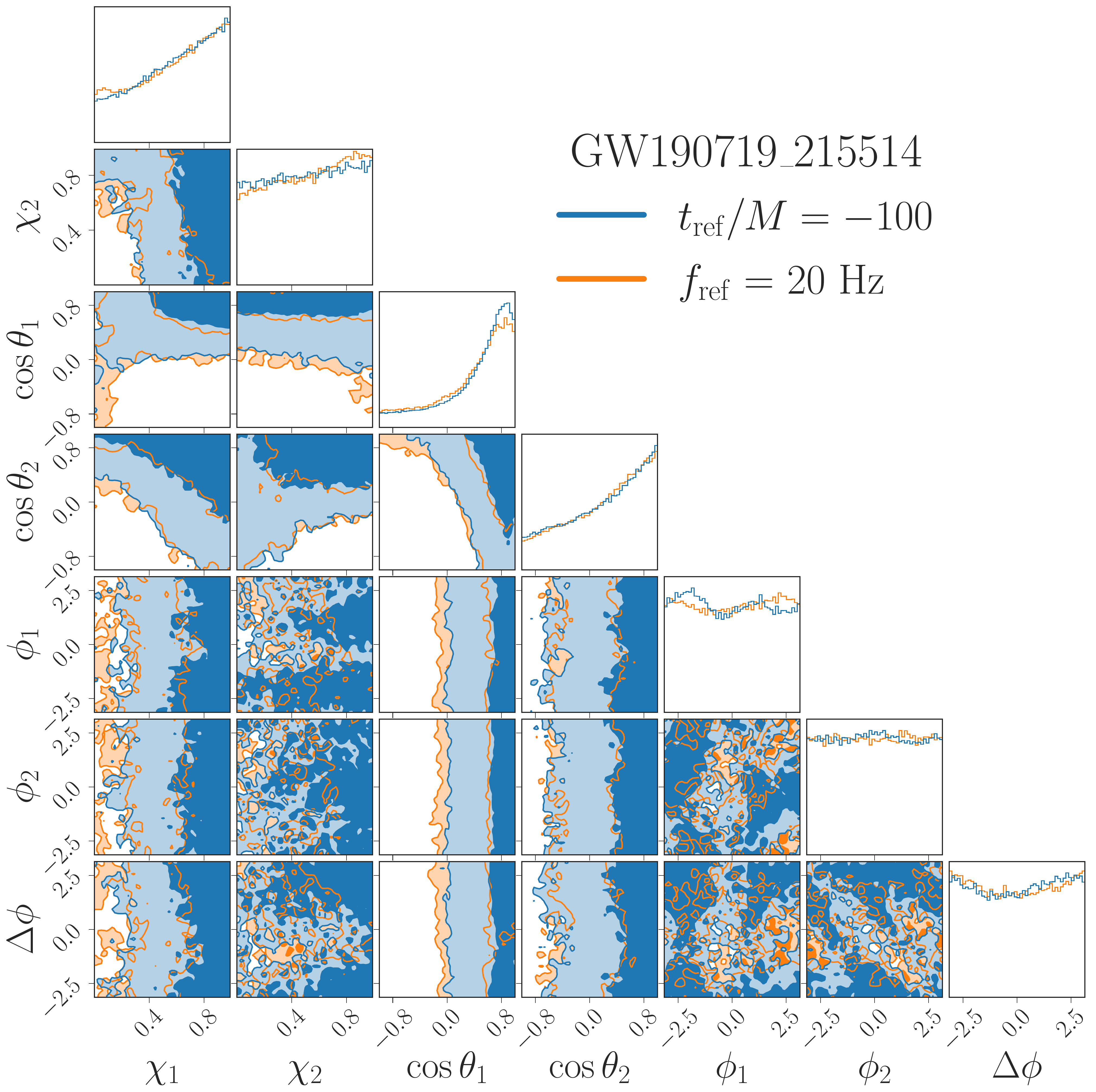}  \hspace{1.5cm}
\includegraphics[width=0.4\textwidth]{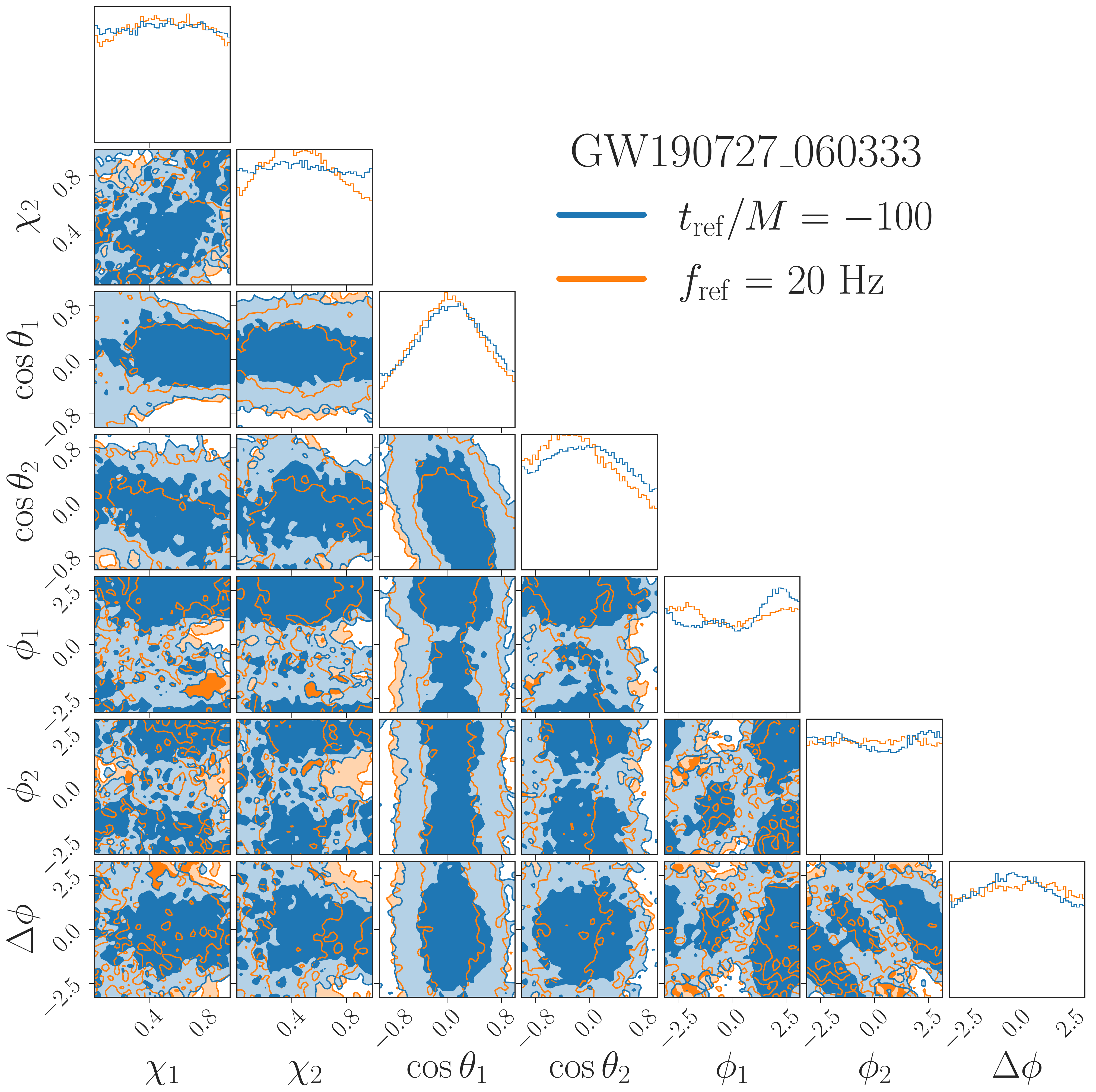}
\caption{
\fullspincaption{4}
}
\label{fig:full_spins_4}
\end{figure*}

\begin{figure*}[p]
\includegraphics[width=0.4\textwidth]{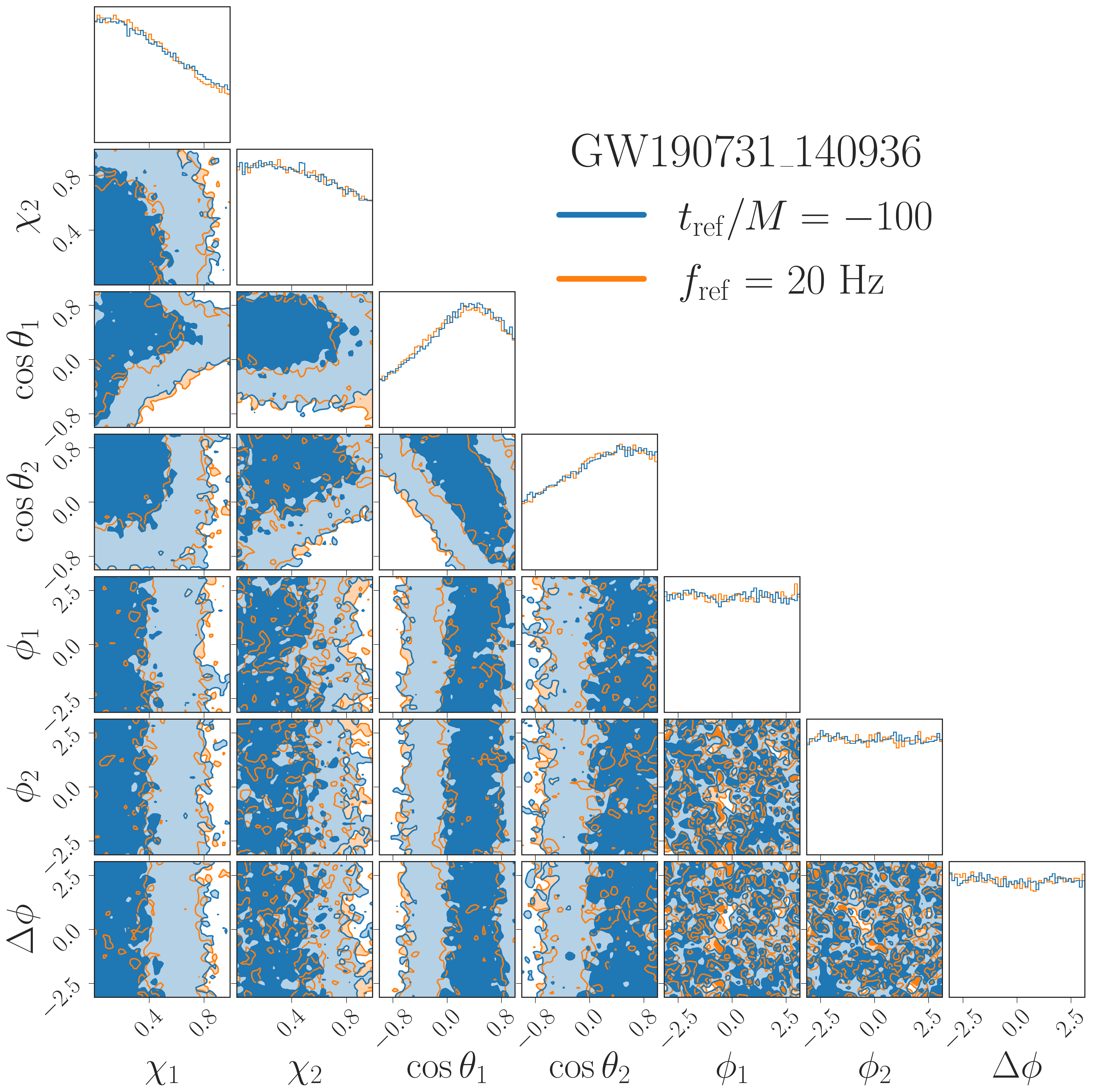}  \hspace{1.5cm}
\includegraphics[width=0.4\textwidth]{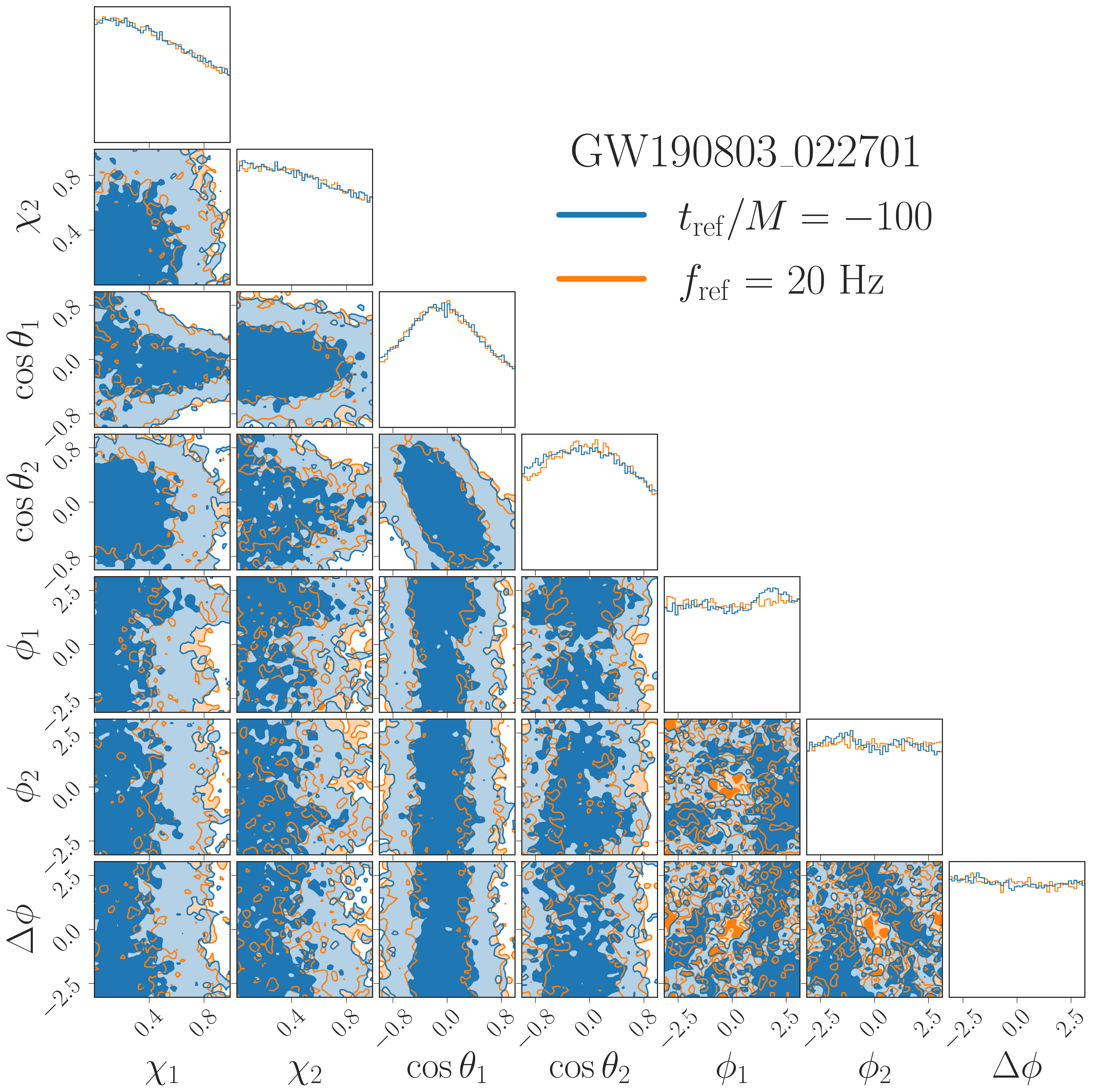}  \\ \vspace{0.2cm}
\includegraphics[width=0.4\textwidth]{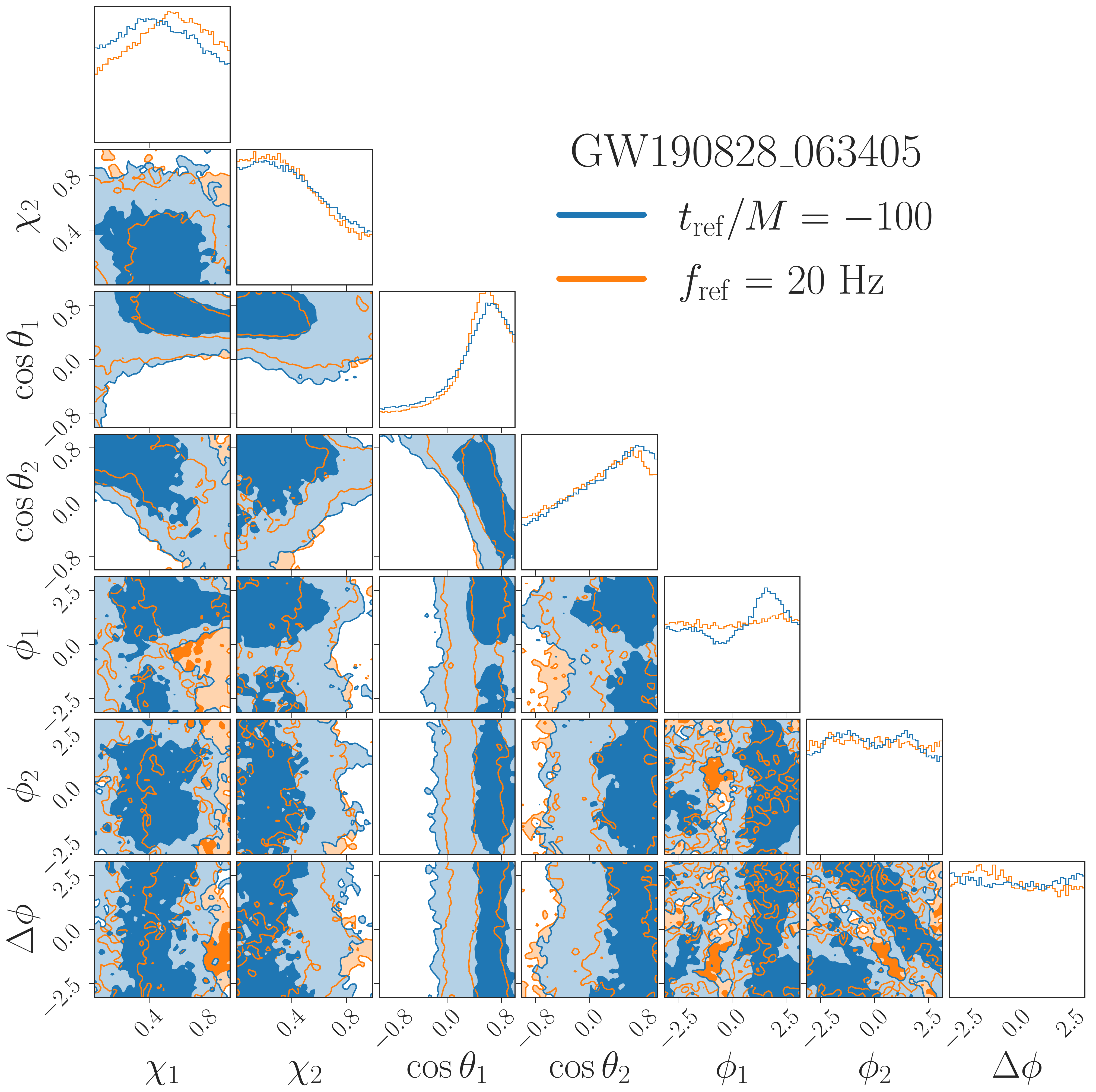}  \hspace{1.5cm}
\includegraphics[width=0.4\textwidth]{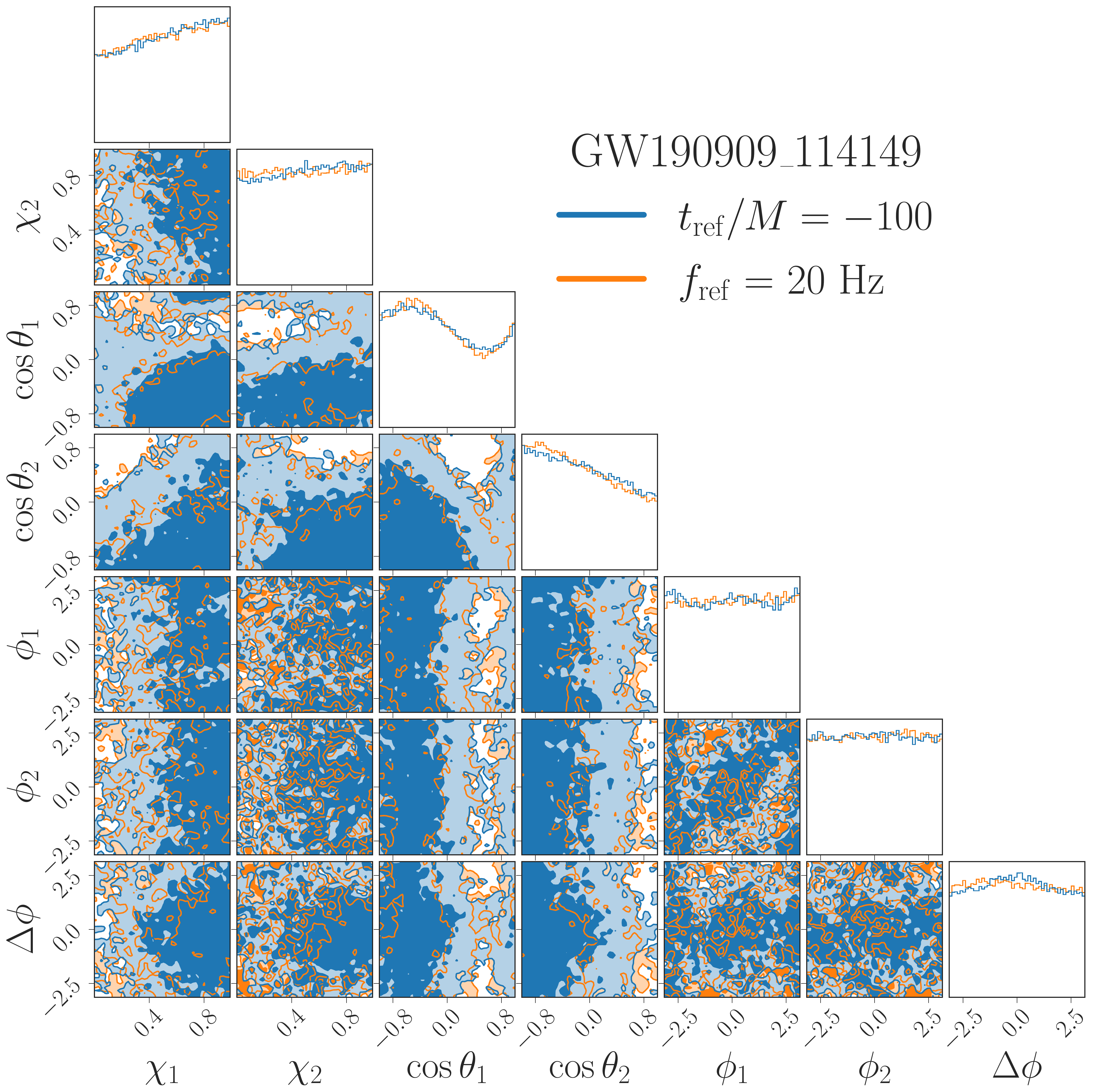}  \\ \vspace{0.2cm}
\includegraphics[width=0.4\textwidth]{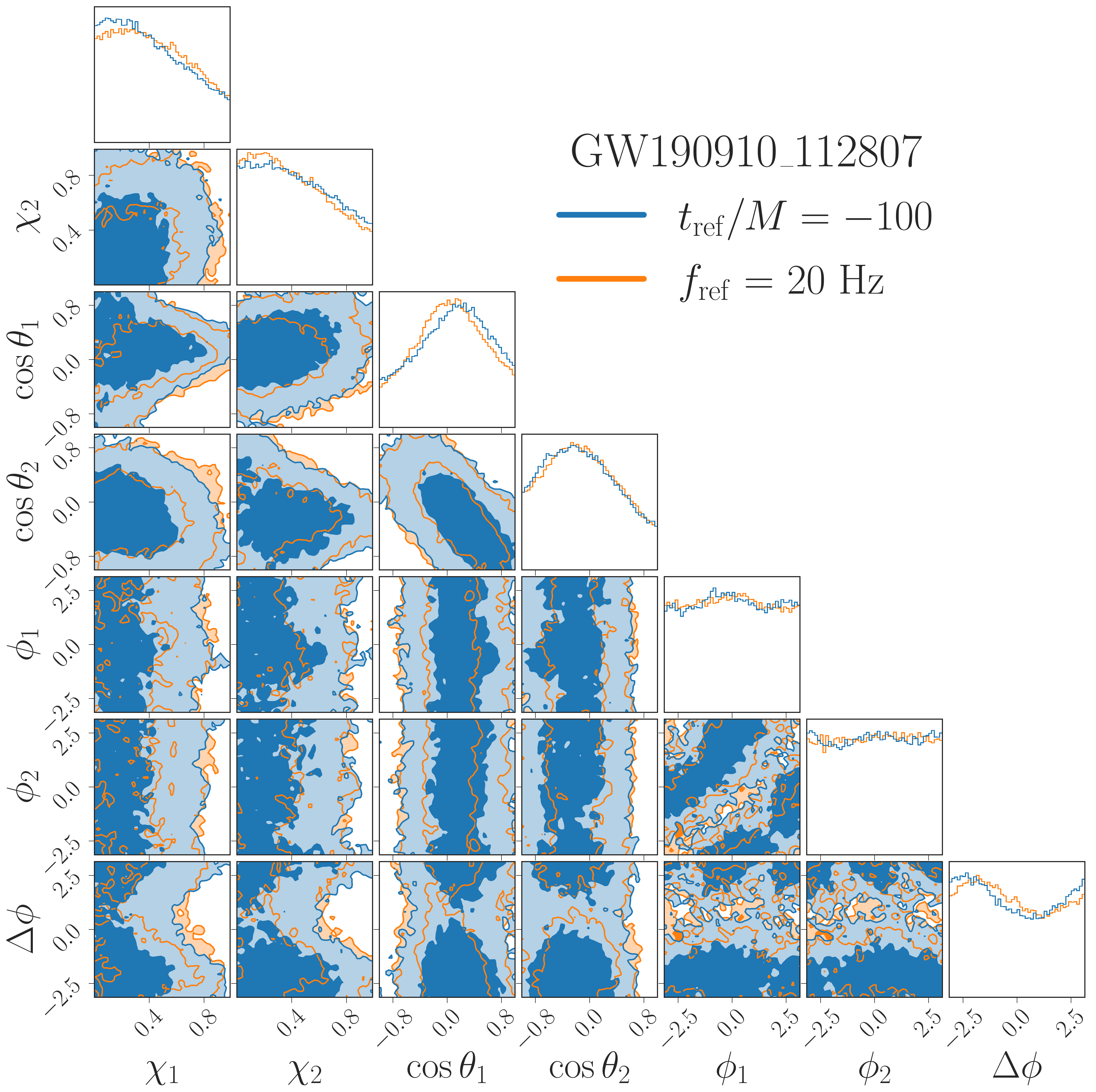}  \hspace{1.5cm}
\includegraphics[width=0.4\textwidth]{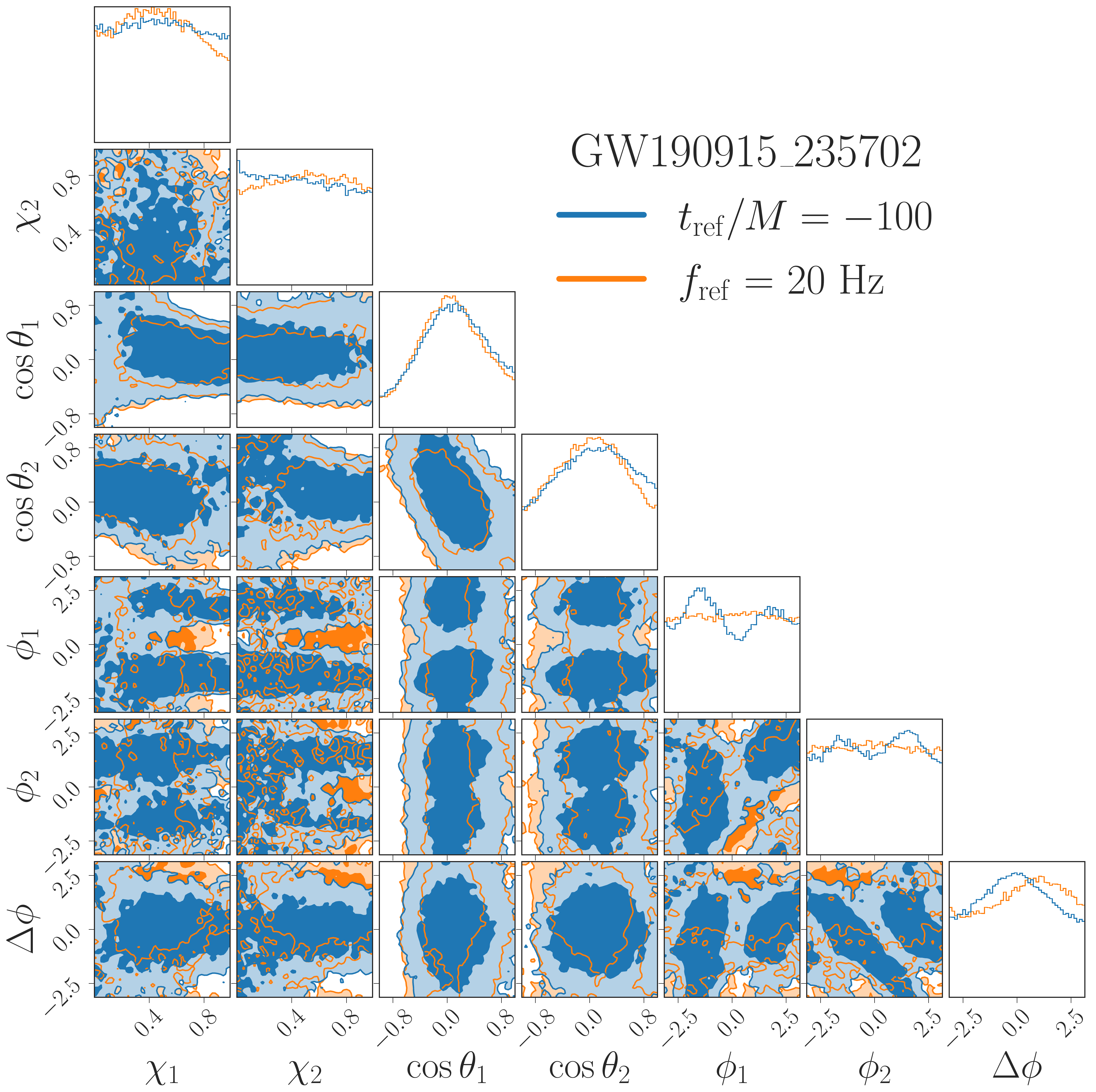}
\caption{
\fullspincaption{5}
}
\label{fig:full_spins_5}
\end{figure*}

\begin{figure}[thb]
\includegraphics[width=0.48\textwidth]{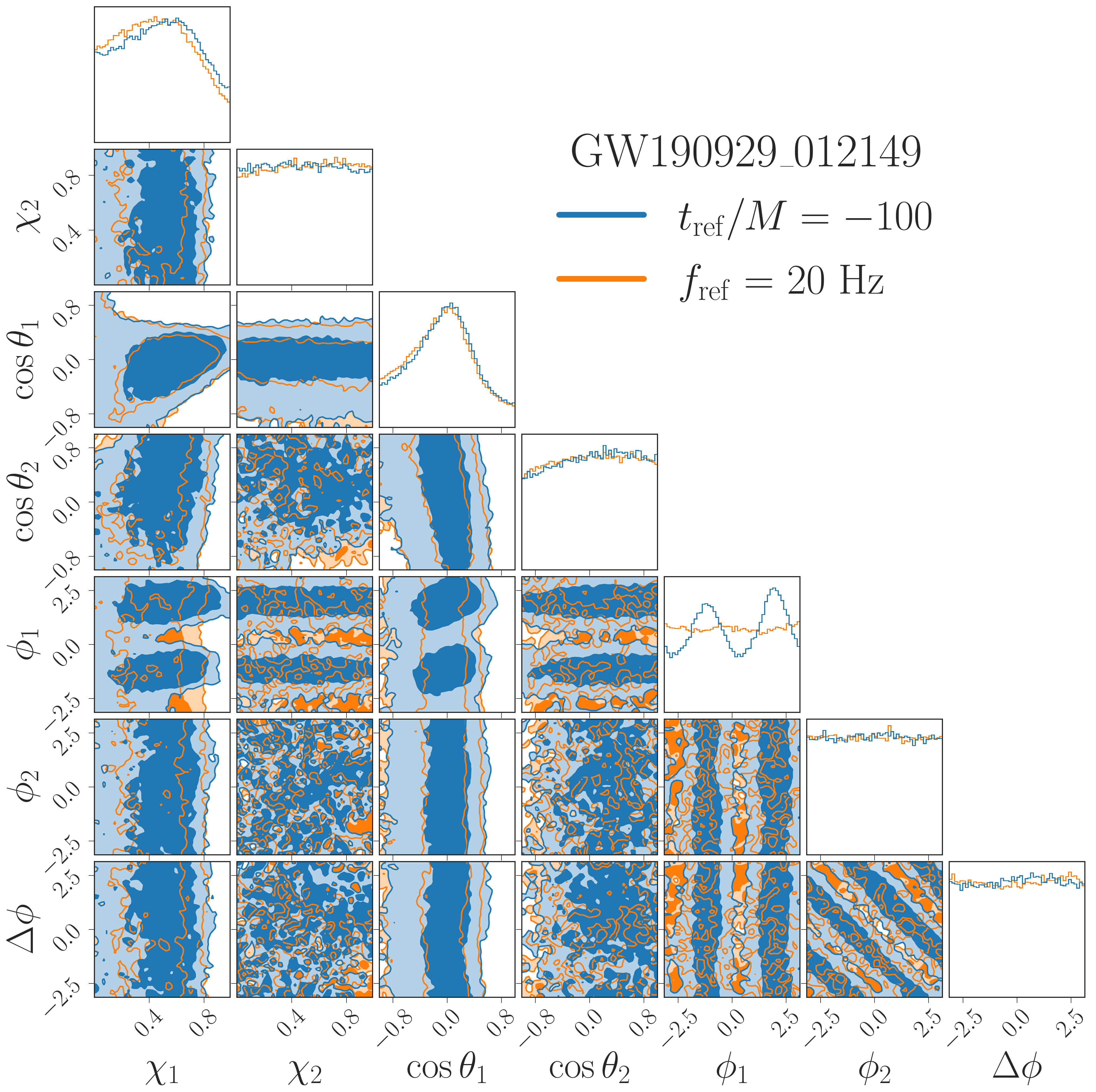}
\caption{
\fullspincaption{6}
}
\label{fig:full_spins_6}
\end{figure}

\section{Full spin posteriors for \NRSur}
\label{sec:app_full_spins}

For completeness, in Figs.~\ref{fig:full_spins_1} -- \ref{fig:full_spins_6} we
show the full spin posteriors for the \numEv events listed in
Tab.~\ref{tab:sur_events}, generated using the \NRSur model at $\trefmHundredM$
and $\frefTwentyHz$. We show the spin magnitudes $\chi_{1,2}$, cosines of the
tilt angles $\cos{\theta_{1,2}}$, and orbital-plane spin angles $\phi_1$,
$\phi_2$, and $\delphi$. The priors on all of these parameters are flat in
their respective ranges (cf. Sec.~\ref{sec:pe_setup}). The spin magnitude and
tilt posteriors are consistent between the two reference points, and are consistent with Ref.~\cite{Abbott:2020niy}.

In the following, we refer exclusively to the spin measurements at
$\trefmHundredM$, and check whether measurements of the orbital-plane spin
angles can be tied to a measurement of $\chi_{1,2}$ and $\cos{\theta_{1,2}}$.
As noted in Sec.~\ref{sec:all_phi_posteriors}, an unambiguous measurement of
the orbital-plane spin angles relies on being able to constrain $\chi_{1,2}$
away from zero, and $\cos{\theta_{1,2}}$ away from $\pm 1$. GW190521
(Fig.~\ref{fig:full_spins_3}) is a good example of this: there is a clear
preference for large $\chi_{1,2}$ and $\cos{\theta_{1,2}} \sim 0$, which likely
enables the $\phi_1$ and $\phi_2$ measurement for this event. On the other
hand, for GW190517\und055101 (Fig.~\ref{fig:full_spins_3}), there is a
preference for large $\chi_{1,2}$, but with $\cos{\theta_{1,2}} \sim 1$.
However, we still see peaks in the $\phi_1$ and $\phi_2$ distributions.
Finally, for GW170818 (Fig.~\ref{fig:full_spins_1}), there is a mild
preference for small $\chi_1$ and a similarly mild preference for large
$\chi_2$, but this is the event with the best $\phi_1$ and $\phi_2$ measurement
(cf. Fig.~\ref{fig:phi_all_comparison}). We conclude that current constraints
on the spin magnitudes and tilts are too broad to look for such correlations
with the orbital-plane spin angles: even for the cases where we see a
preference for small $\chi_{1,2}$ and/or $\cos{\theta_{1,2}} \sim \pm 1$, there
is enough support for large $\chi_{1,2}$ and $\cos{\theta_{1,2}} \sim 0$, that
there can be peaks in the posteriors of the orbital-plane spin angles.

\begin{figure*}[p]
\includegraphics[width=\textwidth]{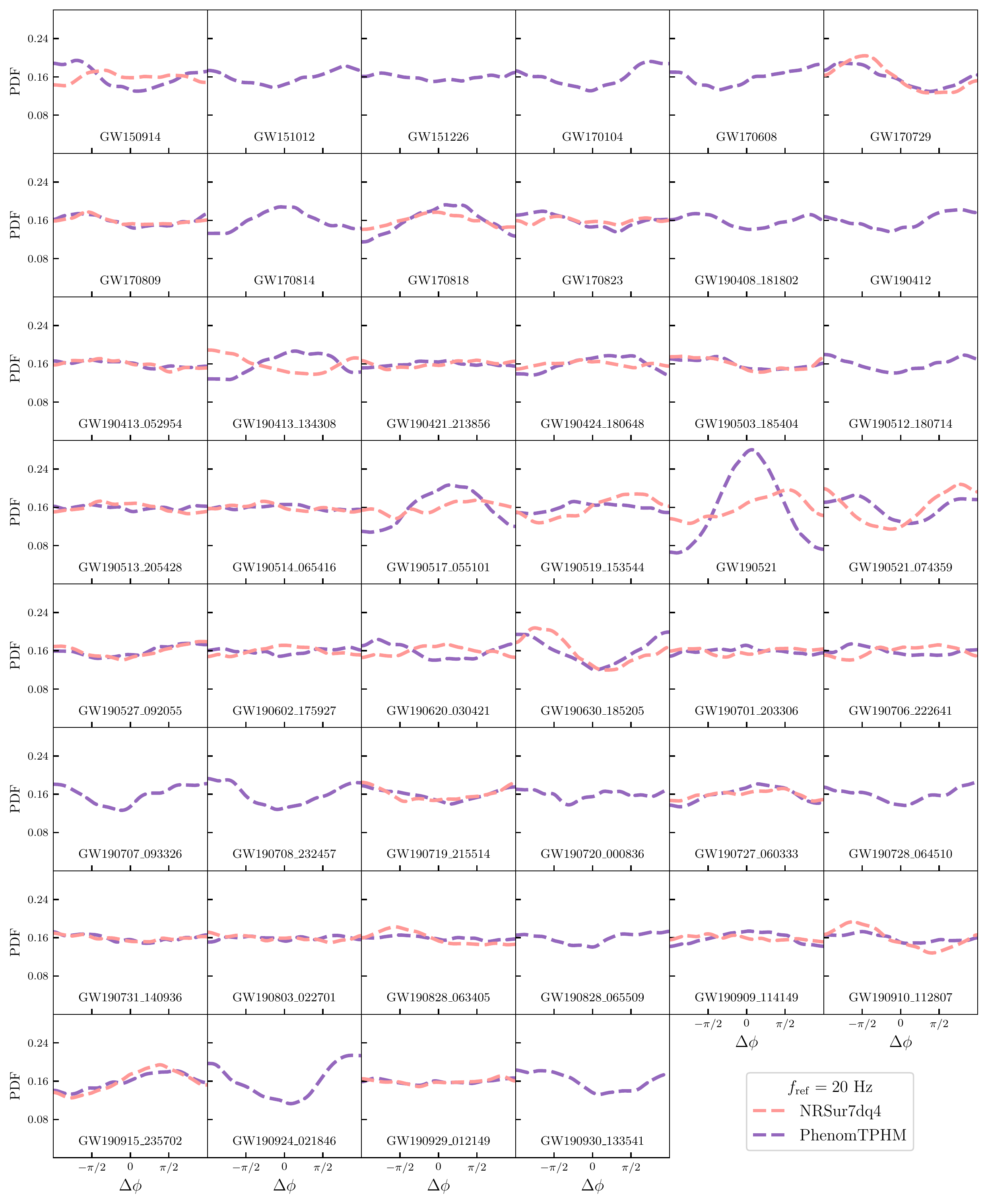}
\caption{
$\delphi$ posteriors at $\frefTwentyHz$ for all 46 GWTC-2 binary BH events,
obtained using the \PhenomT model. We also show the corresponding \NRSur
posterior for the \numEv \NRSur events for comparison.
}
\label{fig:phi_all_comparison_delphi_combined}
\end{figure*}

\section{Results for all GWTC-2 events using \PhenomT}
\label{sec:app_PhenomT_posteriors}

The results in Sec.~\ref{sec:spin_measurements} were restricted to the \numEv
GWTC-2 with $M\gtrsim60M_{\odot}$ due to the length restrictions of \NRSur. The
remaining 15 events with $M\lesssim60M_{\odot}$ are listed in
Tab.~\ref{tab:phenom_events}. For completeness, we now analyze all 46 GWTC-2
binary BH events using \PhenomT. We choose \PhenomT as this model performed
better than \PhenomX in Sec.~\ref{sec:nr_inj}. Once again, for simplicity, we
only consider $\frefTwentyHz$ for \PhenomT. For the 15 events with $M\lesssim
60 M_{\odot}$, we relax the prior constraints described in
Sec.~\ref{sec:pe_setup} to: $5\leq\mathcal{M}\leq400$, $q \leq 20$, and $10\leq
M \leq 400$.

\begin{table}[thb]
\centering
\begin{tabular}{|c|}
\hline
GW events with $M \lesssim 60 M_{\odot}$\\
\hline
\hspace{0.1cm} \pbox{2.5in}{\raggedright
    \vspace{0.2cm}
    GW151012  \myspace \colspace GW151226  \myspace
    GW170104  \myspace \colspace GW170608  \myspace
    GW170814  \myspace \colspace GW190408\und181802
    GW190412  \myspace \colspace GW190512\und180714
    GW190707\und093326 \colspace GW190708\und232457
    GW190720\und000836 \colspace GW190728\und064510
    GW190828\und065509 \colspace GW190924\und021846
    GW190930\und133541
    \vspace{0.17cm} } \\
\hline
\end{tabular}
\caption{
The remaining 15 binary BH events from GWTC-2 that are not included in
Tab.~\ref{tab:sur_events}.
}
\label{tab:phenom_events}
\end{table}

Figure~\ref{fig:phi_all_comparison_delphi_combined} shows $\delphi$ posteriors
for \PhenomT at $\frefTwentyHz$ for all 46 events. We show the corresponding
\NRSur posteriors for the applicable events. For most events, there are no
strong peaks in $\delphi$ for \PhenomT, in agreement with
Fig.~\ref{fig:phi_all_comparison_delphi}. Interestingly, for GW190521,
\PhenomT has a clear peak at $\delphi \sim 0$ which is absent for \NRSur. This
shows that waveform systematics are already important to consider for current
GW events when measuring the orbital-plane spin angles. While further
investigation is necessary to understand the nature of this peak, we note once
again that \PhenomT can have biases in 1D $\delphi$ distributions as shown in
the bottom-left panels of Fig.~\ref{fig:phi_corner_nr_inj_snr30} and
Fig.~\ref{fig:phi_corner_nr_inj_snr45}.
In these cases, the \PhenomT $\delphi$ posterior is more sharply peaked
(compared to \NRSur) but is also biased.

\clearpage

\bibliography{References}

%merlin.mbs apsrev4-1.bst 2010-07-25 4.21a (PWD, AO, DPC) hacked
%Control: key (0)
%Control: author (0) dotless jnrlst
%Control: editor formatted (1) identically to author
%Control: production of article title (0) allowed
%Control: page (1) range
%Control: year (0) verbatim
%Control: production of eprint (0) enabled
\begin{thebibliography}{47}%
\makeatletter
\providecommand \@ifxundefined [1]{%
 \@ifx{#1\undefined}
}%
\providecommand \@ifnum [1]{%
 \ifnum #1\expandafter \@firstoftwo
 \else \expandafter \@secondoftwo
 \fi
}%
\providecommand \@ifx [1]{%
 \ifx #1\expandafter \@firstoftwo
 \else \expandafter \@secondoftwo
 \fi
}%
\providecommand \natexlab [1]{#1}%
\providecommand \enquote  [1]{``#1''}%
\providecommand \bibnamefont  [1]{#1}%
\providecommand \bibfnamefont [1]{#1}%
\providecommand \citenamefont [1]{#1}%
\providecommand \href@noop [0]{\@secondoftwo}%
\providecommand \href [0]{\begingroup \@sanitize@url \@href}%
\providecommand \@href[1]{\@@startlink{#1}\@@href}%
\providecommand \@@href[1]{\endgroup#1\@@endlink}%
\providecommand \@sanitize@url [0]{\catcode `\\12\catcode `\$12\catcode
  `\&12\catcode `\#12\catcode `\^12\catcode `\_12\catcode `\%12\relax}%
\providecommand \@@startlink[1]{}%
\providecommand \@@endlink[0]{}%
\providecommand \url  [0]{\begingroup\@sanitize@url \@url }%
\providecommand \@url [1]{\endgroup\@href {#1}{\urlprefix }}%
\providecommand \urlprefix  [0]{URL }%
\providecommand \Eprint [0]{\href }%
\providecommand \doibase [0]{http://dx.doi.org/}%
\providecommand \selectlanguage [0]{\@gobble}%
\providecommand \bibinfo  [0]{\@secondoftwo}%
\providecommand \bibfield  [0]{\@secondoftwo}%
\providecommand \translation [1]{[#1]}%
\providecommand \BibitemOpen [0]{}%
\providecommand \bibitemStop [0]{}%
\providecommand \bibitemNoStop [0]{.\EOS\space}%
\providecommand \EOS [0]{\spacefactor3000\relax}%
\providecommand \BibitemShut  [1]{\csname bibitem#1\endcsname}%
\let\auto@bib@innerbib\@empty
%</preamble>
\bibitem [{\citenamefont {Aasi}\ \emph {et~al.}(2015)\citenamefont {Aasi} \emph
  {et~al.}}]{TheLIGOScientific:2014jea}%
  \BibitemOpen
  \bibfield  {author} {\bibinfo {author} {\bibfnamefont {J.}~\bibnamefont
  {Aasi}} \emph {et~al.} (\bibinfo {collaboration} {LIGO Scientific}),\
  }\bibfield  {title} {\enquote {\bibinfo {title} {{Advanced LIGO}},}\ }\href
  {\doibase 10.1088/0264-9381/32/7/074001} {\bibfield  {journal} {\bibinfo
  {journal} {Class. Quant. Grav.}\ }\textbf {\bibinfo {volume} {32}},\ \bibinfo
  {pages} {074001} (\bibinfo {year} {2015})},\ \Eprint
  {http://arxiv.org/abs/1411.4547} {arXiv:1411.4547 [gr-qc]} \BibitemShut
  {NoStop}%
%%CITATION = ARXIV:1411.4547;%%
\bibitem [{\citenamefont {Acernese}\ \emph {et~al.}(2015)\citenamefont
  {Acernese} \emph {et~al.}}]{TheVirgo:2014hva}%
  \BibitemOpen
  \bibfield  {author} {\bibinfo {author} {\bibfnamefont {F.}~\bibnamefont
  {Acernese}} \emph {et~al.} (\bibinfo {collaboration} {Virgo}),\ }\bibfield
  {title} {\enquote {\bibinfo {title} {{Advanced Virgo: a second-generation
  interferometric gravitational wave detector}},}\ }\href {\doibase
  10.1088/0264-9381/32/2/024001} {\bibfield  {journal} {\bibinfo  {journal}
  {Class. Quant. Grav.}\ }\textbf {\bibinfo {volume} {32}},\ \bibinfo {pages}
  {024001} (\bibinfo {year} {2015})},\ \Eprint {http://arxiv.org/abs/1408.3978}
  {arXiv:1408.3978 [gr-qc]} \BibitemShut {NoStop}%
%%CITATION = ARXIV:1408.3978;%%
\bibitem [{\citenamefont {Apostolatos}\ \emph {et~al.}(1994)\citenamefont
  {Apostolatos}, \citenamefont {Cutler}, \citenamefont {Sussman},\ and\
  \citenamefont {Thorne}}]{Apostolatos:1994pre}%
  \BibitemOpen
  \bibfield  {author} {\bibinfo {author} {\bibfnamefont {Theocharis~A.}\
  \bibnamefont {Apostolatos}}, \bibinfo {author} {\bibfnamefont {Curt}\
  \bibnamefont {Cutler}}, \bibinfo {author} {\bibfnamefont {Gerald~J.}\
  \bibnamefont {Sussman}}, \ and\ \bibinfo {author} {\bibfnamefont {Kip~S.}\
  \bibnamefont {Thorne}},\ }\bibfield  {title} {\enquote {\bibinfo {title}
  {Spin-induced orbital precession and its modulation of the gravitational
  waveforms from merging binaries},}\ }\href {\doibase
  10.1103/PhysRevD.49.6274} {\bibfield  {journal} {\bibinfo  {journal} {Phys.
  Rev. D}\ }\textbf {\bibinfo {volume} {49}},\ \bibinfo {pages} {6274--6297}
  (\bibinfo {year} {1994})}\BibitemShut {NoStop}%
\bibitem [{\citenamefont {Kidder}(1995)}]{Kidder:1995zr}%
  \BibitemOpen
  \bibfield  {author} {\bibinfo {author} {\bibfnamefont {Lawrence~E.}\
  \bibnamefont {Kidder}},\ }\bibfield  {title} {\enquote {\bibinfo {title}
  {{Coalescing binary systems of compact objects to postNewtonian 5/2 order. 5.
  Spin effects}},}\ }\href {\doibase 10.1103/PhysRevD.52.821} {\bibfield
  {journal} {\bibinfo  {journal} {Phys. Rev. D}\ }\textbf {\bibinfo {volume}
  {52}},\ \bibinfo {pages} {821--847} (\bibinfo {year} {1995})},\ \Eprint
  {http://arxiv.org/abs/gr-qc/9506022} {arXiv:gr-qc/9506022} \BibitemShut
  {NoStop}%
\bibitem [{\citenamefont {Schnittman}(2004)}]{Schnittman:2004vq}%
  \BibitemOpen
  \bibfield  {author} {\bibinfo {author} {\bibfnamefont {Jeremy~D.}\
  \bibnamefont {Schnittman}},\ }\bibfield  {title} {\enquote {\bibinfo {title}
  {{Spin-orbit resonance and the evolution of compact binary systems}},}\
  }\href {\doibase 10.1103/PhysRevD.70.124020} {\bibfield  {journal} {\bibinfo
  {journal} {Phys. Rev. D}\ }\textbf {\bibinfo {volume} {70}},\ \bibinfo
  {pages} {124020} (\bibinfo {year} {2004})},\ \Eprint
  {http://arxiv.org/abs/astro-ph/0409174} {arXiv:astro-ph/0409174} \BibitemShut
  {NoStop}%
\bibitem [{\citenamefont {Varma}\ \emph {et~al.}(2020)\citenamefont {Varma},
  \citenamefont {Isi},\ and\ \citenamefont {Biscoveanu}}]{Varma:2020nbm}%
  \BibitemOpen
  \bibfield  {author} {\bibinfo {author} {\bibfnamefont {Vijay}\ \bibnamefont
  {Varma}}, \bibinfo {author} {\bibfnamefont {Maximiliano}\ \bibnamefont
  {Isi}}, \ and\ \bibinfo {author} {\bibfnamefont {Sylvia}\ \bibnamefont
  {Biscoveanu}},\ }\bibfield  {title} {\enquote {\bibinfo {title} {{Extracting
  the Gravitational Recoil from Black Hole Merger Signals}},}\ }\href {\doibase
  10.1103/PhysRevLett.124.101104} {\bibfield  {journal} {\bibinfo  {journal}
  {Phys. Rev. Lett.}\ }\textbf {\bibinfo {volume} {124}},\ \bibinfo {pages}
  {101104} (\bibinfo {year} {2020})},\ \Eprint
  {http://arxiv.org/abs/2002.00296} {arXiv:2002.00296 [gr-qc]} \BibitemShut
  {NoStop}%
\bibitem [{\citenamefont {Vitale}\ \emph {et~al.}(2014)\citenamefont {Vitale},
  \citenamefont {Lynch}, \citenamefont {Veitch}, \citenamefont {Raymond},\ and\
  \citenamefont {Sturani}}]{Vitale:2014mka}%
  \BibitemOpen
  \bibfield  {author} {\bibinfo {author} {\bibfnamefont {Salvatore}\
  \bibnamefont {Vitale}}, \bibinfo {author} {\bibfnamefont {Ryan}\ \bibnamefont
  {Lynch}}, \bibinfo {author} {\bibfnamefont {John}\ \bibnamefont {Veitch}},
  \bibinfo {author} {\bibfnamefont {Vivien}\ \bibnamefont {Raymond}}, \ and\
  \bibinfo {author} {\bibfnamefont {Riccardo}\ \bibnamefont {Sturani}},\
  }\bibfield  {title} {\enquote {\bibinfo {title} {{Measuring the spin of black
  holes in binary systems using gravitational waves}},}\ }\href {\doibase
  10.1103/PhysRevLett.112.251101} {\bibfield  {journal} {\bibinfo  {journal}
  {Phys. Rev. Lett.}\ }\textbf {\bibinfo {volume} {112}},\ \bibinfo {pages}
  {251101} (\bibinfo {year} {2014})},\ \Eprint {http://arxiv.org/abs/1403.0129}
  {arXiv:1403.0129 [gr-qc]} \BibitemShut {NoStop}%
\bibitem [{\citenamefont {Schmidt}\ \emph {et~al.}(2015)\citenamefont
  {Schmidt}, \citenamefont {Ohme},\ and\ \citenamefont
  {Hannam}}]{Schmidt:2014iyl}%
  \BibitemOpen
  \bibfield  {author} {\bibinfo {author} {\bibfnamefont {P.}~\bibnamefont
  {Schmidt}}, \bibinfo {author} {\bibfnamefont {F.}~\bibnamefont {Ohme}}, \
  and\ \bibinfo {author} {\bibfnamefont {M.}~\bibnamefont {Hannam}},\
  }\bibfield  {title} {\enquote {\bibinfo {title} {{Towards models of
  gravitational waveforms from generic binaries II: Modelling precession
  effects with a single effective precession parameter}},}\ }\href {\doibase
  10.1103/PhysRevD.91.024043} {\bibfield  {journal} {\bibinfo  {journal} {Phys.
  Rev. D}\ }\textbf {\bibinfo {volume} {91}},\ \bibinfo {pages} {024043}
  (\bibinfo {year} {2015})},\ \Eprint {http://arxiv.org/abs/1408.1810}
  {arXiv:1408.1810 [gr-qc]} \BibitemShut {NoStop}%
%%CITATION = ARXIV:1408.1810;%%
\bibitem [{\citenamefont {Biscoveanu}\ \emph
  {et~al.}(2021{\natexlab{a}})\citenamefont {Biscoveanu}, \citenamefont {Isi},
  \citenamefont {Varma},\ and\ \citenamefont {Vitale}}]{Biscoveanu:2021nvg}%
  \BibitemOpen
  \bibfield  {author} {\bibinfo {author} {\bibfnamefont {Sylvia}\ \bibnamefont
  {Biscoveanu}}, \bibinfo {author} {\bibfnamefont {Maximiliano}\ \bibnamefont
  {Isi}}, \bibinfo {author} {\bibfnamefont {Vijay}\ \bibnamefont {Varma}}, \
  and\ \bibinfo {author} {\bibfnamefont {Salvatore}\ \bibnamefont {Vitale}},\
  }\bibfield  {title} {\enquote {\bibinfo {title} {{Measuring the spins of
  heavy binary black holes}},}\ }\href {\doibase 10.1103/PhysRevD.104.103018}
  {\bibfield  {journal} {\bibinfo  {journal} {Phys. Rev. D}\ }\textbf {\bibinfo
  {volume} {104}},\ \bibinfo {pages} {103018} (\bibinfo {year}
  {2021}{\natexlab{a}})},\ \Eprint {http://arxiv.org/abs/2106.06492}
  {arXiv:2106.06492 [gr-qc]} \BibitemShut {NoStop}%
\bibitem [{\citenamefont {Gerosa}\ \emph {et~al.}(2014)\citenamefont {Gerosa},
  \citenamefont {O'Shaughnessy}, \citenamefont {Kesden}, \citenamefont
  {Berti},\ and\ \citenamefont {Sperhake}}]{Gerosa:2014kta}%
  \BibitemOpen
  \bibfield  {author} {\bibinfo {author} {\bibfnamefont {Davide}\ \bibnamefont
  {Gerosa}}, \bibinfo {author} {\bibfnamefont {Richard}\ \bibnamefont
  {O'Shaughnessy}}, \bibinfo {author} {\bibfnamefont {Michael}\ \bibnamefont
  {Kesden}}, \bibinfo {author} {\bibfnamefont {Emanuele}\ \bibnamefont
  {Berti}}, \ and\ \bibinfo {author} {\bibfnamefont {Ulrich}\ \bibnamefont
  {Sperhake}},\ }\bibfield  {title} {\enquote {\bibinfo {title}
  {{Distinguishing black-hole spin-orbit resonances by their gravitational-wave
  signatures}},}\ }\href {\doibase 10.1103/PhysRevD.89.124025} {\bibfield
  {journal} {\bibinfo  {journal} {Phys. Rev. D}\ }\textbf {\bibinfo {volume}
  {89}},\ \bibinfo {pages} {124025} (\bibinfo {year} {2014})},\ \Eprint
  {http://arxiv.org/abs/1403.7147} {arXiv:1403.7147 [gr-qc]} \BibitemShut
  {NoStop}%
\bibitem [{\citenamefont {Trifir\`o}\ \emph {et~al.}(2016)\citenamefont
  {Trifir\`o}, \citenamefont {O'Shaughnessy}, \citenamefont {Gerosa},
  \citenamefont {Berti}, \citenamefont {Kesden}, \citenamefont {Littenberg},\
  and\ \citenamefont {Sperhake}}]{Trifiro:2015zda}%
  \BibitemOpen
  \bibfield  {author} {\bibinfo {author} {\bibfnamefont {Daniele}\ \bibnamefont
  {Trifir\`o}}, \bibinfo {author} {\bibfnamefont {Richard}\ \bibnamefont
  {O'Shaughnessy}}, \bibinfo {author} {\bibfnamefont {Davide}\ \bibnamefont
  {Gerosa}}, \bibinfo {author} {\bibfnamefont {Emanuele}\ \bibnamefont
  {Berti}}, \bibinfo {author} {\bibfnamefont {Michael}\ \bibnamefont {Kesden}},
  \bibinfo {author} {\bibfnamefont {Tyson}\ \bibnamefont {Littenberg}}, \ and\
  \bibinfo {author} {\bibfnamefont {Ulrich}\ \bibnamefont {Sperhake}},\
  }\bibfield  {title} {\enquote {\bibinfo {title} {{Distinguishing black-hole
  spin-orbit resonances by their gravitational wave signatures. II: Full
  parameter estimation}},}\ }\href {\doibase 10.1103/PhysRevD.93.044071}
  {\bibfield  {journal} {\bibinfo  {journal} {Phys. Rev. D}\ }\textbf {\bibinfo
  {volume} {93}},\ \bibinfo {pages} {044071} (\bibinfo {year} {2016})},\
  \Eprint {http://arxiv.org/abs/1507.05587} {arXiv:1507.05587 [gr-qc]}
  \BibitemShut {NoStop}%
\bibitem [{\citenamefont {Afle}\ \emph {et~al.}(2018)\citenamefont {Afle} \emph
  {et~al.}}]{Afle:2018slw}%
  \BibitemOpen
  \bibfield  {author} {\bibinfo {author} {\bibfnamefont {Chaitanya}\
  \bibnamefont {Afle}} \emph {et~al.},\ }\bibfield  {title} {\enquote {\bibinfo
  {title} {{Detection and characterization of spin-orbit resonances in the
  advanced gravitational wave detectors era}},}\ }\href {\doibase
  10.1103/PhysRevD.98.083014} {\bibfield  {journal} {\bibinfo  {journal} {Phys.
  Rev. D}\ }\textbf {\bibinfo {volume} {98}},\ \bibinfo {pages} {083014}
  (\bibinfo {year} {2018})},\ \Eprint {http://arxiv.org/abs/1803.07695}
  {arXiv:1803.07695 [gr-qc]} \BibitemShut {NoStop}%
\bibitem [{\citenamefont {Abbott}\ \emph
  {et~al.}(2021{\natexlab{a}})\citenamefont {Abbott} \emph
  {et~al.}}]{Abbott:2020niy}%
  \BibitemOpen
  \bibfield  {author} {\bibinfo {author} {\bibfnamefont {R.}~\bibnamefont
  {Abbott}} \emph {et~al.} (\bibinfo {collaboration} {LIGO Scientific,
  Virgo}),\ }\bibfield  {title} {\enquote {\bibinfo {title} {{GWTC-2: Compact
  Binary Coalescences Observed by LIGO and Virgo During the First Half of the
  Third Observing Run}},}\ }\href {\doibase 10.1103/PhysRevX.11.021053}
  {\bibfield  {journal} {\bibinfo  {journal} {Phys. Rev. X}\ }\textbf {\bibinfo
  {volume} {11}},\ \bibinfo {pages} {021053} (\bibinfo {year}
  {2021}{\natexlab{a}})},\ \Eprint {http://arxiv.org/abs/2010.14527}
  {arXiv:2010.14527 [gr-qc]} \BibitemShut {NoStop}%
\bibitem [{spi()}]{spinanglespaperfottnoteL}%
  \BibitemOpen
  \href@noop {} {}\bibinfo {howpublished} {More precisely, we use the coorbital
  frame defined in Ref.~\cite{Varma:2019csw}. In this frame, the $z$-axis is
  along the direction that maximises the power in the (2,2) mode, which is
  taken to be the direction of the orbital angular
  momentum~\cite{Boyle:2011gg}. The $x$-axis is along the line of separation
  from the lighter to the heavier BH, and the $y$-axis completes the
  right-handed triad. Note that this frame is defined using the gauge-invariant
  waveform at future null infinity, rather than the gauge-dependent BH
  trajectories.}\BibitemShut {Stop}%
\bibitem [{\citenamefont {Varma}\ \emph {et~al.}(2019)\citenamefont {Varma},
  \citenamefont {Field}, \citenamefont {Scheel}, \citenamefont {Blackman},
  \citenamefont {Gerosa}, \citenamefont {Stein}, \citenamefont {Kidder},\ and\
  \citenamefont {Pfeiffer}}]{Varma:2019csw}%
  \BibitemOpen
  \bibfield  {author} {\bibinfo {author} {\bibfnamefont {Vijay}\ \bibnamefont
  {Varma}}, \bibinfo {author} {\bibfnamefont {Scott~E.}\ \bibnamefont {Field}},
  \bibinfo {author} {\bibfnamefont {Mark~A.}\ \bibnamefont {Scheel}}, \bibinfo
  {author} {\bibfnamefont {Jonathan}\ \bibnamefont {Blackman}}, \bibinfo
  {author} {\bibfnamefont {Davide}\ \bibnamefont {Gerosa}}, \bibinfo {author}
  {\bibfnamefont {Leo~C.}\ \bibnamefont {Stein}}, \bibinfo {author}
  {\bibfnamefont {Lawrence~E.}\ \bibnamefont {Kidder}}, \ and\ \bibinfo
  {author} {\bibfnamefont {Harald~P.}\ \bibnamefont {Pfeiffer}},\ }\bibfield
  {title} {\enquote {\bibinfo {title} {{Surrogate models for precessing binary
  black hole simulations with unequal masses}},}\ }\href {\doibase
  10.1103/PhysRevResearch.1.033015} {\bibfield  {journal} {\bibinfo  {journal}
  {Phys. Rev. Research.}\ }\textbf {\bibinfo {volume} {1}},\ \bibinfo {pages}
  {033015} (\bibinfo {year} {2019})},\ \Eprint
  {http://arxiv.org/abs/1905.09300} {arXiv:1905.09300 [gr-qc]} \BibitemShut
  {NoStop}%
%%CITATION = ARXIV:1905.09300;%%
\bibitem [{\citenamefont {{Kaplan}}(1949)}]{Kaplan:1949isco}%
  \BibitemOpen
  \bibfield  {author} {\bibinfo {author} {\bibfnamefont {S.~A.}\ \bibnamefont
  {{Kaplan}}},\ }\bibfield  {title} {\enquote {\bibinfo {title} {{On Circular
  orbits in Einsteinian Gravitation theory}},}\ }\href@noop {} {\bibfield
  {journal} {\bibinfo  {journal} {ZhETF Pisma Redaktsiiu}\ }\textbf {\bibinfo
  {volume} {19}},\ \bibinfo {pages} {951--952} (\bibinfo {year}
  {1949})}\BibitemShut {NoStop}%
\bibitem [{\citenamefont {Buonanno}\ \emph {et~al.}(2009)\citenamefont
  {Buonanno}, \citenamefont {Iyer}, \citenamefont {Ochsner}, \citenamefont
  {Pan},\ and\ \citenamefont {Sathyaprakash}}]{Buonanno:2009zt}%
  \BibitemOpen
  \bibfield  {author} {\bibinfo {author} {\bibfnamefont {Alessandra}\
  \bibnamefont {Buonanno}}, \bibinfo {author} {\bibfnamefont {Bala}\
  \bibnamefont {Iyer}}, \bibinfo {author} {\bibfnamefont {Evan}\ \bibnamefont
  {Ochsner}}, \bibinfo {author} {\bibfnamefont {Yi}~\bibnamefont {Pan}}, \ and\
  \bibinfo {author} {\bibfnamefont {B.~S.}\ \bibnamefont {Sathyaprakash}},\
  }\bibfield  {title} {\enquote {\bibinfo {title} {{Comparison of
  post-Newtonian templates for compact binary inspiral signals in
  gravitational-wave detectors}},}\ }\href {\doibase
  10.1103/PhysRevD.80.084043} {\bibfield  {journal} {\bibinfo  {journal} {Phys.
  Rev. D}\ }\textbf {\bibinfo {volume} {80}},\ \bibinfo {pages} {084043}
  (\bibinfo {year} {2009})},\ \Eprint {http://arxiv.org/abs/0907.0700}
  {arXiv:0907.0700 [gr-qc]} \BibitemShut {NoStop}%
\bibitem [{\citenamefont {Farr}\ \emph {et~al.}(2014)\citenamefont {Farr},
  \citenamefont {Ochsner}, \citenamefont {Farr},\ and\ \citenamefont
  {O'Shaughnessy}}]{Farr:2014qka}%
  \BibitemOpen
  \bibfield  {author} {\bibinfo {author} {\bibfnamefont {Benjamin}\
  \bibnamefont {Farr}}, \bibinfo {author} {\bibfnamefont {Evan}\ \bibnamefont
  {Ochsner}}, \bibinfo {author} {\bibfnamefont {Will~M.}\ \bibnamefont {Farr}},
  \ and\ \bibinfo {author} {\bibfnamefont {Richard}\ \bibnamefont
  {O'Shaughnessy}},\ }\bibfield  {title} {\enquote {\bibinfo {title} {{A more
  effective coordinate system for parameter estimation of precessing compact
  binaries from gravitational waves}},}\ }\href {\doibase
  10.1103/PhysRevD.90.024018} {\bibfield  {journal} {\bibinfo  {journal} {Phys.
  Rev. D}\ }\textbf {\bibinfo {volume} {90}},\ \bibinfo {pages} {024018}
  (\bibinfo {year} {2014})},\ \Eprint {http://arxiv.org/abs/1404.7070}
  {arXiv:1404.7070 [gr-qc]} \BibitemShut {NoStop}%
\bibitem [{\citenamefont {Abbott}\ \emph {et~al.}(2019)\citenamefont {Abbott}
  \emph {et~al.}}]{LIGOScientific:2018mvr}%
  \BibitemOpen
  \bibfield  {author} {\bibinfo {author} {\bibfnamefont {B.~P.}\ \bibnamefont
  {Abbott}} \emph {et~al.} (\bibinfo {collaboration} {LIGO Scientific,
  Virgo}),\ }\bibfield  {title} {\enquote {\bibinfo {title} {{GWTC-1: A
  Gravitational-Wave Transient Catalog of Compact Binary Mergers Observed by
  LIGO and Virgo during the First and Second Observing Runs}},}\ }\href
  {\doibase 10.1103/PhysRevX.9.031040} {\bibfield  {journal} {\bibinfo
  {journal} {Phys. Rev.}\ }\textbf {\bibinfo {volume} {X9}},\ \bibinfo {pages}
  {031040} (\bibinfo {year} {2019})},\ \Eprint
  {http://arxiv.org/abs/1811.12907} {arXiv:1811.12907 [astro-ph.HE]}
  \BibitemShut {NoStop}%
%%CITATION = ARXIV:1811.12907;%%
\bibitem [{\citenamefont {Abbott}\ \emph
  {et~al.}(2021{\natexlab{b}})\citenamefont {Abbott} \emph
  {et~al.}}]{GWOSC_paper}%
  \BibitemOpen
  \bibfield  {author} {\bibinfo {author} {\bibfnamefont {Rich}\ \bibnamefont
  {Abbott}} \emph {et~al.} (\bibinfo {collaboration} {LIGO Scientific,
  Virgo}),\ }\bibfield  {title} {\enquote {\bibinfo {title} {{Open data from
  the first and second observing runs of Advanced LIGO and Advanced Virgo}},}\
  }\href {\doibase 10.1016/j.softx.2021.100658} {\bibfield  {journal} {\bibinfo
   {journal} {SoftwareX}\ }\textbf {\bibinfo {volume} {13}},\ \bibinfo {pages}
  {100658} (\bibinfo {year} {2021}{\natexlab{b}})},\ \Eprint
  {http://arxiv.org/abs/1912.11716} {arXiv:1912.11716 [gr-qc]} \BibitemShut
  {NoStop}%
\bibitem [{\citenamefont {{LIGO Scientific Collaboration and Virgo
  Collaboration}}(2018)}]{GWOSC:GWTC}%
  \BibitemOpen
  \bibfield  {author} {\bibinfo {author} {\bibnamefont {{LIGO Scientific
  Collaboration and Virgo Collaboration}}},\ }\href@noop {} {\enquote {\bibinfo
  {title} {{GWTC-1}},}\ }\bibinfo {howpublished}
  {\href{https://doi.org/10.7935/82H3-HH23}{https://doi.org/10.7935/82H3-HH23}}
  (\bibinfo {year} {2018})\BibitemShut {NoStop}%
\bibitem [{\citenamefont {{LIGO Scientific Collaboration and Virgo
  Collaboration}}(2020)}]{GWOSC:GWTC-2}%
  \BibitemOpen
  \bibfield  {author} {\bibinfo {author} {\bibnamefont {{LIGO Scientific
  Collaboration and Virgo Collaboration}}},\ }\href@noop {} {\enquote {\bibinfo
  {title} {{GWTC-2}},}\ }\bibinfo {howpublished}
  {\href{https://doi.org/10.7935/99gf-ax93}{https://doi.org/10.7935/99gf-ax93}}
  (\bibinfo {year} {2020})\BibitemShut {NoStop}%
\bibitem [{\citenamefont {Varma}\ \emph {et~al.}(2022)\citenamefont {Varma},
  \citenamefont {Biscoveanu}, \citenamefont {Isi}, \citenamefont {Farr},\ and\
  \citenamefont {Vitale}}]{Varma:2021xbh}%
  \BibitemOpen
  \bibfield  {author} {\bibinfo {author} {\bibfnamefont {Vijay}\ \bibnamefont
  {Varma}}, \bibinfo {author} {\bibfnamefont {Sylvia}\ \bibnamefont
  {Biscoveanu}}, \bibinfo {author} {\bibfnamefont {Maximiliano}\ \bibnamefont
  {Isi}}, \bibinfo {author} {\bibfnamefont {Will~M.}\ \bibnamefont {Farr}}, \
  and\ \bibinfo {author} {\bibfnamefont {Salvatore}\ \bibnamefont {Vitale}},\
  }\bibfield  {title} {\enquote {\bibinfo {title} {{Hints of spin-orbit
  resonances in the binary black hole population}},}\ }\href {\doibase
  10.1103/PhysRevLett.128.031101} {\bibfield  {journal} {\bibinfo  {journal}
  {Phys. Rev. Lett.}\ }\textbf {\bibinfo {volume} {128}},\ \bibinfo {pages}
  {031101} (\bibinfo {year} {2022})},\ \Eprint
  {http://arxiv.org/abs/2107.09693} {arXiv:2107.09693 [astro-ph.HE]}
  \BibitemShut {NoStop}%
\bibitem [{\citenamefont {{Thrane}}\ and\ \citenamefont
  {{Talbot}}(2019)}]{Thrane:2019pe}%
  \BibitemOpen
  \bibfield  {author} {\bibinfo {author} {\bibfnamefont {Eric}\ \bibnamefont
  {{Thrane}}}\ and\ \bibinfo {author} {\bibfnamefont {Colm}\ \bibnamefont
  {{Talbot}}},\ }\bibfield  {title} {\enquote {\bibinfo {title} {{An
  introduction to Bayesian inference in gravitational-wave astronomy: Parameter
  estimation, model selection, and hierarchical models}},}\ }\href {\doibase
  10.1017/pasa.2019.2} {\bibfield  {journal} {\bibinfo  {journal} {Publications
  of the Astronomical Society of Australia}\ }\textbf {\bibinfo {volume}
  {36}},\ \bibinfo {eid} {e010} (\bibinfo {year} {2019})},\ \Eprint
  {http://arxiv.org/abs/1809.02293} {arXiv:1809.02293 [astro-ph.IM]}
  \BibitemShut {NoStop}%
\bibitem [{\citenamefont {Walker}\ \emph {et~al.}(2021)\citenamefont {Walker},
  \citenamefont {Varma},\ and\ \citenamefont {Lovelace}}]{Beth:2021_inprep}%
  \BibitemOpen
  \bibfield  {author} {\bibinfo {author} {\bibfnamefont {Marissa}\ \bibnamefont
  {Walker}}, \bibinfo {author} {\bibfnamefont {Vijay}\ \bibnamefont {Varma}}, \
  and\ \bibinfo {author} {\bibfnamefont {Geoffrey}\ \bibnamefont {Lovelace}},\
  }\bibfield  {title} {\enquote {\bibinfo {title} {{Extending numerical
  relativity surrogate models to near extremal spins}},}\ }\href@noop {} {\
  (\bibinfo {year} {2021})},\ \bibinfo {note} {in preparation}\BibitemShut
  {NoStop}%
\bibitem [{\citenamefont {Smith}\ \emph {et~al.}(2020)\citenamefont {Smith},
  \citenamefont {Ashton}, \citenamefont {Vajpeyi},\ and\ \citenamefont
  {Talbot}}]{Smith:2019ucc}%
  \BibitemOpen
  \bibfield  {author} {\bibinfo {author} {\bibfnamefont {Rory~J.E.}\
  \bibnamefont {Smith}}, \bibinfo {author} {\bibfnamefont {Gregory}\
  \bibnamefont {Ashton}}, \bibinfo {author} {\bibfnamefont {Avi}\ \bibnamefont
  {Vajpeyi}}, \ and\ \bibinfo {author} {\bibfnamefont {Colm}\ \bibnamefont
  {Talbot}},\ }\bibfield  {title} {\enquote {\bibinfo {title} {{Massively
  parallel Bayesian inference for transient gravitational-wave astronomy}},}\
  }\href {\doibase 10.1093/mnras/staa2483} {\bibfield  {journal} {\bibinfo
  {journal} {Mon. Not. Roy. Astron. Soc.}\ }\textbf {\bibinfo {volume} {498}},\
  \bibinfo {pages} {4492--4502} (\bibinfo {year} {2020})},\ \Eprint
  {http://arxiv.org/abs/1909.11873} {arXiv:1909.11873 [gr-qc]} \BibitemShut
  {NoStop}%
\bibitem [{\citenamefont {{Speagle}}(2020)}]{Speagle:2019dynesty}%
  \BibitemOpen
  \bibfield  {author} {\bibinfo {author} {\bibfnamefont {Joshua~S.}\
  \bibnamefont {{Speagle}}},\ }\bibfield  {title} {\enquote {\bibinfo {title}
  {{DYNESTY: a dynamic nested sampling package for estimating Bayesian
  posteriors and evidences}},}\ }\href {\doibase 10.1093/mnras/staa278}
  {\bibfield  {journal} {\bibinfo  {journal} {Monthly Notices of the Royal
  Astronomical Society}\ }\textbf {\bibinfo {volume} {493}},\ \bibinfo {pages}
  {3132--3158} (\bibinfo {year} {2020})},\ \Eprint
  {http://arxiv.org/abs/1904.02180} {arXiv:1904.02180 [astro-ph.IM]}
  \BibitemShut {NoStop}%
\bibitem [{\citenamefont {Romero-Shaw}\ \emph {et~al.}(2020)\citenamefont
  {Romero-Shaw} \emph {et~al.}}]{Romero-Shaw:2020owr}%
  \BibitemOpen
  \bibfield  {author} {\bibinfo {author} {\bibfnamefont {I.~M.}\ \bibnamefont
  {Romero-Shaw}} \emph {et~al.},\ }\bibfield  {title} {\enquote {\bibinfo
  {title} {{Bayesian inference for compact binary coalescences with bilby:
  validation and application to the first LIGO\textendash{}Virgo
  gravitational-wave transient catalogue}},}\ }\href {\doibase
  10.1093/mnras/staa2850} {\bibfield  {journal} {\bibinfo  {journal} {Mon. Not.
  Roy. Astron. Soc.}\ }\textbf {\bibinfo {volume} {499}},\ \bibinfo {pages}
  {3295--3319} (\bibinfo {year} {2020})},\ \Eprint
  {http://arxiv.org/abs/2006.00714} {arXiv:2006.00714 [astro-ph.IM]}
  \BibitemShut {NoStop}%
\bibitem [{\citenamefont {Boyle}\ \emph {et~al.}(2019)\citenamefont {Boyle}
  \emph {et~al.}}]{Boyle:2019kee}%
  \BibitemOpen
  \bibfield  {author} {\bibinfo {author} {\bibfnamefont {Michael}\ \bibnamefont
  {Boyle}} \emph {et~al.},\ }\bibfield  {title} {\enquote {\bibinfo {title}
  {{The SXS Collaboration catalog of binary black hole simulations}},}\ }\href
  {\doibase 10.1088/1361-6382/ab34e2} {\bibfield  {journal} {\bibinfo
  {journal} {Class. Quant. Grav.}\ }\textbf {\bibinfo {volume} {36}},\ \bibinfo
  {pages} {195006} (\bibinfo {year} {2019})},\ \Eprint
  {http://arxiv.org/abs/1904.04831} {arXiv:1904.04831 [gr-qc]} \BibitemShut
  {NoStop}%
%%CITATION = ARXIV:1904.04831;%%
\bibitem [{\citenamefont {Blackman}\ \emph {et~al.}(2017)\citenamefont
  {Blackman}, \citenamefont {Field}, \citenamefont {Scheel}, \citenamefont
  {Galley}, \citenamefont {Ott}, \citenamefont {Boyle}, \citenamefont {Kidder},
  \citenamefont {Pfeiffer},\ and\ \citenamefont
  {Szilágyi}}]{Blackman:2017pcm}%
  \BibitemOpen
  \bibfield  {author} {\bibinfo {author} {\bibfnamefont {Jonathan}\
  \bibnamefont {Blackman}}, \bibinfo {author} {\bibfnamefont {Scott~E.}\
  \bibnamefont {Field}}, \bibinfo {author} {\bibfnamefont {Mark~A.}\
  \bibnamefont {Scheel}}, \bibinfo {author} {\bibfnamefont {Chad~R.}\
  \bibnamefont {Galley}}, \bibinfo {author} {\bibfnamefont {Christian~D.}\
  \bibnamefont {Ott}}, \bibinfo {author} {\bibfnamefont {Michael}\ \bibnamefont
  {Boyle}}, \bibinfo {author} {\bibfnamefont {Lawrence~E.}\ \bibnamefont
  {Kidder}}, \bibinfo {author} {\bibfnamefont {Harald~P.}\ \bibnamefont
  {Pfeiffer}}, \ and\ \bibinfo {author} {\bibfnamefont {Béla}\ \bibnamefont
  {Szilágyi}},\ }\bibfield  {title} {\enquote {\bibinfo {title} {{Numerical
  relativity waveform surrogate model for generically precessing binary black
  hole mergers}},}\ }\href {\doibase 10.1103/PhysRevD.96.024058} {\bibfield
  {journal} {\bibinfo  {journal} {Phys. Rev.}\ }\textbf {\bibinfo {volume}
  {D96}},\ \bibinfo {pages} {024058} (\bibinfo {year} {2017})},\ \Eprint
  {http://arxiv.org/abs/1705.07089} {arXiv:1705.07089 [gr-qc]} \BibitemShut
  {NoStop}%
%%CITATION = ARXIV:1705.07089;%%
\bibitem [{\citenamefont {Estell\'es}\ \emph {et~al.}(2021)\citenamefont
  {Estell\'es}, \citenamefont {Colleoni}, \citenamefont {Garc\'\i{}a-Quir\'os},
  \citenamefont {Husa}, \citenamefont {Keitel}, \citenamefont {Mateu-Lucena},
  \citenamefont {Planas},\ and\ \citenamefont
  {Ramos-Buades}}]{Estelles:2021gvs}%
  \BibitemOpen
  \bibfield  {author} {\bibinfo {author} {\bibfnamefont {H\'ector}\
  \bibnamefont {Estell\'es}}, \bibinfo {author} {\bibfnamefont {Marta}\
  \bibnamefont {Colleoni}}, \bibinfo {author} {\bibfnamefont {Cecilio}\
  \bibnamefont {Garc\'\i{}a-Quir\'os}}, \bibinfo {author} {\bibfnamefont
  {Sascha}\ \bibnamefont {Husa}}, \bibinfo {author} {\bibfnamefont {David}\
  \bibnamefont {Keitel}}, \bibinfo {author} {\bibfnamefont {Maite}\
  \bibnamefont {Mateu-Lucena}}, \bibinfo {author} {\bibfnamefont {Maria
  de~Lluc}\ \bibnamefont {Planas}}, \ and\ \bibinfo {author} {\bibfnamefont
  {Antoni}\ \bibnamefont {Ramos-Buades}},\ }\bibfield  {title} {\enquote
  {\bibinfo {title} {{New twists in compact binary waveform modelling: a fast
  time domain model for precession}},}\ }\href@noop {} {\  (\bibinfo {year}
  {2021})},\ \Eprint {http://arxiv.org/abs/2105.05872} {arXiv:2105.05872
  [gr-qc]} \BibitemShut {NoStop}%
\bibitem [{\citenamefont {Abbott}\ \emph {et~al.}(2020)\citenamefont {Abbott}
  \emph {et~al.}}]{Abbott:2020tfl}%
  \BibitemOpen
  \bibfield  {author} {\bibinfo {author} {\bibfnamefont {R.}~\bibnamefont
  {Abbott}} \emph {et~al.} (\bibinfo {collaboration} {LIGO Scientific,
  Virgo}),\ }\bibfield  {title} {\enquote {\bibinfo {title} {{GW190521: A
  Binary Black Hole Merger with a Total Mass of $150 ~ M_{\odot}$}},}\ }\href
  {\doibase 10.1103/PhysRevLett.125.101102} {\bibfield  {journal} {\bibinfo
  {journal} {Phys. Rev. Lett.}\ }\textbf {\bibinfo {volume} {125}},\ \bibinfo
  {pages} {101102} (\bibinfo {year} {2020})},\ \Eprint
  {http://arxiv.org/abs/2009.01075} {arXiv:2009.01075 [gr-qc]} \BibitemShut
  {NoStop}%
\bibitem [{\citenamefont {Biscoveanu}\ \emph
  {et~al.}(2021{\natexlab{b}})\citenamefont {Biscoveanu}, \citenamefont {Isi},
  \citenamefont {Vitale},\ and\ \citenamefont {Varma}}]{Biscoveanu:2020are}%
  \BibitemOpen
  \bibfield  {author} {\bibinfo {author} {\bibfnamefont {Sylvia}\ \bibnamefont
  {Biscoveanu}}, \bibinfo {author} {\bibfnamefont {Maximiliano}\ \bibnamefont
  {Isi}}, \bibinfo {author} {\bibfnamefont {Salvatore}\ \bibnamefont {Vitale}},
  \ and\ \bibinfo {author} {\bibfnamefont {Vijay}\ \bibnamefont {Varma}},\
  }\bibfield  {title} {\enquote {\bibinfo {title} {{New Spin on LIGO-Virgo
  Binary Black Holes}},}\ }\href {\doibase 10.1103/PhysRevLett.126.171103}
  {\bibfield  {journal} {\bibinfo  {journal} {Phys. Rev. Lett.}\ }\textbf
  {\bibinfo {volume} {126}},\ \bibinfo {pages} {171103} (\bibinfo {year}
  {2021}{\natexlab{b}})},\ \Eprint {http://arxiv.org/abs/2007.09156}
  {arXiv:2007.09156 [astro-ph.HE]} \BibitemShut {NoStop}%
\bibitem [{\citenamefont {Abbott}\ \emph
  {et~al.}(2021{\natexlab{c}})\citenamefont {Abbott} \emph
  {et~al.}}]{Abbott:2020gyp}%
  \BibitemOpen
  \bibfield  {author} {\bibinfo {author} {\bibfnamefont {R.}~\bibnamefont
  {Abbott}} \emph {et~al.} (\bibinfo {collaboration} {LIGO Scientific,
  Virgo}),\ }\bibfield  {title} {\enquote {\bibinfo {title} {{Population
  Properties of Compact Objects from the Second LIGO-Virgo Gravitational-Wave
  Transient Catalog}},}\ }\href {\doibase 10.3847/2041-8213/abe949} {\bibfield
  {journal} {\bibinfo  {journal} {Astrophys. J. Lett.}\ }\textbf {\bibinfo
  {volume} {913}},\ \bibinfo {pages} {L7} (\bibinfo {year}
  {2021}{\natexlab{c}})},\ \Eprint {http://arxiv.org/abs/2010.14533}
  {arXiv:2010.14533 [astro-ph.HE]} \BibitemShut {NoStop}%
\bibitem [{\citenamefont {Finn}(1992)}]{Finn:1992wt}%
  \BibitemOpen
  \bibfield  {author} {\bibinfo {author} {\bibfnamefont {Lee~S.}\ \bibnamefont
  {Finn}},\ }\bibfield  {title} {\enquote {\bibinfo {title} {{Detection,
  measurement and gravitational radiation}},}\ }\href {\doibase
  10.1103/PhysRevD.46.5236} {\bibfield  {journal} {\bibinfo  {journal} {Phys.
  Rev. D}\ }\textbf {\bibinfo {volume} {46}},\ \bibinfo {pages} {5236--5249}
  (\bibinfo {year} {1992})},\ \Eprint {http://arxiv.org/abs/gr-qc/9209010}
  {arXiv:gr-qc/9209010} \BibitemShut {NoStop}%
\bibitem [{\citenamefont {Cutler}\ and\ \citenamefont
  {Flanagan}(1994)}]{Cutler:1994ys}%
  \BibitemOpen
  \bibfield  {author} {\bibinfo {author} {\bibfnamefont {Curt}\ \bibnamefont
  {Cutler}}\ and\ \bibinfo {author} {\bibfnamefont {Eanna~E.}\ \bibnamefont
  {Flanagan}},\ }\bibfield  {title} {\enquote {\bibinfo {title} {{Gravitational
  waves from merging compact binaries: How accurately can one extract the
  binary's parameters from the inspiral wave form?}}}\ }\href {\doibase
  10.1103/PhysRevD.49.2658} {\bibfield  {journal} {\bibinfo  {journal} {Phys.
  Rev. D}\ }\textbf {\bibinfo {volume} {49}},\ \bibinfo {pages} {2658--2697}
  (\bibinfo {year} {1994})},\ \Eprint {http://arxiv.org/abs/gr-qc/9402014}
  {arXiv:gr-qc/9402014} \BibitemShut {NoStop}%
\bibitem [{\citenamefont {{LIGO Scientific
  Collaboration}}(2018)}]{aLIGODesignNoiseCurve}%
  \BibitemOpen
  \bibfield  {author} {\bibinfo {author} {\bibnamefont {{LIGO Scientific
  Collaboration}}},\ }\href@noop {} {\emph {\bibinfo {title} {Updated Advanced
  LIGO sensitivity design curve}}},\ \bibinfo {type} {Tech. Rep.}\ (\bibinfo
  {year} {2018})\ \bibinfo {note}
  {\url{https://dcc.ligo.org/LIGO-T1800044/public}}\BibitemShut {NoStop}%
\bibitem [{\citenamefont {Ma}\ \emph {et~al.}(2021)\citenamefont {Ma},
  \citenamefont {Giesler}, \citenamefont {Varma}, \citenamefont {Scheel},\ and\
  \citenamefont {Chen}}]{Ma:2021znq}%
  \BibitemOpen
  \bibfield  {author} {\bibinfo {author} {\bibfnamefont {Sizheng}\ \bibnamefont
  {Ma}}, \bibinfo {author} {\bibfnamefont {Matthew}\ \bibnamefont {Giesler}},
  \bibinfo {author} {\bibfnamefont {Vijay}\ \bibnamefont {Varma}}, \bibinfo
  {author} {\bibfnamefont {Mark~A.}\ \bibnamefont {Scheel}}, \ and\ \bibinfo
  {author} {\bibfnamefont {Yanbei}\ \bibnamefont {Chen}},\ }\bibfield  {title}
  {\enquote {\bibinfo {title} {{Universal features of gravitational waves
  emitted by superkick binary black hole systems}},}\ }\href {\doibase
  10.1103/PhysRevD.104.084003} {\bibfield  {journal} {\bibinfo  {journal}
  {Phys. Rev. D}\ }\textbf {\bibinfo {volume} {104}},\ \bibinfo {pages}
  {084003} (\bibinfo {year} {2021})},\ \Eprint
  {http://arxiv.org/abs/2107.04890} {arXiv:2107.04890 [gr-qc]} \BibitemShut
  {NoStop}%
\bibitem [{\citenamefont {Cram{\'e}r}(1999)}]{Cramer1999_Fisher}%
  \BibitemOpen
  \bibfield  {author} {\bibinfo {author} {\bibfnamefont {H.}~\bibnamefont
  {Cram{\'e}r}},\ }\href {https://books.google.co.in/books?id=CRTKKaJO0DYC}
  {\emph {\bibinfo {title} {Mathematical Methods of Statistics}}},\ Princeton
  Mathematical Series\ (\bibinfo  {publisher} {Princeton University Press},\
  \bibinfo {year} {1999})\BibitemShut {NoStop}%
\bibitem [{\citenamefont {Radhakrishna~Rao}(1945)}]{Rao1945_Fisher}%
  \BibitemOpen
  \bibfield  {author} {\bibinfo {author} {\bibfnamefont {C.}~\bibnamefont
  {Radhakrishna~Rao}},\ }\bibfield  {title} {\enquote {\bibinfo {title}
  {Information and the accuracy attainable in the estimation of statistical
  parameters},}\ }\href@noop {} {\bibfield  {journal} {\bibinfo  {journal}
  {Bull. Calcutta Math. Soc.}\ }\textbf {\bibinfo {volume} {37}},\ \bibinfo
  {pages} {81--91} (\bibinfo {year} {1945})}\BibitemShut {NoStop}%
\bibitem [{\citenamefont {Vallisneri}(2008)}]{Vallisneri:2007ev}%
  \BibitemOpen
  \bibfield  {author} {\bibinfo {author} {\bibfnamefont {Michele}\ \bibnamefont
  {Vallisneri}},\ }\bibfield  {title} {\enquote {\bibinfo {title} {{Use and
  abuse of the Fisher information matrix in the assessment of
  gravitational-wave parameter-estimation prospects}},}\ }\href {\doibase
  10.1103/PhysRevD.77.042001} {\bibfield  {journal} {\bibinfo  {journal} {Phys.
  Rev. D}\ }\textbf {\bibinfo {volume} {77}},\ \bibinfo {pages} {042001}
  (\bibinfo {year} {2008})},\ \Eprint {http://arxiv.org/abs/gr-qc/0703086}
  {arXiv:gr-qc/0703086} \BibitemShut {NoStop}%
\bibitem [{\citenamefont {{SXS Collaboration}}()}]{SXSCatalog}%
  \BibitemOpen
  \bibfield  {author} {\bibinfo {author} {\bibnamefont {{SXS Collaboration}}},\
  }\href@noop {} {\enquote {\bibinfo {title} {The {SXS} collaboration catalog
  of gravitational waveforms},}\ }\bibinfo {note}
  {\url{http://www.black-holes.org/waveforms}}\BibitemShut {NoStop}%
\bibitem [{\citenamefont {Mroue}\ \emph {et~al.}(2013)\citenamefont {Mroue}
  \emph {et~al.}}]{Mroue:2013xna}%
  \BibitemOpen
  \bibfield  {author} {\bibinfo {author} {\bibfnamefont {Abdul~H.}\
  \bibnamefont {Mroue}} \emph {et~al.},\ }\bibfield  {title} {\enquote
  {\bibinfo {title} {{Catalog of 174 Binary Black Hole Simulations for
  Gravitational Wave Astronomy}},}\ }\href {\doibase
  10.1103/PhysRevLett.111.241104} {\bibfield  {journal} {\bibinfo  {journal}
  {Phys. Rev. Lett.}\ }\textbf {\bibinfo {volume} {111}},\ \bibinfo {pages}
  {241104} (\bibinfo {year} {2013})},\ \Eprint {http://arxiv.org/abs/1304.6077}
  {arXiv:1304.6077 [gr-qc]} \BibitemShut {NoStop}%
%%CITATION = ARXIV:1304.6077;%%
\bibitem [{\citenamefont {Pratten}\ \emph {et~al.}(2021)\citenamefont {Pratten}
  \emph {et~al.}}]{Pratten:2020ceb}%
  \BibitemOpen
  \bibfield  {author} {\bibinfo {author} {\bibfnamefont {Geraint}\ \bibnamefont
  {Pratten}} \emph {et~al.},\ }\bibfield  {title} {\enquote {\bibinfo {title}
  {{Computationally efficient models for the dominant and subdominant harmonic
  modes of precessing binary black holes}},}\ }\href {\doibase
  10.1103/PhysRevD.103.104056} {\bibfield  {journal} {\bibinfo  {journal}
  {Phys. Rev. D}\ }\textbf {\bibinfo {volume} {103}},\ \bibinfo {pages}
  {104056} (\bibinfo {year} {2021})},\ \Eprint
  {http://arxiv.org/abs/2004.06503} {arXiv:2004.06503 [gr-qc]} \BibitemShut
  {NoStop}%
\bibitem [{\citenamefont {Ossokine}\ \emph {et~al.}(2020)\citenamefont
  {Ossokine} \emph {et~al.}}]{Ossokine:2020kjp}%
  \BibitemOpen
  \bibfield  {author} {\bibinfo {author} {\bibfnamefont {Serguei}\ \bibnamefont
  {Ossokine}} \emph {et~al.},\ }\bibfield  {title} {\enquote {\bibinfo {title}
  {{Multipolar Effective-One-Body Waveforms for Precessing Binary Black Holes:
  Construction and Validation}},}\ }\href {\doibase
  10.1103/PhysRevD.102.044055} {\bibfield  {journal} {\bibinfo  {journal}
  {Phys. Rev. D}\ }\textbf {\bibinfo {volume} {102}},\ \bibinfo {pages}
  {044055} (\bibinfo {year} {2020})},\ \Eprint
  {http://arxiv.org/abs/2004.09442} {arXiv:2004.09442 [gr-qc]} \BibitemShut
  {NoStop}%
\bibitem [{\citenamefont {Collaboration}\ and\ \citenamefont
  {Collaboration}()}]{GW_open_science_center}%
  \BibitemOpen
  \bibfield  {author} {\bibinfo {author} {\bibfnamefont {LIGO~Scientific}\
  \bibnamefont {Collaboration}}\ and\ \bibinfo {author} {\bibfnamefont {Virgo}\
  \bibnamefont {Collaboration}},\ }\bibfield  {title} {\enquote {\bibinfo
  {title} {{Gravitational Wave Open Science Center}},}\ }\href@noop {} {\
  }\bibinfo {note} {\url{https://www.gw-openscience.org}}\BibitemShut {NoStop}%
\bibitem [{\citenamefont {Boyle}\ \emph {et~al.}(2011)\citenamefont {Boyle},
  \citenamefont {Owen},\ and\ \citenamefont {Pfeiffer}}]{Boyle:2011gg}%
  \BibitemOpen
  \bibfield  {author} {\bibinfo {author} {\bibfnamefont {Michael}\ \bibnamefont
  {Boyle}}, \bibinfo {author} {\bibfnamefont {Robert}\ \bibnamefont {Owen}}, \
  and\ \bibinfo {author} {\bibfnamefont {Harald~P.}\ \bibnamefont {Pfeiffer}},\
  }\bibfield  {title} {\enquote {\bibinfo {title} {{A geometric approach to the
  precession of compact binaries}},}\ }\href {\doibase
  10.1103/PhysRevD.84.124011} {\bibfield  {journal} {\bibinfo  {journal} {Phys.
  Rev.}\ }\textbf {\bibinfo {volume} {D84}},\ \bibinfo {pages} {124011}
  (\bibinfo {year} {2011})},\ \Eprint {http://arxiv.org/abs/1110.2965}
  {arXiv:1110.2965 [gr-qc]} \BibitemShut {NoStop}%
%%CITATION = ARXIV:1110.2965;%%
\end{thebibliography}%

\end{document}